\documentclass[%
a4paper,
prd,
twocolumn,
reprint,
superscriptaddress,
preprintnumbers,
nofootinbib,
nobibnotes,
amsmath,
amssymb,
aps,
floatfix
]{revtex4-2}
\usepackage{array}[=2016-10-06]
\usepackage[T1]{fontenc}
\usepackage[english]{babel}
\usepackage{booktabs}
\usepackage{amsfonts,mathrsfs,bm}
\usepackage{color,graphics,graphicx,xcolor}
\usepackage{dcolumn,multirow}
\usepackage{placeins}
\definecolor{xlinkcolor}{cmyk}{1,1,0,0}
\usepackage[bookmarks=true,
unicode=true,
pdfnewwindow=true,
colorlinks=true,
linkcolor=xlinkcolor, citecolor=xlinkcolor, filecolor=xlinkcolor, urlcolor=xlinkcolor,
final=true
]{hyperref}
\newcommand{\black}{\color{black}}
\newcommand{\Msun}{M_\odot}
\newcommand{\Rsun}{R_\odot}
\newcommand{\Lsun}{L_\odot}
\newcommand{\dd}{\mathrm{d}}
\newcommand{\gae}{g_{ae}}
\newcommand{\gag}{g_{a\gamma}}
\newcommand{\mfp}{\lambda_{\rm mfp}}
\newcommand{\eps}{\varepsilon}

\newcommand{\Gyr}{\,\mathrm{Gyr}}
\newcommand{\Earth}{\oplus}

\begin{document}

\preprint{INR-TH-2026-014}

\title[Searches for new particle physics by methods of classical and neutrino astronomy]
{Searches for new particle physics by methods of classical and neutrino astronomy}
\author{Sergey\,V.\,Troitsky}
\thanks{Presented at the Scientific session of the Physics Division of the Russian Academy of Sciences, ``Astrophysics, cosmology and space physics'', Moscow, September 30, 2026.}
\affiliation{Institute for Nuclear
Research of the Russian Academy of Sciences, 60th October Anniversary Prospect 7a, Moscow 117312, Russia}
\affiliation{Lomonosov Moscow State University, 1 Leninskie Gory, Building 2, Moscow 119991, Russia}
\thanks{e-mail: st@inr.ac.ru}
\begin{abstract}
Selected astrophysical methods for testing particle-physics models are reviewed. Stellar structure and evolution constrain additional channels of energy loss and transport; high-energy photon propagation probes photon mixing with new particles and modifications of electron–positron pair production; large-volume neutrino telescopes are used to search for magnetic monopoles, neutrinos from superheavy dark-matter decay, and non-standard neutrino propagation. Recent applications of these methods are discussed, involving the tip of the red-giant branch in the color–luminosity diagram, gamma-ray burst 221009A, and data from the Baikal Gigaton Volume Detector (Baikal-GVD) and the IceCube experiment. In each case, particle-physics parameters are inferred through a forward model that accounts for the response to new physics, propagation, detection, and nuisance parameters.
\end{abstract}

\maketitle
\tableofcontents
\section{Introduction}
\label{sec:introduction}

The Standard Model (SM) does not account for neutrino masses, dark matter, or the cosmological matter--antimatter asymmetry.  Astrophysical observations test its extensions at energies, densities, fields, baselines, and exposures that are inaccessible in the laboratory \cite{Raffelt1996,Troitsky2012UFN}.  The measured quantity is, however, generally not a Lagrangian parameter.  It is a stellar luminosity, an attenuated spectrum, an event rate, or a flavor ratio; its interpretation requires a model of the relevant astrophysical system.

Both discoveries and exclusion limits follow this logic. The solar-neutrino problem, first revealed by the chlorine experiment and subsequently confirmed by Kamiokande and the gallium experiments \cite{Davis1968Solar,Hirata1989Solar,Anselmann1992GALLEX,Abdurashitov1994SAGE}, culminated in direct evidence for flavor conversion from SNO \cite{Ahmad2002SNO}. The neutrino observations associated with supernova 1987A \cite{Hirata1987SN,Bionta1987SN,Alekseev1987SN,Aglietta1987LSD} subsequently provided constraints on neutrino properties and on additional free-streaming particles \cite{ArnettRosner1987,RaffeltSeckel1988}.
The examples below concern three complementary classes of observables: stellar structure and evolution; propagation of high-energy photons; and direct or indirect new-physics searches with large-volume neutrino telescopes.  Axion-like particles (ALPs) and Lorentz-invariance violation (LIV) will be used as representative propagation scenarios.  The scope is selective; emphasis is placed on the relation between a microscopic interaction and the measured observable.

\section{Stars and new physics}
\label{sec:stars-new-physics}

Stellar limits on weakly coupled particles follow from modifications of energy loss or transport.  Main-sequence stars respond through the nuclear thermostat; electron-degenerate cores and white dwarfs have a much weaker pressure response to energy losses.  The relevant observables are therefore stage-dependent: solar neutrino fluxes and helioseismology, the tip of the red-giant branch (TRGB), horizontal-branch (HB) lifetimes, and white-dwarf cooling.  We restrict the discussion to new energy-flow channels; modifications of stellar structure, nuclear reaction rates, or the equation of state require separate treatments
\cite{Raffelt1996,Kippenhahn2012,CaputoRaffelt2024}.

\subsection{Stellar structure and new physics}
\label{subsec:structure}

\subsubsection{The minimal stellar framework: where new physics enters}

The dynamical, Kelvin--Helmholtz, and nuclear time scales are approximately
\begin{equation}
 t_{\rm dyn}\sim(G\bar\rho)^{-1/2},\qquad
 t_{\rm KH}\sim\frac{GM_\star^2}{R_\star L_\star},\qquad
 t_{\rm nuc}\sim\frac{qM_{\rm fuel}}{L_\star},
 \label{eq:stellar_times}
\end{equation}
where $G$ is Newton's constant; $\bar\rho$, $M_\star$, $R_\star$, and $L_\star$ are the mean density, mass, radius, and luminosity, $M_{\rm fuel}$ is the available fuel mass, and $q$ is the released energy per unit mass.  \black Here $L_\star$ denotes the stellar photon luminosity at the surface. \black  Usually, $t_{\rm dyn}\ll t_{\rm KH}\ll t_{\rm nuc}$, so stellar evolution is a sequence of nearly hydrostatic configurations.  In spherical symmetry,
\begin{align}
 \frac{\dd M_r}{\dd r} &= 4\pi r^2\rho,\qquad
 \frac{\dd P}{\dd r} = -\frac{G M_r\rho}{r^2}, \label{eq:hydro_map}\\
 \frac{\dd L_r}{\dd r} &=4\pi r^2\rho\,
 \bigl(\eps_{\rm nuc}+\eps_{\rm grav}-\eps_\nu-\eps_x\bigr),
 \label{eq:energy_map}
\end{align}
where $r$ is the radial coordinate, $M_r$ and $L_r$ are the enclosed mass and luminosity, $P$, $\rho$, and $T$ are pressure, density, and temperature, and $\eps_{\rm nuc}$, $\eps_{\rm grav}$, $\eps_\nu$, and $\eps_x$ are the specific rates of nuclear release, gravitational heating, standard-neutrino
energy loss, and non-standard energy loss, respectively. The system of equations describing a star is completed by an equation of state, composition equations, and an energy-transport prescription.  In the radiative region,
\begin{equation}
 \nabla_{\rm rad}\equiv\left(\frac{\dd\ln T}{\dd\ln P}\right)_{\rm rad}
 =\frac{3\kappa P L_r}{16\pi a_{\rm rad} c G M_r T^4},
 \label{eq:nablarad}
\end{equation}
where $\kappa$ is the total opacity, $a_{\rm rad}=\pi^2/15$ is the radiation constant in natural units $\hbar=c=k_{\rm B}=1$, and $c$ is the speed of light.  Convection occurs when $\nabla_{\rm rad}$ exceeds the appropriate adiabatic gradient.  Consequently, additional cooling or conductivity changes $L_r$, the temperature gradient, and possibly convective boundaries.  A bound on $\eps_x$ must therefore be obtained from an evolved stellar model, not by comparing exotic and photon luminosities at fixed structure.

\subsubsection{Self-regulation and its breakdown}

For a non-degenerate ideal gas, $P\simeq\rho k_{\rm B}T/(\mu m_u)$, where $k_{\rm B}$ is the Boltzmann constant, $\mu$ is the mean molecular weight, and $m_u$ is the atomic mass unit.  Extra cooling induces contraction and, by the virial theorem, heating.  If
\begin{equation}
 \eps_{\rm nuc}\propto \rho^{a_\rho}T^{b_T},
 \label{eq:nuclear_scaling}
\end{equation}
even a small temperature change modifies nuclear power. This thermostat, operating in a stellar core during hydrogen-burning main-sequence stage, converts any exotic energy loss into correlated changes of the central temperature, composition profile, neutrino fluxes, and lifetime. For a perturbation about a calibrated model, one can decompose
\begin{equation}
 \frac{\delta\eps_{\rm nuc}}{\eps_{\rm nuc}}
 \simeq a_\rho\frac{\delta\rho}{\rho}
       +b_T\frac{\delta T}{T}.
 \label{eq:thermostat_linear}
\end{equation}
A large $b_T$ implies that a loss luminosity $L_x\ll L_\star$ may be compensated by a much smaller relative change of the core temperature. This response does not hide the new channel: quantities with a stronger temperature dependence than the total nuclear power, notably individual neutrino fluxes, can change appreciably.  The observable is therefore the self-consistent perturbed model rather than $L_x$ alone.

For non-relativistic degenerate electrons,
\begin{equation}
 P_e=\frac{(3\pi^2)^{2/3}\hbar^2}{5m_e}n_e^{5/3}
 \propto\rho^{5/3},
 \label{eq:degpressure}
\end{equation}
where $n_e$ and $m_e$ are the electron number density and mass.  The leading pressure is independent of $T$; losses thus produce no comparably efficient thermostat.  This property controls the sensitivity of degenerate red-giant cores and cooling white dwarfs to new physics.

\subsubsection{Free streaming, trapping, and the maximum-impact regime}

For a mean free path $\mfp\gg R_\star$, a new particle escapes freely and contributes to Eq.~\eqref{eq:energy_map} through $\eps_x$.  For $\mfp\ll R_\star$, it is trapped and transports energy diffusively.  Transport is maximal when $\mfp$ is comparable to the local temperature scale height. Plasma mixing and other in-medium effects can also affect these processes \cite{HardyLasenby2017,DolanDarkPhoton2024}.

These regimes require different calculations.  In the free-streaming limit, new particles add the associated energy losses to the star's luminosity,
\begin{equation}
 L_x=4\pi\int_0^{R_\star}r^2\rho(r)\eps_x(r)\,\dd r .
 \label{eq:exotic_luminosity}
\end{equation}
In the diffusive limit, the new species instead contributes a conductive flux $F_x=-K_x\,\dd T/\dd r$, where $K_x$ is its thermal conductivity.  The mixed regime depends on the transport mean free path and the stellar geometry; extending a local free-streaming emissivity into the trapped regime may overestimate the cooling.

The escape condition is more precisely expressed through
\begin{equation}
 \tau_x(E,r)=\int_r^{R_\star}\frac{\dd r'}{\lambda_x(E,r')},
 \label{eq:stellar_optical_depth}
\end{equation}
where $\lambda_x(E,r)$ is the energy- and radius-dependent mean free path. Emission from a shell is weighted approximately by $\exp[-\tau_x]$.  In an optically thick region, detailed balance relates emission and absorption, and the new species approaches a thermal distribution.  The same microscopic interaction therefore sets production, escape and diffusive transport.

\subsubsection{The Sun as a precision stellar laboratory}
The standard solar model (SSM) evolves a one-solar-mass star to
$t_\odot\simeq4.57\Gyr$. Its initial helium mass fraction $Y$, heavy-element mass fraction $Z$, and mixing-length parameter are calibrated to reproduce $\Lsun$, $\Rsun$, and the surface composition,
\begin{equation}
\begin{aligned}
 L_\star(t_\odot)&=L_\odot,\qquad R_\star(t_\odot)=R_\odot,\\
 (Z/X)_{\rm surf}&=(Z/X)_{\rm obs}.
\end{aligned}
 \label{eq:ssm_calibration}
\end{equation}
Here $X=1-Y-Z$ is the hydrogen mass fraction. The model predicts $T(r)$, $\rho(r)$, the convective boundary, the surface helium abundance, and solar-neutrino fluxes \cite{Serenelli2016,Vinyoles2017}. Helioseismic measurements constrain the sound-speed profile, $c_s^2=\Gamma_1 P/\rho$, where $\Gamma_1=(\partial\ln P/\partial\ln\rho)_{\rm ad}$; acoustic modes also constrain the radiative interior and convective boundary \cite{Buldgen2019}.
Including $\eps_x$ changes the calibrated initial composition and subsequent diffusion history; comparison at fixed inputs would incorrectly attribute part of the usual solar-model calibration to new physics.
In the nearly ideal-gas radiative interior,
\begin{equation}
 \frac{\delta c_s^2}{c_s^2}\simeq
 \frac{\delta T}{T}-\frac{\delta\mu}{\mu}
 +\frac{\delta\Gamma_1}{\Gamma_1}.
 \label{eq:sound_perturbation}
\end{equation}
An exotic transport process changes $T(r)$ directly and the composition profile indirectly through the calibrated evolutionary history.  The two terms cannot be varied independently in the SSM.  Frequency-separation ratios and inversion kernels supply complementary radial weightings; a single central temperature parameter is insufficient for a precision limit.

Hydrogen burning goes through the net reaction
\begin{equation}
 4p\rightarrow{}^4{\rm He}+2e^++2\nu_e+26.73\ {\rm MeV}.
 \label{eq:hburning}
\end{equation}
In the proton--proton (pp) chains, it proceeds dominantly through weak $p+p$ fusion and subsequent light-nucleus reactions.  In the carbon--nitrogen--oxygen (CNO) cycle, C, N, and O nuclei act as catalysts.  Near the solar central temperature, the rates may be represented schematically in the form of Eq.~\eqref{eq:nuclear_scaling} as
\begin{equation}
\begin{aligned}
 \eps_{pp}&\propto\rho X^2T^{\nu_{pp}},
 &\nu_{pp}&\simeq4,\\
 \eps_{\rm CNO}&\propto\rho X X_{\rm CNO}T^{\nu_{\rm CNO}},
 &\nu_{\rm CNO}&\simeq17\text{--}20,
\end{aligned}
 \label{eq:ppcno}
\end{equation}
where $X_{\rm CNO}$ is the total catalytic CNO mass fraction.  The exponents are local approximations.  The pp chains provide more than 99\% of the present solar power; the CNO fraction grows rapidly with temperature and dominates hydrogen burning in more massive main-sequence stars
\cite{Kippenhahn2012,Derbin2014}.

The pp-chain entrance reaction, $p+p\to d+e^++\nu_e$, fixes the
main-sequence time scale through its weak matrix element.  The pp-I termination is $^3{\rm He}+{}^3{\rm He}\to{}^4{\rm He}+2p$; capture of $^4$He instead opens the pp-II and pp-III branches and produces $^7$Be and $^8$B neutrinos. In the dominant CNO-I cycle, the slow reaction $^{14}{\rm N}(p,\gamma){}^{15}{\rm O}$ controls the rate, while beta decays of $^{13}$N and $^{15}$O generate neutrinos.  The catalysts are restored after each cycle, but their abundance enters Eq.~\eqref{eq:ppcno}.  The branches therefore provide distinct temperature and composition diagnostics
\cite{Derbin2014}.

The neutrino branches probe the core differently.  For controlled SSM variations, $\Phi({}^{7}{\rm Be})\sim T_c^{10}$ and $\Phi({}^{8}{\rm B})\sim T_c^{20\text{--}25}$ are useful estimates, whereas the pp flux is constrained by $\Lsun$.  Precision fits use logarithmic response coefficients and all measured fluxes rather than a universal temperature power. The solar luminosity constraint is
\begin{equation}
 \frac{L_\odot}{4\pi d_\odot^2}=\sum_i Q_i\Phi_i,
 \label{eq:solar_luminosity_constraint}
\end{equation}
where $d_\odot$ is the distance from the Earth to the Sun, $\Phi_i$ is neutrino flux component $i$,
and $Q_i$ is the thermal energy supplied to the star per emitted neutrino of
that component.  This relation assumes quasi-stationary nuclear burning and
negligible non-photonic loss; a new-physics fit must modify the energy balance.

Interpretation is limited by the solar-composition problem.  The low-metallicity photospheric abundances \cite{Asplund2021} disagree with helioseismic SSM predictions. The higher metallicity of Ref.~\cite{Magg2022}, obtained with one-dimensional rather than three-dimensional modeling, gives better helioseismic agreement. Composition directly affects opacity \cite{VillanteSerenelli2020}.

CNO neutrinos add nearly linear sensitivity to the core C+N abundance after
temperature-sensitive pp-chain fluxes constrain the thermal state.  Borexino
established the CNO signal \cite{Borexino2022,Borexino2023}; global analyses
combine it with other solar and terrestrial data and, optionally, the solar
luminosity constraint \cite{GonzalezGarcia2024}.  Alternative background models
can shift the inferred rate \cite{Bezrukov2023}.  Solar limits on new physics
must consequently fit helioseismology and several neutrino fluxes within the
same calibrated model.

The main degeneracy follows from Eq.~\eqref{eq:nablarad}: an opacity
perturbation, a change of the metal mixture, and an additional transport
channel may produce similar temperature gradients.  Their radial and
temperature dependences differ.  Helioseismology probes the profile and the
convective boundary, whereas $^8$B and CNO neutrinos weight the innermost core.
Joint fits test whether one interaction can reproduce all diagnostics with a
single set of solar calibration parameters.
For small changes, this dependence is organized with logarithmic response
coefficients,
\begin{equation}
 \delta\ln O_i=\sum_j \alpha_{ij}\,\delta\ln p_j,
 \qquad
 \alpha_{ij}=\frac{\partial\ln O_i}{\partial\ln p_j},
 \label{eq:solar_response_matrix}
\end{equation}
where $O_i$ is a neutrino or seismic observable and $p_j$ is a model input such
as opacity, abundance or reaction rate.  A new interaction generally does not
correspond to varying one existing $p_j$: its radial source term must first be
added to the evolution.  The response matrix then propagates standard-input
covariances and identifies degeneracies.

For a small additional emissivity about a calibrated reference model,
\begin{equation}
 \delta\ln O_i=
 \int_0^{R_\star}\dd r\,\mathscr K_i^{(x)}(r)\,
 \delta\eps_x(r),
 \label{eq:stellar_response_kernel}
\end{equation}
where $\mathscr K_i^{(x)}$ is the response kernel of observable $O_i$, including
recalibration and subsequent evolution.  It is not simply the local weight in
Eq.~\eqref{eq:exotic_luminosity}.  A neutrino-flux kernel is concentrated in
the core, while seismic inversions and the convective-boundary radius probe
different regions.  For transport, rather than loss, the kernel may change
sign because energy is redistributed between radii.

If the microscopic rate is
$\eps_x=g_x^{\,p}\widehat\eps_x$ at small coupling $g_x$, Eq.~\eqref{eq:stellar_response_kernel}
gives
\begin{equation}
 \delta\ln O_i=g_x^{\,p}I_i+\mathcal O(g_x^{\,2p}),\qquad
 I_i=\int_0^{R_\star}\dd r\,\mathscr K_i^{(x)}(r)
 \widehat\eps_x(r).
 \label{eq:stellar_coupling_response}
\end{equation}
Here, $p$ is fixed by the production amplitude, $\widehat\eps_x$ is the
coupling-independent rate, and $I_i$ is the integrated response. This scaling fails near trapping, resonance boundaries, or evolutionary
thresholds such as helium ignition.

For a vector of observables $\bm O$, a consistent bound follows from the
profile likelihood, schematically
\begin{equation}
\begin{aligned}
 \Delta\bm O&=\bm O_{\rm obs}-\bm O_{\rm th}(g_x,\bm\eta),\\
 \chi^2(g_x,\bm\eta)&=
 \Delta\bm O^{T}\bm V^{-1}\Delta\bm O
 +\chi^2_{\rm aux}(\bm\eta),
\end{aligned}
 \label{eq:stellar_profile_likelihood}
\end{equation}
where $\bm V$ is the observational covariance matrix, $\bm\eta$ denotes
stellar-model nuisance parameters, and $\chi^2_{\rm aux}$ contains their
external constraints.  Profiling common opacity, composition, distance, and
nuclear-rate uncertainties is not equivalent to adding independent errors to
each observable.

\begin{figure}[!tbp]
  \centering
  \black\includegraphics[width=0.99\columnwidth,trim={22 0 0 0}, clip]{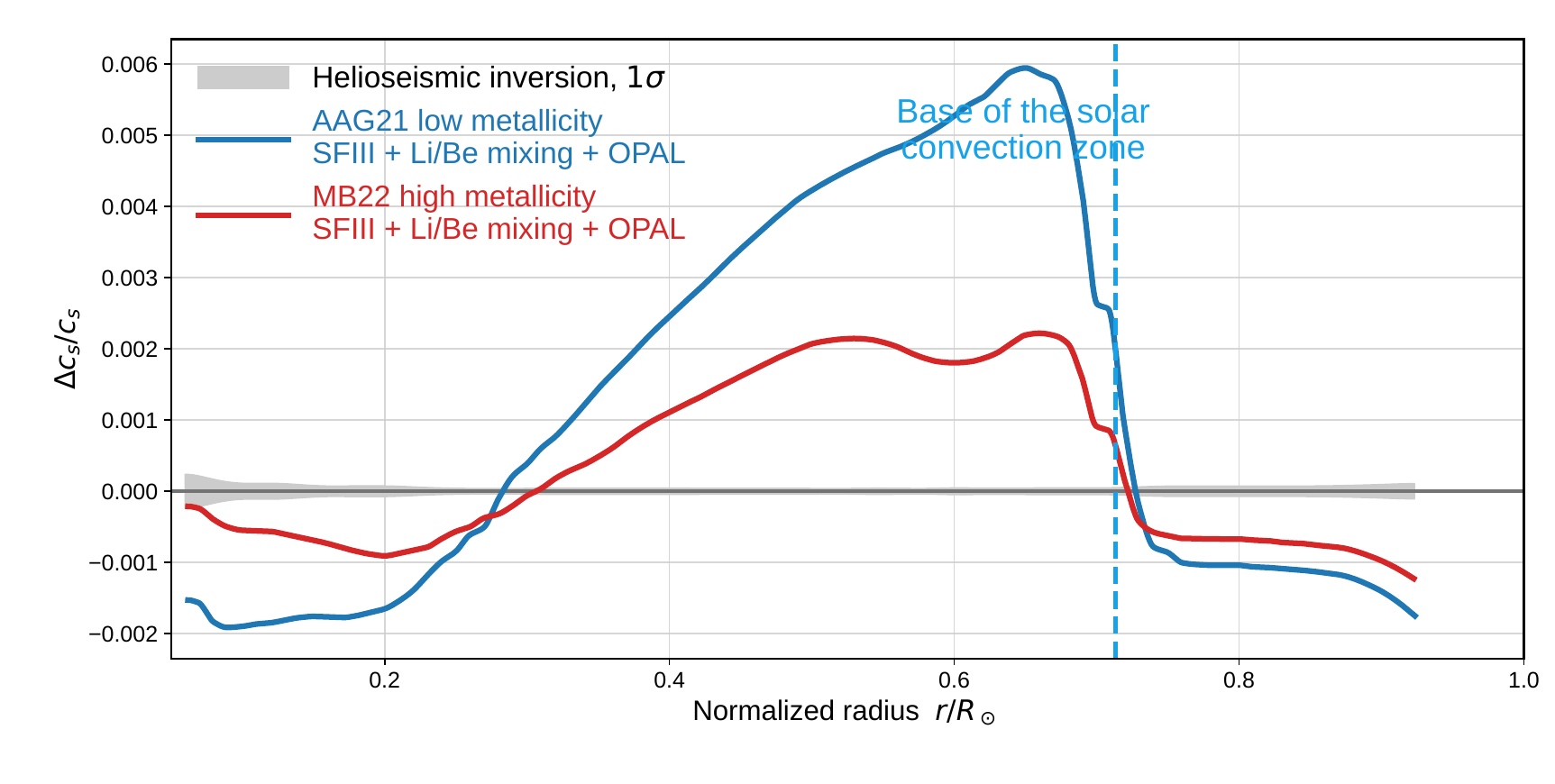}\black\\[-6pt]
  \includegraphics[width=0.99\columnwidth,trim={52 0 0 0}, clip]{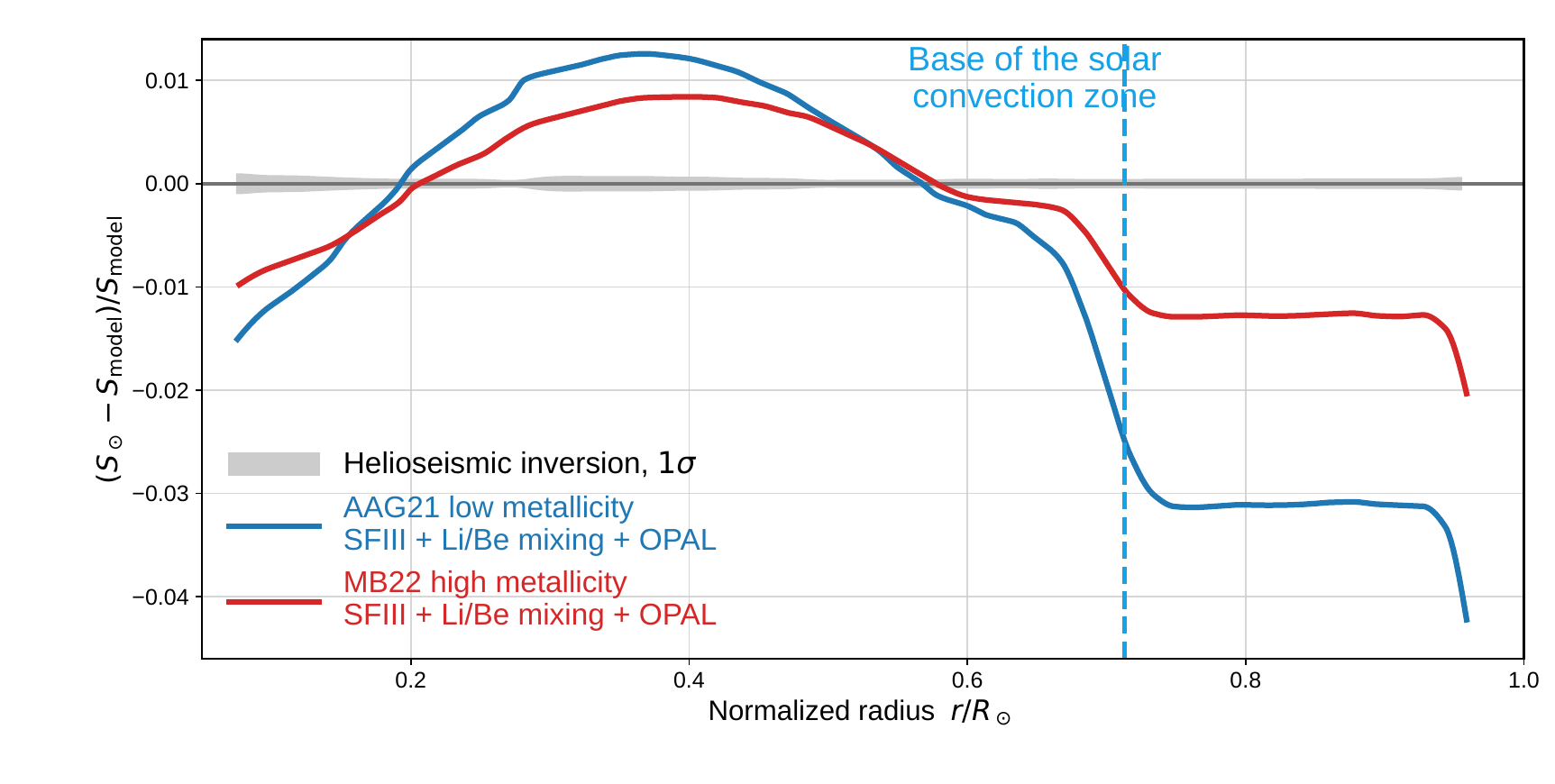}\\[-0pt]
  \includegraphics[width=1.00\columnwidth,trim={7 0 0 0}, clip]{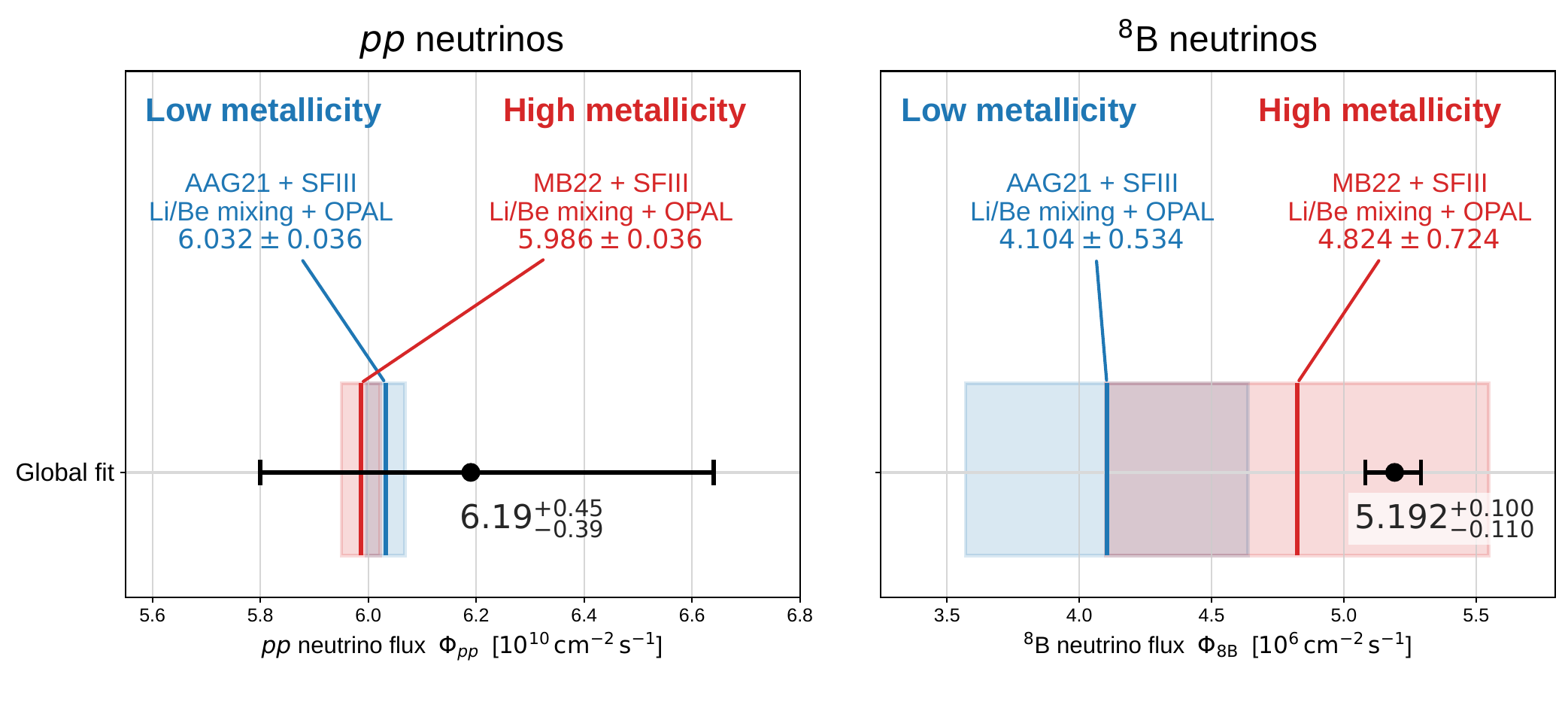}\\[0pt]
  \includegraphics[width=0.99\columnwidth,trim={130 0 0 0}, clip]{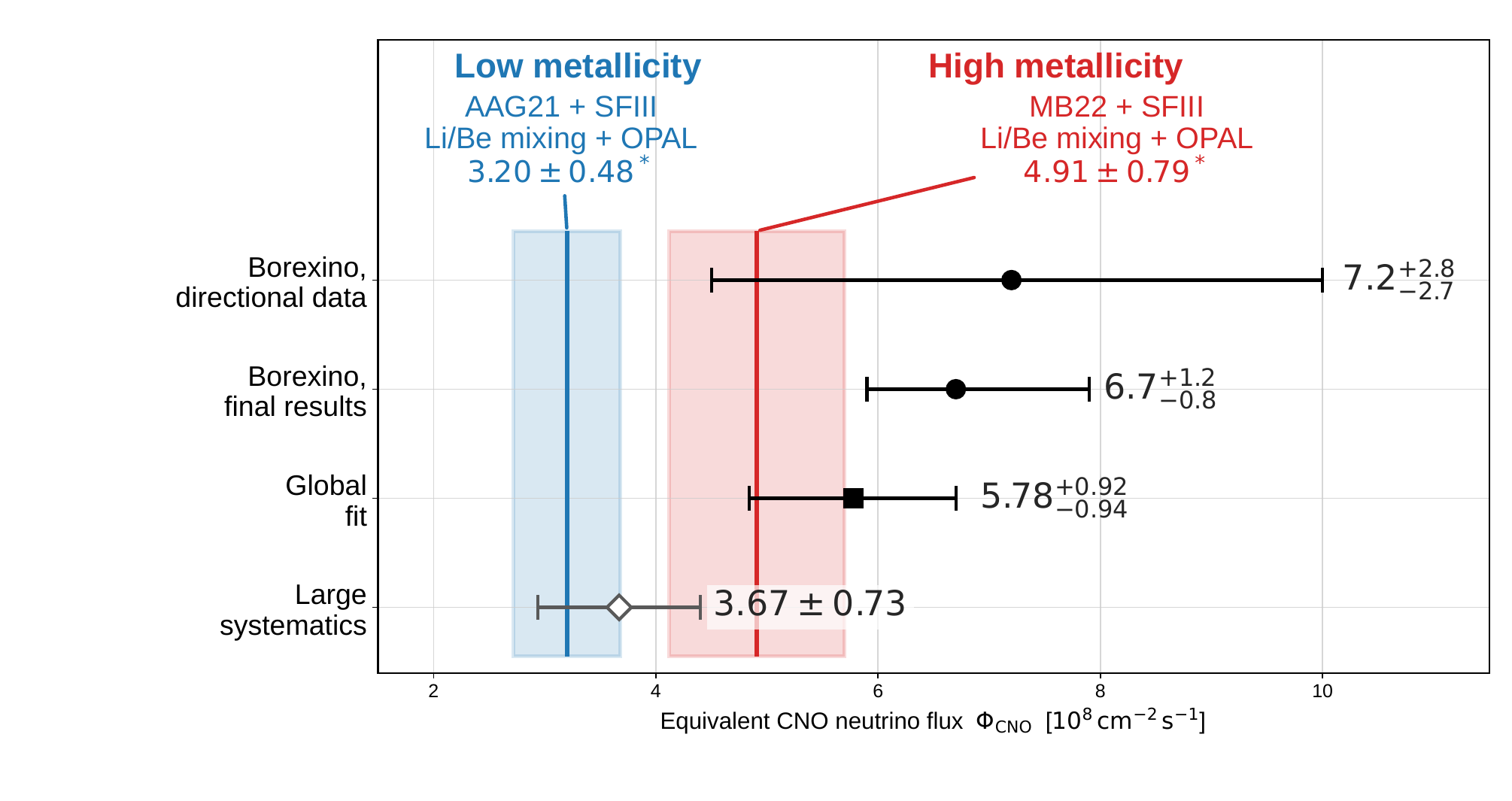}
  \caption{Solar-interior diagnostics. From top to bottom: relative differences of the sound speed \black($\Delta c_s/c_s$)\black and entropy proxy $S=P/\rho^{\Gamma_1}$ from helioseismic inversions for models using the low-metallicity AAG21 \cite{Asplund2021}
  and high-metallicity MB22 \cite{Magg2022} compositions, Solar Fusion III rates, Li/Be mixing, and Opacity Project at Livermore (OPAL) opacities \cite{Sandron2026}; predicted
  $pp$ and $^8$B neutrino fluxes compared with the global determination of Ref.~\cite{GonzalezGarcia2024}; and the corresponding total CNO neutrino fluxes. The lower two panels use values from Refs.~\cite{Sandron2026,GonzalezGarcia2024,Bezrukov2023}.}
  \label{fig:solar-complementarity}
\end{figure}

Different microscopic models populate different parts of this diagnostic
space.  Free-streaming axions add a centrally concentrated $L_x$; weakly
interacting massive particles captured by the Sun can transport rather than
remove energy; kinetically mixed vector particles may be produced resonantly at a
plasma-frequency crossing.
Figure~\ref{fig:solar-complementarity} compares the helioseismic and neutrino
diagnostics for two recent abundance mixtures.  It illustrates their
complementarity rather than providing a new solar-model fit.

\subsection{Stellar evolution and new physics}
\label{subsec:evolution}

\subsubsection{Evolutionary clocks}

In a population of co-evolving stars, the number of stars in a short evolutionary phase is proportional, after completeness corrections, to its duration.  Old Galactic globular clusters are useful because stars leaving the hydrogen-burning non-degenerate phase (main sequence) have similar initial masses, near $0.8\,\Msun$, and their color--magnitude diagrams (CMDs) contain several such clocks.  Additional cooling changes ignition thresholds and fuel-consumption rates and hence endpoint luminosities and star-count ratios. Writing $\dot N_{\rm off}$ for the rate at which stars leave the main sequence, the number in a phase $j$ is approximately
\begin{equation}
 N_j\simeq \dot N_{\rm off}t_j\mathcal C_j,
 \label{eq:star_counts}
\end{equation}
where $t_j$ is its duration and $\mathcal C_j$ is the selection completeness.  Ratios
of neighboring phases cancel $\dot N_{\rm off}$ and part of the initial-mass
function and distance dependence.  Endpoint luminosities instead test the
core mass or temperature at ignition.  The two classes of observable thus
probe different responses to the same local emissivity.

For a non-resonant reaction between nuclei of charges $Z_1e$ and $Z_2e$, the
thermal rate contains the Gamow integral
\begin{equation}
\begin{aligned}
 \langle\sigma v\rangle={}&
 \left(\frac{8}{\pi\mu_{12}}\right)^{1/2}(k_{\rm B}T)^{-3/2}\\
 &{}\times\int_0^\infty \dd E\,S(E)
 \exp\left[-\frac{E}{k_{\rm B}T}
 -\left(\frac{E_G}{E}\right)^{1/2}\right],
\end{aligned}
 \label{eq:gamow_rate}
\end{equation}
where $e$ is the electron charge, $\alpha$ is the fine-structure constant, $\mu_{12}$ is the reduced mass, $S(E)$ is the astrophysical factor, and $E_G=2\mu_{12}c^2(\pi\alpha Z_1Z_2)^2$ is the Gamow energy.  Competition between Maxwell suppression and Coulomb tunneling produces a narrow effective energy range.  The triple-$\alpha$ process instead proceeds through the $^8$Be and Hoyle-state resonances.  A structural temperature shift therefore changes hydrogen-shell and helium-core burning by different amounts; one universal temperature exponent is inadequate.

\subsubsection{Low-mass stars: the red-giant branch and the helium flash}

After core hydrogen exhaustion, shell burning increases the mass $M_c$ of an
electron-degenerate helium core.  Equations~\eqref{eq:degpressure} and
hydrostatic equilibrium give approximately
\begin{equation}
 P_e\propto \rho^{5/3},\qquad R_c\propto M_c^{-1/3}.
\end{equation}
Here $R_c$ is the core radius.
The red-giant-branch (RGB) luminosity consequently increases steeply with
$M_c$.  Additional core cooling delays helium ignition, increases $M_c$, and
brightens the TRGB \cite{Raffelt1996,Viaux2013,Straniero2020,Fung2024}.
The core thermal balance may be represented schematically as
\begin{equation}
 C_c\frac{\dd T_c}{\dd t}=H_{\rm comp}-L_\nu-L_x-L_{\rm cond},
 \label{eq:rgb_core_thermal}
\end{equation}
where $C_c$ is the total core heat capacity, $H_{\rm comp}$ is compressional
heating, $L_\nu$ is the standard-neutrino luminosity, and $L_{\rm cond}$ is
the conductive energy flow through the core
boundary.  Hydrogen-shell burning increases $M_c$ and $H_{\rm comp}$, while
standard plasma-neutrino emission and $L_x$ cool the center.  Helium ignites
when the evolving $(\rho_c,T_c)$ trajectory reaches the triple-$\alpha$
runaway region.

Helium ignition is controlled by
\begin{equation}
 3\alpha\rightarrow{}^{12}{\rm C}^{*}\rightarrow{}^{12}{\rm C}+\gamma,
 \qquad \eps_{3\alpha}\propto\rho^2Y^3T^{\nu_{3\alpha}},
 \label{eq:triplealpha}
\end{equation}
where $\nu_{3\alpha}\sim40$ near
$10^8$~K.  Degeneracy prevents immediate expansion, producing the helium
flash; expansion after degeneracy is lifted stabilizes core helium burning.

\subsubsection{Horizontal-branch stars and population ratios}

Additional loss from the non-degenerate helium-burning core increases the
fuel-consumption rate and shortens the HB lifetime.  The standard observable is
\begin{equation}
 R\equiv\frac{N_{\rm HB}}{N_{\rm RGB}},
 \label{eq:Rparameter}
\end{equation}
where $N_{\rm HB}$ and $N_{\rm RGB}$ are HB and upper-RGB counts. The ratio $R_2=N_{\rm AGB}/N_{\rm HB}$, where AGB denotes
the asymptotic giant branch, also depends on convective-core boundary physics
\cite{Dolan2022R2}.  Multiple populations and completeness must be included in
precision star-count analyses.
Schematically, if $\black q_{3\alpha}\black$ is the energy released per unit helium mass,
\begin{equation}
 t_{\rm HB}\sim
 \frac{\black q_{3\alpha}\black M_{\rm He,burn}}
 {L_\gamma+L_\nu+L_x},
 \label{eq:hb_lifetime}
\end{equation}
where $M_{\rm He,burn}$ is the helium mass consumed and $L_\gamma$, $L_\nu$
and $L_x$ are the photon, standard-neutrino and exotic luminosities.  Core
mixing changes both numerator and denominator and is therefore a leading
correlated systematic.

\subsubsection{White dwarfs and other degenerate objects}

Electron degeneracy also supports white dwarfs, while their thermal reservoir is
mainly ionic.  Exotic emission therefore accelerates cooling; observables are
the luminosity function and the secular drift $\dot P$ of pulsation periods
\cite{MillerBertolami2014,Giannotti2016}.  Massive stars and compact remnants
probe higher temperatures and densities but introduce larger nuclear and
transport uncertainties \cite{HegerWoosley2002,Woosley2017}.
The cooling equation can be written schematically as
\begin{equation}
 L_\gamma+L_\nu+L_x=-\frac{\dd E_{\rm th}}{\dd t}+L_{\rm lat}+L_{\rm grav},
 \label{eq:wd_cooling}
\end{equation}
where $E_{\rm th}$ is the thermal energy and $L_{\rm lat}$ and $L_{\rm grav}$
describe latent heat and chemical or gravitational energy release.  The
luminosity function measures the residence time $\dd t/\dd L_\gamma$ over a
population; pulsating objects measure the cooling rate of individual stars.
Crystallization, envelope composition and the star-formation history must be
fitted together with $L_x$.
For a pulsation mode of period $P_k$, the secular drift is approximately
\begin{equation}
 \frac{\dot P_k}{P_k}\simeq
 -a_k\frac{\dot T_c}{T_c}+b_k\frac{\dot R_\star}{R_\star},
 \label{eq:wd_period_drift}
\end{equation}
where $a_k$ and $b_k$ are mode-dependent structure coefficients.  Contraction
is small for a cool white dwarf, so an additional negative $\dot T_c$ generally
increases $\dot P_k$.  Mode trapping and envelope stratification control the
conversion from a measured drift to a core-cooling rate.

In massive stars, radiation pressure, neutrino losses and short burning stages
replace electron degeneracy as the controlling physics.  Pair creation lowers
the first adiabatic exponent; dynamical instability occurs when a sufficiently
large core region approaches $\Gamma_1<4/3$.  Additional cooling changes the
carbon-to-oxygen ratio, core entropy and the masses at which pulsational or full
pair instability occurs \cite{HegerWoosley2002,Woosley2017}.  Remnant-mass and
explosion constraints consequently require a coupled calculation of nuclear
rates, convection, mass loss and the exotic emissivity.  They probe higher
temperatures than globular-cluster stars, but with larger stellar-systematic
uncertainties.

\subsubsection{From a microscopic emissivity to an evolutionary observable}

A stellar constraint has the structure
\begin{equation}
\begin{gathered}
 \text{interaction}\ \longrightarrow\ \eps_x(\rho,T,Y_i)\\
 \longrightarrow\ \text{stellar response}\ \longrightarrow\ \text{observable}.
\end{gathered}
 \label{eq:chain}
\end{equation}
The local rate must be integrated over an evolving stellar model.

For a process $i\to f+x$, the energy emitted per unit volume and time has the
kinetic-theory form
\begin{equation}
\begin{aligned}
 Q_x={}&\int \dd\Pi_i\,\dd\Pi_f\,\dd\Pi_x (2\pi)^4
 \delta^{(4)}(P_i-P_f-p_x)\\
 &{}\times E_x|{\cal A}|^2 f_i(1\pm f_f),
\end{aligned}
 \label{eq:kinetic_emissivity}
\end{equation}
with products over all initial and final particles understood.  Here, $Q_x$ is
the volumetric emissivity, $E_x$ is the emitted-particle energy,
$\dd\Pi_j=\dd^3p_j/[(2\pi)^3 2E_j]$ is invariant phase space,
${\cal A}$ is the in-medium amplitude, $P_i$ and $P_f$ are the total initial
and final four-momenta, $f_i$ and $f_f$ are occupation factors,
and the sign accounts for Bose enhancement or Pauli blocking.  The specific
emissivity in Eq.~\eqref{eq:energy_map} is $\eps_x=Q_x/\rho$.  Screening,
thermal masses and dispersion relations enter both ${\cal A}$ and phase space;
vacuum cross sections are therefore generally insufficient
\cite{Raffelt1996,CaputoRaffelt2024}.

The dominant channel depends on both coupling and environment.  For instance, ALPs $a$ are produced predominanltly in the following processes:
\begin{align}
 \gamma+(\black Z_{\rm nuc}\black,A)&\to a+(\black Z_{\rm nuc}\black,A) &&\text{(Primakoff)},\\
 e+(\black Z_{\rm nuc}\black,A)&\to e+(\black Z_{\rm nuc}\black,A)+a &&\text{(bremsstrahlung)},\label{eq:brem}\\
 \gamma+e&\to e+a &&\text{(Compton-like)}.
 \label{eq:stellar_alp_channels}
\end{align}
Here, $(\black Z_{\rm nuc}\black,A)$ denotes a nuclei with the charge $\black Z_{\rm nuc}\black$ and mass number $A$. In a non-degenerate plasma, the Primakoff emissivity scales as
\begin{equation}
 Q_a^{\rm P}=\frac{\gag^2T^7}{4\pi}
 \black\mathcal G_{\rm P}\black\!\left(\frac{\kappa_s}{T}\right),
 \qquad
 \kappa_s^2\simeq\frac{4\pi\alpha}{T}
 \sum_j n_jZ_j^2,
 \label{eq:primakoff_emissivity}
\end{equation}
in natural units, where $\kappa_s$ is the screening wave number, 
$n_j$ and $Z_j$ are the number density and charge of plasma species $j$, and
$\black\mathcal G_{\rm P}\black$ is a dimensionless screening function.  The Primakoff
rate is therefore proportional to $\gag^2$ and is efficient in non-degenerate main-sequence and horizontal-branch cores; electron bremsstrahlung is proportional to the square
of the ALP--electron coupling $\gae$
and dominates in the dense RGB core and in white dwarfs.  For particles that
mix with photons, longitudinal and transverse plasma modes may convert
resonantly where their in-medium dispersion relation crosses the new-particle
mode.  Consequently, the same vacuum interaction can be tested by different
stars through different production mechanisms.

For the axion--electron interaction, written in derivative form as
\begin{equation}
 \mathcal L_{ae}=\frac{\gae}{2m_e}(\partial_\mu a)
 \bar\psi_e\gamma^\mu\gamma_5\psi_e,
 \label{eq:Lae}
\end{equation}
where $\psi_e$ is the electron field,
$\partial_\mu$ is the four-gradient, and $\gamma^\mu$, $\gamma_5$ are Dirac
matrices,
electron--ion bremsstrahlung (\ref{eq:brem}) is the dominant production process in the dense, partially degenerate helium
core near the TRGB.  Pauli blocking restricts electron--hole excitations to a shell of width $\sim T$ around the Fermi surface, giving
\begin{equation}
 \eps_a\propto \gae^2T^4.
\end{equation}
Numerically, for a red giant close to the TRGB ($T\simeq10^8$~K and $\rho\simeq2\times10^5\,\mathrm{g\,cm^{-3}}$),
\begin{equation}
 \eps_a^{\rm TRGB}\sim (1\text{--}2)\,
 \left(\frac{\gae}{10^{-13}}\right)^2
 \left(\frac{T}{10^8\,\mathrm K}\right)^4
 \mathrm{erg\,g^{-1}\,s^{-1}},
 \label{eq:epsaxion_num}
\end{equation}
up to composition and screening corrections.

Table~\ref{tab:regimes} summarizes the main logic for the evolutionary stages most often used in particle-physics applications.

\begin{table*}[t]
\caption{Representative stellar regimes and their response to an additional
weakly interacting free-streaming channel. The table is schematic; the
dominant microscopic production process depends on the interaction under
consideration.}
\label{tab:regimes}
{\footnotesize
\setlength{\tabcolsep}{2.2pt}
\begin{ruledtabular}
\begin{tabular}{
p{0.18\textwidth}
p{0.22\textwidth}
p{0.30\textwidth}
p{0.23\textwidth}
}
Stellar regime &
Thermodynamic response &
Main new-physics effect &
Representative observable \\
\hline
Main sequence / Sun &
Nuclear thermostat &
Mild contraction/heating or new transport &
$\nu$ fluxes, helioseismology, lifetime \\

RGB degenerate He core &
Pressure weakly depends on $T$ &
Delayed He ignition, larger $M_c$ &
TRGB luminosity \\

Horizontal branch &
Stable He burning &
Larger fuel-consumption rate &
$R=N_{\rm HB}/N_{\rm RGB}$ \\

White dwarf &
Degeneracy-supported remnant &
Faster cooling &
Luminosity function, $\dot P$ \\

Massive late burning &
Radiation pressure and pairs &
Shifted stability/burning conditions &
Remnant masses, explosions \\
\end{tabular}
\end{ruledtabular}
}
\end{table*}

\subsection{From the TRGB to an axion--electron bound}
\label{subsec:trgb-case}
Let us illustrate the practical implementation of the TRGB method following recent Ref.~\cite{Gontcharov2026TRGB}, which combines homogeneous photometry of 24 Galactic globular clusters with cluster-specific stellar-evolution calculations to obtain constraints on $g_{ae}$. The brightest observed red giant is accepted if it is non-variable or has a trustable estimate of the average fluxes derived from observations at many epochs. 

For cluster $i$, the forward prediction is
\begin{equation}
 M_{{\rm bol},i}^{\rm th}=
 M_{{\rm bol},i}^{\rm th}(\gae;\bm\theta_i,\bm\eta),
 \label{eq:forward_trgb}
\end{equation}
where $\bm\theta_i$ contains the cluster parameters and $\bm\eta$ denotes
shared nuisance parameters. A representative star for every cluster was
evolved with Modules for Experiments in Stellar Astrophysics (MESA) \cite{Paxton2011,Jermyn2023,Gontcharov2026TRGB}, varying
$g_{13}\equiv\gae/10^{-13}$ over the interval $0\leq g_{13}\leq1$.
The masses, ages and compositions are constrained by the cluster-specific
isochrone fits; no universal TRGB shift or additional broad marginalization
over these fitted parameters is imposed.

The adopted bolometric luminosities are obtained from the fitted
isochrones of a Bag of Stellar Tracks and Isochrones (BaSTI) and the Dartmouth
Stellar Evolution Database (DSED). Non-variable candidates use the available
\textit{Gaia}, Hubble Space Telescope and terrestrial optical photometry; accepted
variables use phase-averaged photometry from multiple epochs of terrestrial observations. 

The brightest giant retained after excluding $k$ even brighter but poorly measured variable candidates has not yet reached the true RGB tip (the same is true even for $k=0$). Let $M_{(k+1)}$ be its bolometric magnitude, and define
\begin{equation}
 \delta_k=M_{(k+1)}-M_{\rm tip}\geq0,
 \qquad M_{\rm tip}=M_{(k+1)}-\delta_k .
 \label{eq:trgb_order_statistic}
\end{equation}
The observed count $N_{\rm obs}$ includes retained giants within 2.5~mag below
this star. This counting window is displaced relative to the true tip, so
$N_{\rm obs}$ differs from the unknown number $n$ in the true upper-RGB
interval. A time-weighted reference BaSTI track gives the offset density
$f(\delta)$ and cumulative distribution $F(\delta)$ in that interval. Monte-Carlo simulations determine the correction $\delta_k$ and its strongly non-Gaussian uncertainties for a given $N_{\rm obs}$. Other uncertainties include cluster-specific distance, reddening, composition, age and optical-method systematics, together with shared stellar-physics theoretical uncertainties. These include thermal-neutrino losses,
conductive and radiative opacities, and nuclear rates. The joint likelihood combines these individual and correlated uncertainties by Monte-Carlo propagation. 

\begin{figure}[!tbp]
  \centering
  (a)\\[-1pt]
  \includegraphics[width=0.99\linewidth]{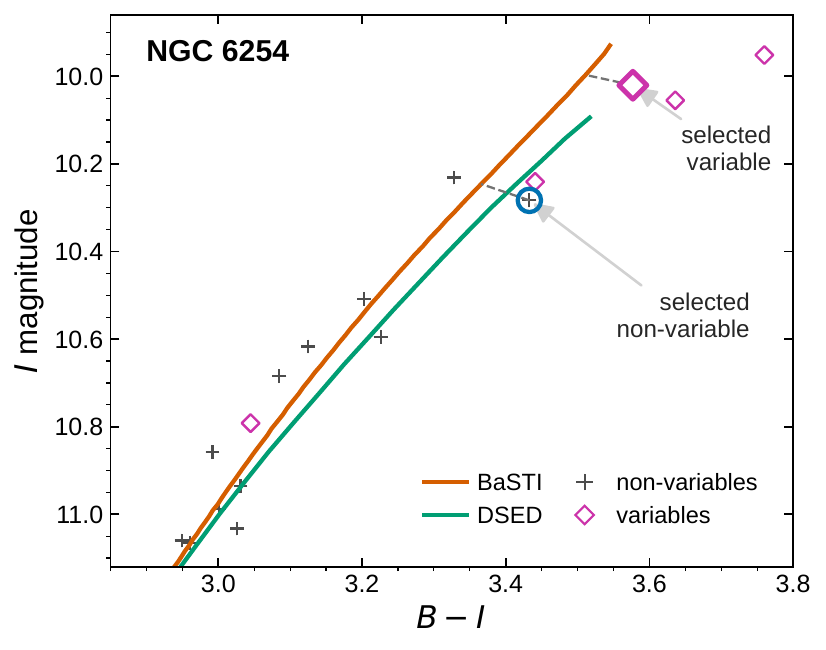}\\[-0pt]
  (b)\\[-1pt]
  \includegraphics[width=0.99\linewidth, trim={15 10 0 0}, clip]{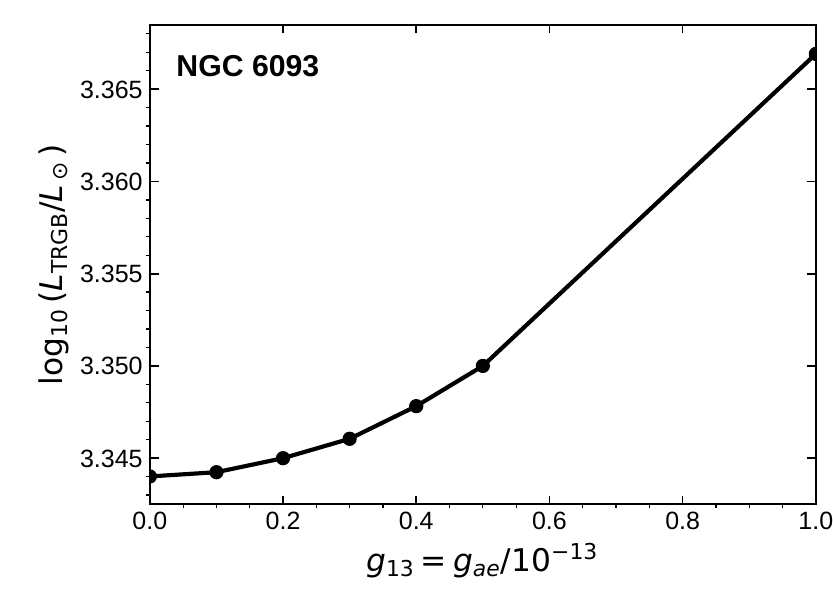}\\[-0pt]
  (c)\\[-1pt]
  \includegraphics[width=0.99\linewidth, trim={15 10 0 0}, clip]{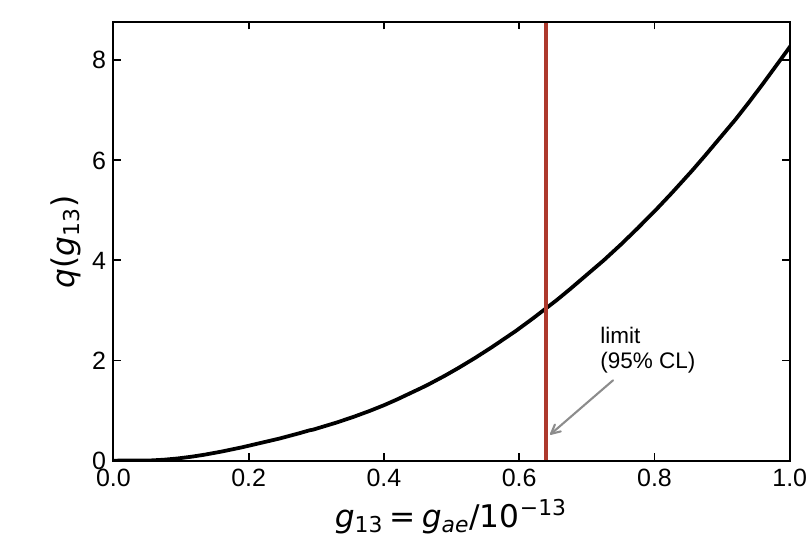}
  \caption{TRGB inference in the revised analysis of Ref.~\cite{Gontcharov2026TRGB}. Panel (a) shows the upper RGB of NGC~6254 with BaSTI and DSED isochrones, panel (b) the MESA TRGB-luminosity response of the canonical NGC~6093 model as a function of $g_{13}$, and panel (c) the baseline 24-cluster physical profile-likelihood scan; the vertical line marks the adopted 95\% CL limit.}
  \label{fig:trgb-inference}
\end{figure}

The physical upper limit \cite{Gontcharov2026TRGB} is obtained by the power-constrained profile-likelihood 
method \cite{Cowan:2011PCL}, with the calibration based
on background-only pseudoexperiments. The best fit is
consistent with no additional cooling, and the limit reads as
\begin{equation}
 \gae<6.4\times10^{-14}\qquad (95\%\ \mathrm{CL}),
 \label{eq:gae_bound}
\end{equation}
cf.\ Fig.~\ref{fig:trgb-inference}. It is obtained entirely within
the calculated MESA range and applies to freely escaping light ALPs with
$m_a\lesssim1$~keV; larger masses require different stellar-evolution calculations.

\section{High-energy photon propagation as a probe of new physics}
\label{sec:gamma-propagation}

High-energy photons undergo electron--positron pair production on diffuse radiation backgrounds. The Breit--Wheeler cross section is \cite{Nikishov1962,GouldSchreder1967}
\begin{align}
 \sigma_{\gamma\gamma}(s)&=\frac{3\sigma_{\rm T}}{16}(1-\beta_e^2)
 \left[(3-\beta_e^4)\ln\frac{1+\beta_e}{1-\beta_e}\right.\notag\\
 &\left.\hspace{5em}{}-2\beta_e(2-\beta_e^2)\right],\notag\\
 \beta_e&=\left(1-\frac{4m_e^2}{s}\right)^{1/2},
 \label{eq:breit_wheeler}
\end{align}
where $\sigma_{\rm T}$ is the Thomson cross section and $\beta_e$ is the
electron speed in the center-of-momentum frame; $s=2E\epsilon(1-\cos\theta)$ is the Mandelstam invariant. Pair production becomes possible at $s\geq 4m_e^2$. The cross section strongly peaks at $s\sim 8m_e^2$. Thus $E\sim 1$~TeV photons primarily probe optical and near-infrared extragalactic background light (EBL), whereas $E\sim 300$~TeV ones create pairs mostly on the cosmic microwave background (CMB). Together, these radiation fields produce strongly energy-dependent intergalactic gamma-ray absorption.  

Cosmological redshift affects both the gamma-ray energy and the evolving EBL spectrum. For a source at redshift $z_s$, the optical depth is
\begin{align}
 \tau(E_0,z_s) = &\int_0^{z_s} \dd z\,\frac{\dd\ell}{\dd z}
 \int_{-1}^{1} \frac{\dd u}{2}(1-u)
 \int_{\epsilon_{\rm th}}^{\infty}\dd\epsilon\,
 n_\gamma(\epsilon,z) \nonumber\\
 &\times \sigma_{\gamma\gamma}\!\left[s=2E_0(1+z)\epsilon(1-u)\right] ,
 \label{eq:tau}
\end{align}
where $E_0$ is the observed energy, $u=\cos\theta$, $\ell$ is proper path
length, $n_\gamma$ is the differential background-photon number density, and
$\epsilon_{\rm th}$ is the angle-dependent threshold.  The observed and
intrinsic fluxes are, in the absence of secondary emission, related by
\begin{equation}
 F_{\rm obs}(E_0)=F_{\rm int}(E_0)\exp[-\tau(E_0,z_s)] .
 \label{eq:attenuation}
\end{equation}
Equivalently, in terms of the local pair-production mean free path
\begin{equation}
 \lambda_{\gamma\gamma}^{-1}(E,z)=
 \int_{-1}^{1}\frac{\dd u}{2}(1-u)
 \int_{\epsilon_{\rm th}}^\infty\dd\epsilon\,
 n_\gamma(\epsilon,z)
 \sigma_{\gamma\gamma}[2E\epsilon(1-u)] ,
 \label{eq:gamma_mean_free_path}
\end{equation}
the optical depth is
\begin{equation}
 \tau(E_0,z_s)=
 \int_0^{z_s}\frac{c\,\dd z}{(1+z)H(z)}
 \lambda_{\gamma\gamma}^{-1}[E_0(1+z),z]
 \label{eq:gamma_tau_compact}
\end{equation}
and we use $\dd\ell/\dd z=c/[(1+z)H(z)]$, where $H(z)$ is the Hubble rate.  This form separates the local interaction
calculation from cosmological propagation and is convenient when either the
cross section or the background spectrum is modified.

For practical implementations, it is important to remember that the cross section is broad above threshold; consequently a measured multi-TeV bin samples an interval of EBL wavelengths and redshifts.  Energy dispersion must be folded before comparing a steeply attenuated spectrum with data, because upward migration from more numerous lower-energy photons can otherwise mimic excess transmission. The EBL dominates at TeV energies and carries model uncertainty \cite{SaldanaLopez2021,MAGIC2019,Banerjee2026}; the CMB dominates at hundreds of TeV and is accurately known.  Consequently, a TeV transparency excess is degenerate with the source spectrum and EBL, whereas a cosmological photon at several hundred TeV tests a less adjustable attenuation prediction.

Hard deabsorbed blazar spectra and optical-depth-correlated residuals motivated an anomalous-transparency hypothesis \cite{HornsMeyer2012,breaks}.  Some analyses, e.g.\ of the Fermi Large Area Telescope (Fermi-LAT) and Major Atmospheric
Gamma Imaging Cherenkov telescopes (MAGIC), were consistent with standard attenuation \cite{MAGIC2019}, while selected high-optical-depth samples retained smaller tensions \cite{breaks2}.  Tests must therefore compare a specified propagation model with source, EBL and detector nuisance parameters.

New-physics tests with the high-energy photon propagation include models in which either gamma-rays bypass the absorber without changing pair-production kinematics (e.g.\ photon--ALP conversion), or the kinematics and dynamics are fundamentally changed at high energies (e.g.\ LIV).  

\subsection{Photon--ALP mixing: bypassing the absorber}
\label{subsec:alp}

For a pseudoscalar ALP $a$, the interaction relevant here is
\begin{equation}
 \mathcal{L}_{a\gamma}= -\frac{1}{4}\gag aF_{\mu\nu}\widetilde F^{\mu\nu}
 =\gag a\,\bm E\!\cdot\!\bm B .
 \label{eq:alpL}
\end{equation}
Here, $\gag$ is the ALP--photon coupling, $F_{\mu\nu}$ is the electromagnetic field tensor,
$\widetilde F^{\mu\nu}$ its dual, and $\bm E$ and $\bm B$ are
the electric and magnetic fields.  

An external transverse magnetic field mixes one photon polarization with the ALP \cite{RaffeltStodolsky1988}. For propagation along coordinate $y$, let $\Psi=(A_1,A_2,a)^T$ contain the two photon polarization amplitudes and the ALP amplitude. Choosing $A_1$ perpendicular and $A_2$ parallel to the transverse magnetic field, the propagation equation is
\begin{equation}
 i\frac{\dd\Psi}{\dd y}=\mathcal{H}_{\rm mix}(E,y)\Psi,
 \label{eq:mixing}
\end{equation}
where, up to an irrelevant term proportional to the identity,
\begin{equation}
 \mathcal{H}_{\rm mix}=
 \begin{pmatrix}
  \Delta_\perp & 0 & 0\\
  0 & \Delta_\parallel & \Delta_{a\gamma}\\
  0 & \Delta_{a\gamma} & \Delta_a
 \end{pmatrix}.
 \label{eq:mixingH}
\end{equation}
The matrix elements are
\begin{equation}
 \begin{split}
 \Delta_{a\gamma}&=\frac{\gag B_T}{2},\qquad
 \Delta_a=-\frac{m_a^2}{2E},\\
 \Delta_{\rm pl}&=-\frac{\omega_{\rm pl}^2}{2E},\\
 \Delta_\perp&=\Delta_{\rm pl}+2\Delta_{\rm QED},\\
 \Delta_\parallel&=\Delta_{\rm pl}
                  +\frac{7}{2}\Delta_{\rm QED},
 \end{split}
 \label{eq:deltas}
\end{equation}
where $B_T$ is the transverse magnetic field, $m_a$ is the
ALP mass, and $\omega_{\rm pl}$ is the plasma frequency.

The QED vacuum-birefringence contribution, often quoted as 
\begin{equation}
 \Delta_{\rm QED}=
 \frac{\alpha E}{45\pi}
 \left(\frac{B_T}{B_{\rm cr}}\right)^2,
 \qquad B_{\rm cr}=\frac{m_e^2}{e},
\end{equation}
applies for $B_T\ll B_{\rm cr}$ and
\begin{equation}
 \black\chi_\gamma\black\equiv\frac{E}{m_e}\frac{B_T}{B_{\rm cr}}
 \ll 1.
\end{equation}
At $\black\chi_\gamma\black\gtrsim 1$, the QED contribution no longer
scales as $B_T^2 E$, and the full energy-dependent
polarization operator, including its absorptive part,
must be used \cite{FairbairnRashbaTroitsky2011,Troitsky2017}.

Mixing is large when $\Delta_{a\gamma}$ dominates
the diagonal splittings and the path supports an
appreciable phase. The mass and plasma terms dominate
at low energy; an approximately energy-independent
strong-mixing range may follow. 
For one homogeneous domain of length $d_B$, neglecting absorption,
\begin{align}
 P_{\gamma\to a}&=
 \left(\frac{2\Delta_{a\gamma}}{\Delta_{\rm osc}}\right)^2
 \sin^2\left(\frac{\Delta_{\rm osc}d_B}{2}\right),\notag\\
 \Delta_{\rm osc}^2&=(\Delta_\parallel-\Delta_a)^2
 +4\Delta_{a\gamma}^2.
 \label{eq:alp_probability}
\end{align}
The low-energy boundary of strong mixing is set approximately by
\begin{equation}
 E_{\rm crit}\simeq\frac{|m_a^2-\omega_{\rm pl}^2|}{2\gag B_T}.
 \label{eq:alp_critical_energy}
\end{equation}
At higher energies within the $\chi_\gamma\ll1$ regime, the QED term $\Delta_{\rm QED}\propto E$ can again suppress the transitions; beyond this regime the full energy-dependent QED contribution is required. 

In non-uniform fields, Eq.~\eqref{eq:mixing} must be integrated
numerically.
For propagation problems, the density-matrix form is often preferable, because the photon state is not measured coherently and absorption makes the evolution non-unitary.  Schematically,
\begin{equation}
 i\frac{\dd\varrho}{\dd y}=\mathcal{H}_{\rm eff}\varrho-\varrho\mathcal{H}_{\rm eff}^{\dagger},
 \label{eq:densitymatrix}
\end{equation}
where $\varrho$ is the density matrix and the imaginary photon entries of the
effective Hamiltonian $\mathcal H_{\rm eff}$ encode pair-production losses.
Photons may convert to ALPs in a
source or host field, propagate without EBL/CMB attenuation, and reconvert in
the Galactic magnetic field \cite{Troitsky2017,GalantiRoncadelli2022,Troitsky2024}.
Depending on parameters, intergalactic conversion may be possible.

For an initially unpolarized photon beam with no ALP component,
$$
 \varrho_0={\rm diag}(1/2,1/2,0).
$$
Absorption over a local mean free path $\lambda_{\gamma\gamma}$ enters the
photon diagonal elements of $\mathcal H_{\rm eff}$ through
$-i/(2\lambda_{\gamma\gamma})$.  For a stochastic field, the likelihood should be marginalized over field realizations; substituting the mean survival probability into a nonlinear count likelihood is not generally equivalent.

The transfer matrix predicts energy-dependent irregularities, polarization effects, and transitions between weak- and strong-mixing regimes, rather than a constant rescaling of optical depth.  It depends crucially on the magnetic-field geometry.  Several large-scale Galactic models fit current rotation-measure
and polarization data, while fields in other galaxies and galaxy clusters are much worse known.  Limits obtained for a fixed field realization can therefore overstate the inference precision
\cite{LibanovTroitsky2020}.

The ALP mixing makes the Universe more transparent for high-energy photons in the ALP parameter region with very light masses and 
$\gag\sim10^{-11}$--$10^{-10}\,\mathrm{GeV}^{-1}$.  This benchmark is of order the recent constraints on the coupling. The extended European
Organization for Nuclear Research (CERN) Axion Solar Telescope (CAST) run
gives $\gag<5.8\times10^{-11}\,\mathrm{GeV}^{-1}$ at 95\% CL for
$m_a\lesssim0.02$ eV \cite{CAST2024}.  Stellar evolution and magnetic-white-dwarf observations provide complementary, partly model-dependent constraints
\cite{Ayala2014,Troitsky2025Stars,Gontcharov2026Clusters,Dessert2022}.

\subsection{Lorentz-invariance violation: modifying pair production}
\label{subsec:liv}

A different possibility is that the high-energy dispersion relation itself departs slightly from the relativistic form.  A common phenomenological parametrization for photons is
\begin{equation}
 E^2-p^2 = s_n\frac{E^{n+2}}{E_{{\rm LIV},n}^{\,n}},
 \qquad s_n=\pm1,
 \label{eq:livdisp}
\end{equation}
where $p$ is photon momentum, $n$ is the order of the correction,
$E_{{\rm LIV},n}$ is its characteristic scale, and $s_n$ fixes the sign.
Analogous terms may occur for electrons.  For the subluminal photon sign,
the pair-production threshold and phase space can shift toward reduced
absorption \cite{Addazi2022}.

A threshold shift is not a complete opacity calculation: LIV also changes the
phase space and, in a specified theory, the matrix element.  The
effective-field-theory calculation of Ref.~\cite{Rubtsov2012}, the exact
quadratic-LIV calculation of Ref.~\cite{Vrban2022}, and the optical-depth
prescriptions of Refs.~\cite{Carmona2024,SatuninTroitsky2026} show why replacing
only the standard threshold or Mandelstam variable is insufficient.
In a specified LIV theory the cross section follows from
\begin{equation}
\begin{aligned}
 \sigma_{\rm LIV}&=\frac{1}{\black\mathcal K_{\rm in}\black}
 \int \dd\Pi_+\dd\Pi_-\,(2\pi)^4\\
 &\quad\times\delta^{(4)}(p_\gamma+p_b-p_+-p_-)
 |{\cal A}_{\rm LIV}|^2,
\end{aligned}
 \label{eq:liv_cross_section}
\end{equation}
where $\black\mathcal K_{\rm in}\black$ is the initial-state flux factor, $p_\gamma$ and $p_b$ are
the high-energy and background photon four-momenta, and $p_\pm$ and
$\dd\Pi_\pm$ are the electron/positron four-momenta and phase-space elements.
Both the energy-conserving delta function and ${\cal A}_{\rm LIV}$ may differ
from their Lorentz-invariant forms.  The result must then replace
$\sigma_{\gamma\gamma}$ inside Eq.~\eqref{eq:tau}; a shifted lower integration
limit alone is not equivalent.
In the preferred cosmological frame, the LIV cross section need not be a
function of one variable $s$: it may depend separately on the hard-photon
energy, target energy, and collision angle.  After the delta functions are
integrated, the remaining phase-space integral contains the Jacobian of the
modified energy-conservation condition.  Its allowed endpoints can move and
merge as $E/E_{{\rm LIV},n}$ increases.  This change of phase-space weight is
the origin of differences between a threshold-only prescription and the exact
quadratic calculation \cite{Vrban2022,Carmona2024}.

The result is conditional on the particle sectors included in the effective
theory.  Modifying only photon dispersion while retaining standard electron
dispersion and vertices is one model; correlated photon- and electron-sector
coefficients define another.  A numerical implementation should recover
Eq.~\eqref{eq:breit_wheeler} as $E_{{\rm LIV},n}\to\infty$, vanish outside the
modified phase space, and use the same assumptions in photon propagation and
air-shower calculations.

The LIV scale is not a universal observable.  Time-of-flight measurements
constrain the group velocity; transparency constrains interaction kinematics
and dynamics.  Their relation is theory-dependent: for instance, the Large High Altitude
Air Shower Observatory (LHAASO) light curve of
GRB~221009A limits energy-dependent photon speeds \cite{LHAASOLIV2024}, but
does not directly bound every opacity-changing model.

The same operators can modify Bethe--Heitler pair production and air-shower
development.  Galactic gamma rays constrain quadratic LIV
\cite{Rubtsov2012,Satunin2021}; photon subshowers in hadronic cascades give
stronger conditional limits \cite{Martynenko2025}.  Linear dimension-five
photon operators are tightly restricted by vacuum birefringence
\cite{MyersPospelov2003,Goetz2013}.  An opacity fit is viable only if its
parameters satisfy these independent constraints.
The sign of the correction determines which otherwise forbidden process opens.
A sufficiently superluminal photon has positive effective mass squared and can
decay, $\gamma\to e^+e^-$, in vacuum; observation of a photon of energy
$E_{\gamma,\max}$ requires its decay threshold to exceed that energy.  A
superluminal electron instead permits vacuum Cherenkov radiation
$e^-\to e^-\gamma$ above a model-dependent threshold.  Subluminal photons do
not decay but can suppress background pair production and modify showers.
These statements also depend on electron-sector coefficients, so setting them
to zero is a physical assumption rather than a choice of notation
\cite{Addazi2022,Rubtsov2012}.

In an effective field theory, rotationally invariant dimension-five photon
operators have opposite signs for the two helicities.  The accumulated
polarization rotation is proportional to a high power of energy and distance,
which makes polarimetry of cosmological transients more sensitive than an
opacity fit for that operator class \cite{MyersPospelov2003,Goetz2013}.
Quadratic dimension-six modifications need not be birefringent and therefore
require the complementary threshold, cross-section and shower tests discussed
above.

\subsection{Gamma-ray burst 221009A: an opacity test with a single transient}
\label{subsec:grb}

Gamma-ray burst (GRB) 221009A, at redshift $z\simeq0.151$, was detected beyond
10 TeV.  The initial LHAASO circular reported photons up to about 18 TeV; the published analysis found more than 140 photons above 3 TeV in the 230--900-s interval
\cite{LHAASO2023}.  The highest reconstructed energy is spectrum-dependent;
the robust result is a statistically rich spectrum beyond 10 TeV.

A higher-energy event was found by Carpet at the Baksan Neutrino Observatory
of the Institute for Nuclear Research of the Russian Academy of Sciences.
Carpet-2 reported a photon-like shower near 251 TeV \cite{Carpet-ATel}. A reconstruction using the new Carpet-3 detector data gave $E_{\gamma}=300^{+43}_{-38}$~TeV and strengthened the photon-like classification \cite{Carpet2025}.  The event arrived 4536 s after the GRB trigger, when the source was already leaving the LHAASO field of view.  This single-event nature must remain explicit in any interpretation: it is a potentially extraordinary datum, not a high-statistics spectral bin in the usual sense.

The two energy regimes probe qualitatively different absorbers.  For $E\sim10$--20 TeV and $z=0.151$, the optical depth depends significantly on the optical/infrared EBL; even then, standard models predict strong attenuation, and deabsorption requires a hard intrinsic spectrum.  At $E\sim300$ TeV, the relevant target field is predominantly the CMB, for which the expected attenuation over a cosmological distance is overwhelming.  EBL normalization cannot rescue such a photon.  The price is statistical rather than astrophysical: the claim rests on one air shower and on its spatial and temporal association with the burst.

The fluence fit of Ref.~\cite{SatuninTroitsky2026} combines the LHAASO Water
Cherenkov Detector Array (WCDA), KM2A, and Carpet-3 data.  It uses Gaussian
likelihoods at high counts and Poisson likelihoods in sparse bins.  WCDA data are in a range where attenuation is small and anchor the intrinsic spectrum before propagation; this prevents the
highest-energy event from determining the fit in isolation.
For a reconstructed-energy bin $j$, the forward prediction has the form
\begin{equation}
 \mu_j=\int \dd E\,{\cal R}_j(E)\,
 \black A_{\rm eff}(E)\black
 \black\frac{\dd\mathcal F_{\rm int}}{\dd E}(E;\bm\eta_s)\black
 P_{\gamma\gamma}(E;\bm\theta_{\rm NP},\bm\eta_p),
 \label{eq:gamma_forward_model}
\end{equation}
where $\mu_j$ is the expected count, ${\cal R}_j$ is the energy-response
kernel, \black $A_{\rm eff}$ is the effective area averaged, where needed, over the
transient observation, and $\dd\mathcal F_{\rm int}/\dd E$ is the intrinsic
time-integrated differential fluence; in contrast, $F_{\rm int}$ in
Eqs.~\eqref{eq:attenuation} and~\eqref{eq:opacity_residual} denotes a
spectral flux. \black $\bm\eta_s$ contains source-spectrum nuisance
parameters, $\bm\theta_{\rm NP}$ contains new-physics parameters, and
$\bm\eta_p$ contains propagation nuisances such as EBL and magnetic fields.
For the Carpet bin, the Poisson factor depends on $\mu_j$ and the estimated
background.


Within the assumptions of that analysis, the ALP fit gives approximately
\begin{equation}
 m_a\simeq5.2\times10^{-7}\ {\rm eV},\qquad
 \gag\simeq6\times10^{-11}\ {\rm GeV}^{-1},
\end{equation}
and improves the fit more strongly than the quadratic subluminal-LIV model, whose preferred scale is $E_{{\rm LIV},2}\sim4\times10^{12}$ GeV \cite{SatuninTroitsky2026}.  The reason is instructive.  The LIV model mainly removes the high-energy suppression and therefore produces a rather direct recovery of the intrinsic spectrum.  Photon--ALP mixing can also reshape the multi-TeV spectrum, including the region around the onset of strong absorption.  In other words, the discriminating information is not only the existence of the Carpet-3 point but the spectral bridge between the WCDA data and that point.

\begin{figure}[!t]
  \centering
  \includegraphics[width=\columnwidth]{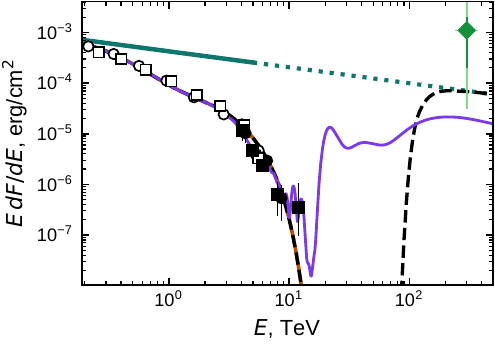}
  \caption{Combined fluence spectrum of GRB~221009A, based on Ref.~\cite{SatuninTroitsky2026}.  Black symbols are LHAASO data and the green diamond is the Carpet-3 event.  The thick solid curve is the intrinsic power-law spectrum (dotted where extrapolated), the orange curve shows the standard EBL+CMB expectation, the lighter violet solid curve illustrates the best-fit photon--ALP scenario, and the dashed black curve presents the best-fit quadratic subluminal-LIV scenario. Both ALP and LIV mechanisms restore transmission at high energy, but they predict different spectral morphology in the 5--100 TeV interval.}
  \label{fig:grbfit}
\end{figure}

It is instructive to compare this fit with Ref.~\cite{GalantiRoncadelli2026}, which finds an ALP-only interpretation of the same observations less favorable and proposes combining ALPs to explain the LHAASO data with LIV to account for the Carpet-3 event. However, Refs.~\cite{Galanti2023,GalantiRoncadelli2026} restrict ALPs to $m_a\simeq10^{-11}$--$10^{-7}$~eV and $\gag\simeq(3$--$5)\times10^{-12}\,\mathrm{GeV}^{-1}$, a region where substantial mixing near $E_\gamma\sim1$~TeV can occur in an intergalactic magnetic field. They adopt $m_a=10^{-10}$~eV and $\gag=4\times10^{-12}\,\mathrm{GeV}^{-1}$ as a benchmark. By contrast, Ref.~\cite{SatuninTroitsky2026} scans a broad ALP parameter range, with preferred values of the coupling over an order of magnitude larger.

At the Carpet energy this difference is amplified by high-energy dispersion.  In the host-galaxy model of Ref.~\cite{Troitsky2024}, the CMB refractive term and QED vacuum polarization remain subdominant to photon--ALP mixing at $E=251$ TeV only in the approximate field interval
\begin{equation}
  1.3\,g_{11}^{-1}\lesssim \frac{B}{\mu\mathrm{G}}
  \lesssim 4.2\,g_{11},
  \qquad
  g_{11}\equiv \frac{\gag}{10^{-11}\,\mathrm{GeV}^{-1}} .
  \label{eq:highEwindow}
\end{equation}
\black Here $B$ denotes the host-field amplitude in that model; its transverse
component $B_T$ enters the mixing Hamiltonian~\eqref{eq:mixingH}. \black
For $g_{11}=6$, Eq.~\eqref{eq:highEwindow} gives
$0.22\lesssim B/(\mu\mathrm{G})\lesssim25$.  However, for $g_{11}=0.4$, it would require simultaneously $B\gtrsim3.3\,\mu$G and $B\lesssim1.7\,\mu$G.  At small $B$, the CMB refractive term suppresses mixing; at large $B$ the QED term, proportional to $B^2E$, does so.  High-energy transparency is therefore not determined by $\gag B$ alone.

\begin{figure}[!t]
  \centering
  \includegraphics[width=\columnwidth]{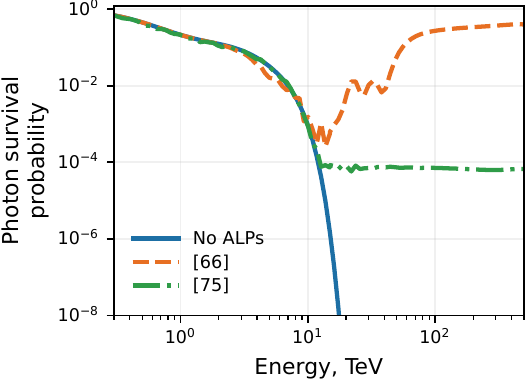}
  \caption{Photon survival probability for the benchmark models of Refs.~\cite{SatuninTroitsky2026,GalantiRoncadelli2026}. Host, Milky-Way and intergalactic fields and all absorption/dispersion terms are identical; only $m_a$ and $\gag$ differ.}
  \label{fig:stgrcommon}
\end{figure}

Figure~\ref{fig:stgrcommon} compares the benchmarks \cite{SatuninTroitsky2026,GalantiRoncadelli2026} in identical fields: the host model of Ref.~\cite{Troitsky2024} at $y_0=0$, $\varphi=80^\circ$, the axisymmetric-spiral (ASS) Galactic field of Ref.~\cite{Pshirkov2011}, and the extragalactic domain model of Ref.~\cite{GalantiRoncadelli2026} with $B_{\rm ext}=1$ nG and domain lengths 0.2--10 Mpc.  Absorption and plasma, QED, and CMB dispersion are also identical; only $(m_a,\gag)$ changes.  The survival probabilities are
\begin{equation}
\begin{array}{c|cc}
 E & \text{$P_{\gamma\gamma}$, Ref.~\cite{SatuninTroitsky2026}} & \text{$P_{\gamma\gamma}$, Ref.~\cite{GalantiRoncadelli2026}} \\ \hline
 15~\mathrm{TeV}  & 6.4\times10^{-4} & 8.5\times10^{-5} \\
 300~\mathrm{TeV} & 3.7\times10^{-1} & 6.3\times10^{-5} ,
\end{array}
\label{eq:commonfieldnumbers}
\end{equation}
so that the difference at 300 TeV is nearly four orders of magnitude even after the magnetic fields have been made identical.  In the same realization, the host produces an ALP fraction $P_a\simeq0.44$ for parameters \cite{SatuninTroitsky2026} but only $P_a\simeq4.7\times10^{-3}$ for the benchmark \cite{GalantiRoncadelli2026}.  Switching the extragalactic field from 1 nG to zero changes the 300-TeV result by less than $0.2\%$ in this test, consistent with the observation of Ref.~\cite{GalantiRoncadelli2026} that CMB-induced photon dispersion suppresses intergalactic photon--ALP oscillations above a few TeV.

The statistical treatments also differ.  Ref.~\cite{SatuninTroitsky2026} fits the full time-integrated spectrum; the 5--10-TeV KM2A structure contributes appreciably to the ALP preference, while the one-event Carpet bin has limited leverage. Ref.~\cite{GalantiRoncadelli2026} instead require the low-coupling region to yield a sufficient one-day Carpet count.  They find $N_\gamma\lesssim\mathcal O(10^{-4})$, below their $N_\gamma\geq0.0513$ criterion at 95\% CL.

The results address different parameter domains and statistical questions.  A high-coupling ALP solution fits the complete fluence spectrum, whereas the restricted low-coupling benchmark gives an insufficient 300-TeV count.  Magnetic-field uncertainties affect both results but are not relevant for their difference.  Neither constitutes a discovery claim.  The LIV best-fit region of Ref.~\cite{SatuninTroitsky2026} is also in tension with air-shower constraints \cite{Satunin2021,Martynenko2025}.

The arrival of the Carpet-3 event 4536~s after the burst may reflect source processes, chance coincidence or propagation effects such as LIV; connecting it to transparency requires a specified LIV theory \cite{OfengeimPiran2025}.
A robust analysis must separate event classification from source association, constrain the intrinsic spectrum with the least-absorbed data, use a complete transfer model, treat EBL, field and detector uncertainties as nuisance parameters, and test the preferred parameters against independent constraints.  

\subsection{Correlations with BL~Lac type objects at ultra-high energies}
\label{subsec:bllac}

A related test uses correlations between cosmic rays of ultra-high energy ($\sim10^{19}$~eV) and distant BL Lacertae (BL Lac) objects \cite{Gorbunov2004,HiRes:BL}. For a charged primary of charge $Ze$ crossing a turbulent field with root-mean-square strength $B_{\rm rms}$, coherence length $\ell_B$, and distance $D_B$, the small-angle random-walk estimate is
\begin{equation}
 \delta_{\rm rms}\simeq
 \frac{ZeB_{\rm rms}}{E}\sqrt{D_B\ell_B},
 \label{eq:uhecr_deflection}
\end{equation}
where $E$ is the particle energy.  Correlations at the instrumental angular scale suggest that the primaries responsible for approximately 2\% of the events are neutral and originate at cosmological distances. Standard photons would then be strongly attenuated according to Eq.~\eqref{eq:tau}; other neutral Standard Model particles face difficulties of production, propagation or detection.

The Telescope Array Collaboration has recently revisited these correlations with a larger, independent surface-detector data set.  Preliminary results in event samples designed to test the earlier HiRes signal suggest an excess associated with BL Lac objects \cite{KudenkoTA2025}. The result is not yet a final collaboration measurement, and individual surface-detector events are not identified as photons.  It therefore motivates continued analysis but does not establish new propagation physics.

\subsection{What would constitute a robust signal?}
\label{subsec:method}

The central difficulty in using gamma-ray opacity to search for new physics is that attenuation is inferred, not directly observed.  The measured quantity is a source spectrum folded with detector response.  New propagation physics is then inferred from a mismatch among four ingredients: the source emission, the background radiation field, the propagation law, and the detector/event classification.  For this reason, the most valuable future observations are not necessarily the single highest-energy photons.  They are data sets that overconstrain these ingredients.
For diagnostic purposes one may define a deabsorption residual
\begin{equation}
 R_\gamma(E,z)=
 \frac{F_{\rm obs}(E,z)}{F_{\rm int}(E,z)}e^{\tau(E,z)}.
 \label{eq:opacity_residual}
\end{equation}
Standard propagation gives $R_\gamma=1$, but $F_{\rm int}$ is inferred from a
source model.  A single residual is therefore not a model-independent
observable.  A population likelihood can test whether residuals depend on
optical depth after controlling for source class, spectral curvature,
variability and detector selection.

A particularly clean strategy is to span several optical-depth regimes with the same transient or source class.  Below the strong-absorption region, the spectrum constrains the source.  Across $\tau\sim(1$--10), EBL absorption and ALP-induced spectral structure can be tested.  At hundreds of TeV, CMB absorption makes the standard prediction essentially parameter-free but event statistics become sparse.  Simultaneous spectral and temporal coverage from wide-field air-shower arrays, together with pointed Cherenkov observations, is therefore more powerful than either alone.

Different new-physics explanations also require different systematics to improve.  For ALPs, the limiting astrophysics is often magnetism: better Galactic-field reconstructions and, where possible, information about the host environment can turn a broad family of transfer functions into a predictive model.  For LIV, the limiting issue is consistency: the same effective operators must be confronted with gamma-ray attenuation, arrival-time dispersion, polarization/birefringence, photon decay or splitting where applicable, and air-shower development.  A model that solves one propagation plot while failing one of these mandatory cross-checks is not a viable explanation.

Finally, population tests remain essential.  A universal change of pair-production kinematics should recur systematically with energy and redshift.  Photon--ALP conversion instead depends on the magnetic environment and can vary strongly from one line of sight to another.  This difference turns source-to-source scatter from a nuisance into a discriminator.  The coming combination of deeper EBL measurements, long time baselines, wide-field multi-TeV surveys and better-characterized magnetic fields can therefore test not only whether the Universe is unexpectedly transparent, but \emph{why} any departure from standard attenuation would occur.
More specifically, in LIV scenarios that reduce gamma-ray absorption, the energy scale for the recovery of transparency is governed by the modified pair-production kinematics and the universal LIV coefficients, with additional dependence on the target-photon background.  ALP conversion predicts critical
energies tied to $m_a$, $\gag$, plasma density and field strength, together with
line-of-sight scatter and possible oscillatory structure.  Secondary photons
from intergalactic cascades instead correlate with charged-particle deflection,
source activity time and angular extension.  These dependences permit a joint
spectral, temporal and morphological test rather than an opacity-only fit.

\section{Neutrino astronomy and large-volume neutrino telescopes}
\label{sec:neutrino}

Large-volume water- and ice-Cherenkov arrays are continuously operating particle detectors, constructed as neutrino observatories. We consider here three examples  of their use for new-physics searches: direct optical detection of magnetic monopoles; neutrinos from superheavy dark matter (SHDM) decays; and tests of nonstandard neutrino properties with three-flavor propagation of astrophysical neutrinos.  The corresponding signals use the same instruments but different response functions and backgrounds. For broader reviews of high-energy neutrino astrophysics and its source problem, see e.g.\ Refs.~\cite{Troitsky2021UFN,Troitsky2023UFN,Spiering2026}.

Optical modules sample Cherenkov photons from charged particles produced in the
medium.  For a neutrino flux, the expected number of selected events in class
$\black h\black$ is
\begin{equation}
 \black N_h=t_{\rm live}\sum_{\alpha}\sum_{s=\nu, \bar\nu}\int \dd E_\nu\,\dd\Omega\,
 A_{{\rm eff},h}^{\alpha,s}(E_\nu,\Omega)
 \frac{\dd\Phi_{\alpha,s}}{\dd E_\nu\,\dd\Omega},
 \label{eq:telescope_counts}
\end{equation}
where $t_{\rm live}$ is the detector livetime and \black
$A_{{\rm eff},h}^{\alpha,s}$ is the effective area for flavor $\alpha$,
particle type $s$ and event class $h$, retaining the different neutrino and
antineutrino responses. \black  The effective area includes the
interaction probability, Earth absorption, detector geometry, triggering,
reconstruction and selection.  It is therefore specific to a spectrum,
direction and event topology; an instrumented volume cannot by itself be
converted to a particle-flux limit.
For an incident neutrino traversing column density $\mathcal X(\vartheta)$, a leading
attenuation factor is
\begin{equation}
 P_{\rm surv}(E_\nu,\vartheta)=
 \exp[-N_A\sigma_{\nu N}(E_\nu)\mathcal X(\vartheta)],
 \label{eq:earth_absorption}
\end{equation}
where $\vartheta$ is zenith angle, $N_A$ is Avogadro's constant, and
$\sigma_{\nu N}$ is the neutrino--nucleon cross section, which grows with energy: above $E_\nu\sim 100$~TeV, the Earth gradually becomes opaque for neutrinos.  Neutral-current scattering and $\nu_\tau$ regeneration redistribute energy and require numerical transport.  Above the PeV range, this direction dependence  changes the neutrino-flux response essentially.

The same neutrino telescope supports three logically distinct measurements considered below: direct light from a monopole, neutrinos from SHDM decay, and the flavor composition of the astrophysical neutrino flux.  Their backgrounds and response functions differ even though the hardware is common.

\subsection{Magnetic monopoles in large-volume neutrino telescopes}
\label{sec:neutrino:monopoles}

A hypothetical magnetic monopole crossing the instrumented volume would produce light detectable by the same equipment as that from a neutrino interaction.  The detector response separates into two velocity regimes.  Relativistic or mildly relativistic monopoles are detected through direct or indirect Cherenkov emission and, below threshold, through luminescence.  Massive grand-unified-theory (GUT) relics moving at Galactic velocities may instead produce a sequence of baryon decays through the Rubakov mechanism. The two searches require different triggers, time windows and model assumptions.

\subsubsection{Magnetic-monopole models}

Single-valuedness of an electric-charge wave function in a monopole field gives the Dirac quantization condition \cite{Dirac1931}
\begin{equation}
  e \black g_{\cal M}\black = \frac{k\hbar c}{2},
  \qquad k\in\mathbb{Z},
\end{equation}
or, for the smallest magnetic charge,
\begin{equation}
  g_{\rm D}=\frac{e}{2\alpha},
  \qquad
  \frac{g_{\rm D}}{e}\simeq 68.5 ,
  \label{eq:DiracCharge}
\end{equation}
in the particle-physics units. Here, $\black g_{\cal M}\black$ is magnetic charge, $\black g_{\rm D}\black$ is its smallest value, $k$ is an integer number, and $\hbar$ is the reduced Planck constant. The very large effective charge \eqref{eq:DiracCharge} makes a monopole an unusually ionizing particle. 

't~Hooft--Polyakov monopoles are finite-energy topological solitons of spontaneously broken non-Abelian gauge theories \cite{tHooft1974,Polyakov1974}.  Their mass is model-dependent; GUT-scale solutions are commonly discussed at masses of order $10^{14}$--$10^{17}\,\mathrm{GeV}$ or above, while intermediate symmetry-breaking stages can produce substantially lighter ``intermediate-mass'' monopoles.

Topological stability permits a relic monopole population.  Galactic and extragalactic fields accelerate magnetic charge, with a mass-dependent Lorentz factor: light and intermediate-mass monopoles can become relativistic, whereas canonical GUT-scale monopoles remain slow \cite{Wick2003,IceCubeSubRel2025}.  The same acceleration extracts energy from large-scale magnetic fields, leading to the Parker and extended Parker bounds on the cosmic monopole flux \cite{Parker1970,TurnerParkerBogdan1982}.  Direct limits from neutrino telescopes, in contrast, do not require assumptions about Galactic-field evolution. For an isotropic population at a fixed speed, the monopole number density $n_{\cal M}$, speed $v$, and differential flux $\Phi_{\cal M}$ per unit solid angle obey $\Phi_{\cal M}=n_{\cal M}v/(4\pi)$. The Lorentz factor after acceleration in a magnetic field is
\begin{equation}
 \gamma_{\cal M}=1+\frac{\Delta E_{\cal M}}{m_{\cal M}c^2}
 =1+\frac{g_{\cal M}}{m_{\cal M}c^2}
 \int\bm B\cdot\dd\bm\ell,
 \label{eq:monopole_acceleration}
\end{equation}
where $m_{\cal M}$ is the monopole mass and $\Delta E_{\cal M}$ the work done along its trajectory. Field reversals reduce the coherent integral. Thus the speed depends on the mass and magnetic-field history, and flux limits are commonly given as functions of $\beta=v/c$.

\subsubsection{Neutrino telescopes as natural monopole detectors}

Fast-monopole identification by Cherenkov light was proposed in Refs.~\cite{Tompkins1965,Tompkins1965Second}.  Large and comparatively uniform light yield distinguishes monopole's emission from  stochastic radiation of a high-energy muon.

For a particle with electric charge $ze$, the Frank--Tamm photon yield is proportional to $z^{2}$.  For magnetic charge $\black g_{\cal M}\black$ in a dielectric medium, the effective charge is proportional to $\black g_{\cal M}\black n_{\rm ref}$. For one Dirac charge,
\begin{equation}
  \frac{\dd^{2}N_{\gamma}}{\dd x\,\dd\lambda}
  =
  \frac{2\pi\alpha}{\lambda^{2}}
  \left(\frac{g_{\rm D}n_{\rm ref}}{e}\right)^{2}
  \left(1-\frac{1}{\beta^{2}n_{\rm ref}^{2}}\right),
  \label{eq:MonopoleFrankTamm}
\end{equation}
where $N_\gamma$ is the number of photons, $x$ is path length, $\lambda$ is wavelength, and $n_{\rm ref}$ is the refractive index \cite{Tompkins1965}.  Above the threshold $\beta>1/n_{\rm ref}$, water or ice with $n_{\rm ref}\simeq1.33$ gives about $8.3\times10^{3}$ times more \emph{direct} Cherenkov photons per unit length than a relativistic singly charged minimum-ionizing track \cite{BaikalFast2008,IceCubeRel2016}.  This large factor explains why the effective area for monopoles can substantially exceed the geometrical cross-section of the instrumented volume.

Below the threshold for direct monopole Cherenkov emission, several other mechanisms remain available.  Energetic $\delta$-electrons knocked out by the monopole can themselves exceed the electron Cherenkov threshold.  Detailed calculations use the quantum monopole--electron scattering cross section of Ref.~\cite{KazamaYangGoldhaber1977}; in water the indirect-Cherenkov channel becomes useful at monopole velocities of order $\beta\gtrsim0.5$ \cite{BaikalFast2008,IceCubeRel2016}.  At still lower velocities, excitation and subsequent de-excitation of the medium can produce luminescence.  The latter is experimentally attractive because it has no Cherenkov threshold, but its absolute yield is much more dependent on the composition, temperature and microscopic properties of the medium \cite{Pollmann2017}.  The Baikal collaboration discussed all three mechanisms already in the 1990s: direct Cherenkov radiation for relativistic monopoles, $\delta$-electron Cherenkov light at intermediate velocities, and water luminescence at lower velocities \cite{BaikalMonopoles1996}.

Therefore, in the search for relativistic monopoles with a neutrino telescope, the dominant emission mechanism changes with $\beta$, while the experimental observable remains an anomalously bright, geometrically track-like light pattern.

\subsubsection{Slow GUT monopoles and the Rubakov effect}

For a superheavy monopole bound to the Galactic halo, the characteristic velocity is $\beta\sim10^{-3}$; ordinary Cherenkov radiation is then ineffective.  GUT monopoles may instead catalyze baryon-number violation through the Rubakov mechanism \cite{Rubakov1981,Rubakov1982,Callan1982,Callan1983}.  A benchmark channel is
\begin{equation}
  {\cal M}+p\rightarrow {\cal M}+e^{+}+\pi^{0},
  \label{eq:RCbenchmark}
\end{equation}
where ${\cal M}$ is the monopole.  It survives the reaction; the positron and $\pi^0$ initiate an electromagnetic cascade with an energy of order 1~GeV.

The catalysis cross section depends on the GUT model, target nucleus and velocity.  It is conventionally parameterized as
\begin{equation}
 \sigma_{\rm cat}(\beta)
 \sim \frac{\sigma_{0}}{\beta}\,f_{\rm nuc}(\beta),
 \label{eq:cat-xsec}
\end{equation}
where $\sigma_0$ is the normalization and $f_{\rm nuc}(\beta)$ describes low-velocity
and nuclear corrections \cite{ArafuneFukugita1983}.  The Baikal analysis used
$\sigma_{\rm cat}\simeq0.17\sigma_0/\beta^2$ in its relevant velocity range
\cite{BaikalMonopoles1996}; present limits are therefore stated for specified
values of $\sigma_{\rm cat}$ \cite{IceCubeSlow2014}.

A slow monopole produces a line of GeV-scale flashes with a mean separation $\lambda_{\rm cat}=1/(n_{\rm t}\sigma_{\rm cat})$, where $n_{\rm t}$ is the target number density. At $\beta=10^{-3}$, the time needed to cross a kilometre is about 3.3~ms.
The data-acquisition (DAQ) system must therefore retain hit or rate information
on millisecond scales.  Unlike the relativistic search, the conversion of a
null result to a flux limit depends explicitly on $\sigma_{\rm cat}$.
For a visible track segment of length $d_{\rm vis}$, the mean number of
catalyzed decays is
\begin{equation}
 \bar N_{\rm cat}=\frac{d_{\rm vis}}{\lambda_{\rm cat}}
 =n_{\rm t}\sigma_{\rm cat}d_{\rm vis}.
 \label{eq:catalysis_multiplicity}
\end{equation}
If the target is approximately uniform, the separation distribution is
exponential and the probability of no catalysis in the segment is
$e^{-\bar N_{\rm cat}}$.  Detector efficiency additionally depends on whether
individual flashes exceed threshold and whether their space--time sequence is
retained by the trigger. The search of slow monopoles with a neutrino telescope was proposed in Ref.~\cite{Rubakov1983MonopoleDetection} in the context of
the Deep Underwater Muon and Neutrino Detector (DUMAND).

\subsubsection{Comparison and results}
\label{subsec:monopole-light-yield}

\begin{figure*}[t]
  \centering
  \includegraphics[width=\textwidth,trim={35 0 0 0}, clip]{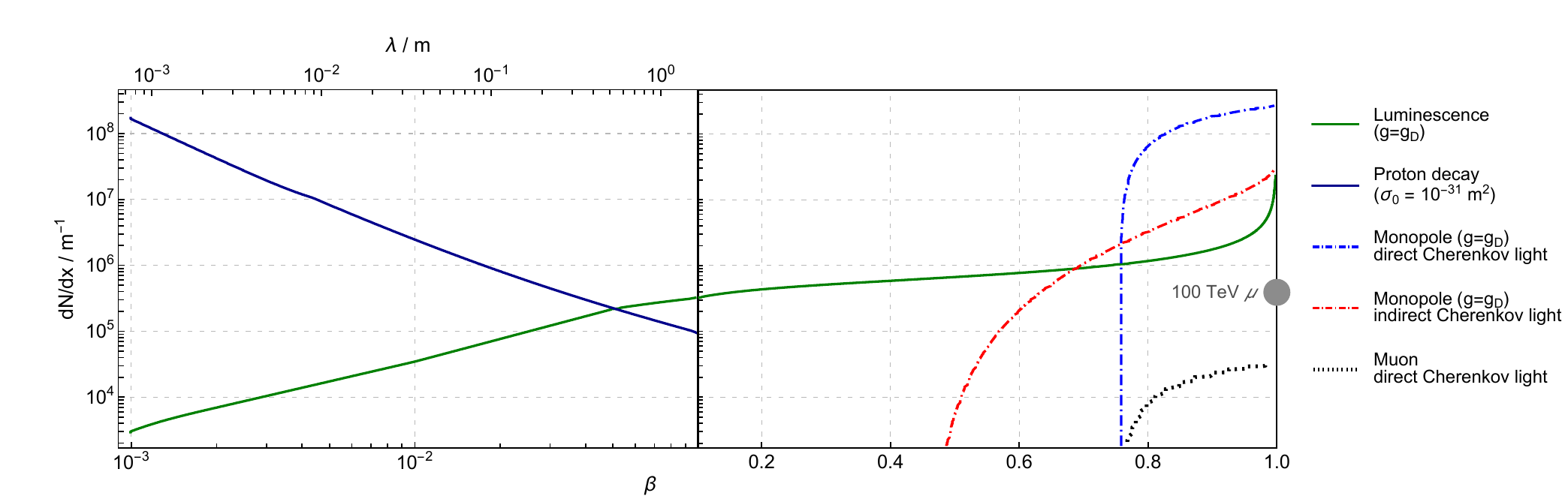}
  \caption{
  Light yield per unit path length for a monopole carrying one Dirac charge.
  The monopole curves and the dotted bare-muon direct-Cherenkov reference are taken from Ref.~\cite{IceCubeSubRel2025}. The contributions to the total light yield include luminescence, light from monopole-catalyzed proton decay through the Rubakov mechanism for the benchmark normalization $\sigma_{0}=10^{-31}\,\mathrm{m}^{2}$, direct monopole Cherenkov radiation, and indirect Cherenkov radiation from energetic $\delta$-electrons. The upper scale represents the characteristic separation of catalyzed decays for the benchmark proton-decay model used in the source figure.  The gray point at $\beta\simeq1$ represents an approximate estimate of the \emph{mean total} light yield of a $100\,\mathrm{TeV}$ muon after including radiative energy losses, see the text.
  }
  \label{fig:monopole-light-yield}
\end{figure*}

Figure~\ref{fig:monopole-light-yield} 
summarizes the relevant light-production mechanisms. The monopole curves are taken from Fig.~1 of the recent IceCube study of sub-relativistic monopoles \cite{IceCubeSubRel2025}.  We have added a gray reference point for a realistic $100\,\mathrm{TeV}$ muon to make the comparison with the ``bare-muon'' Cherenkov curve more physically informative.

The steep blue direct-Cherenkov curve is the most robust part of the figure:
once $\beta>1/n_{\rm ref}$, its normalization follows from the Dirac charge and
electrodynamics.  The indirect-Cherenkov curve extends the sensitivity to lower
velocities because the secondary electrons, rather than the monopole itself,
radiate.  The luminescence curve reaches lower $\beta$ but carries larger
medium-dependent uncertainties.  
The proton-decay curve represents light generation by repeated baryon-number-violating reactions catalyzed by the monopole's GUT core.
The two search regimes are summarized in Table~\ref{tab:fast-slow-monopoles}.
\begin{table*}[t]
\centering
\begin{tabular}{p{0.22\textwidth}p{0.34\textwidth}p{0.34\textwidth}}
\toprule
 & Fast / relativistic monopole & Slow GUT relic \\
\midrule
Typical velocity &
$\beta\gtrsim0.5$; direct Cherenkov for $\beta>1/n_{\rm ref}$ &
$\beta\sim10^{-3}$ (model- and halo-dependent) \\
Dominant optical signal &
Direct Cherenkov; $\delta$-electron Cherenkov; luminescence at lower $\beta$ &
Sequence of cascades from catalyzed baryon decays \\
Characteristic time scale &
Ordinary relativistic-track scale, $\mu$s or less across the detector &
Millisecond scale across a kilometre \\
Main theoretical normalization &
Dirac charge and electrodynamics; comparatively robust above threshold &
Catalysis cross section $\sigma_{\rm cat}$; strongly model-dependent \\
Background issue &
Very bright atmospheric-muon events and stochastic muon energy losses &
Noise-rate excursions and accidental time-extended hit patterns \\
Natural analysis strategy &
Brightness, track geometry, smoothness, direction &
Space--time consistency of repeated flashes along a slow trajectory \\
\bottomrule
\end{tabular}
\caption{Schematic comparison of the two principal optical monopole searches
in large-volume water/ice neutrino telescopes.}
\label{tab:fast-slow-monopoles}
\end{table*}

The dotted curve gives only the direct light of a unit-charge muon.  The gray point includes its mean radiative losses at $E_\mu=100$~TeV.  It was estimated using the PROPOSAL energy-loss calculation and the Geant4 cascade light yield gives
approximately $3.9\times10^5$ photons m$^{-1}$ in the (300--500)-nm band \cite{Alameddine2024Proposal,RadelWiebuschMuon2012,
RadelWiebuschCascade2013}.  This point sets the scale of the optical signal rather than a detector-level prediction: real high-energy muons emit light in stochastic cascades, whereas direct monopole Cherenkov emission is much brighter and smoother. For slow monopoles, the effective area is $A_{\cal M}=A_{\cal M}(\beta,\sigma_{\rm cat})$; for fast ones it is determined mainly by brightness, geometry and event selection.

Early Girlyanda and NT36 arrays in Lake Baikal implemented independent slow-particle triggers based on rate increases over ($0.1$--$1$)~ms \cite{BezrukovBaikal1990,BaikalMonopoles1996}. The later NT200 array also searched for relativistic monopoles; the combined 1998--2003 result was $\Phi_{\cal M}(\beta=1)<4.6\times10^{-17}\,\mathrm{cm^{-2}\,s^{-1}\,sr^{-1}}$ at 90\% CL \cite{BaikalFast2008}.

The eight-year IceCube analysis for $0.750<\beta<0.995$ reached limits below
\begin{equation}
  2\times10^{-19}\,
  \mathrm{cm^{-2}\,s^{-1}\,sr^{-1}}
\end{equation}
over most of that range \cite{IceCubeRel2022}; Astronomy with a Neutrino
Telescope and Abyss environmental RESearch (ANTARES) obtained limits of order
$10^{-18}\,\mathrm{cm^{-2}\,s^{-1}\,sr^{-1}}$ with different water-optics
systematics \cite{ANTARESFullMonopoles2026}.  The published IceCube slow-monopole
limits are approximately ($10^{-18}$--$10^{-17})\,
\mathrm{cm^{-2}\,s^{-1}\,sr^{-1}}$, depending on $\sigma_{\rm cat}$
\cite{IceCubeSlow2014}. 
\subsubsection{Future reach: rate triggers and larger arrays}
\label{subsec:monopole-future}

The slow channel requires millisecond-scale rate information.  A Baikal
Gigaton Volume Detector (Baikal-GVD)
fiber-optic DAQ prototype operated a trigger when the 1-ms counting rate of an
optical module exceeded its mean by four standard deviations; the same method
was tested on a full-size string \cite{BaikalGVDStatus2018,Allakhverdyan2021}.
Rate readout is also applicable to collective signals from a Galactic
core-collapse supernova \cite{Kondratyev2024Supernova}.  The time and spatial
filters differ, but both applications require continuous rate data.  

No full-array Baikal-GVD monopole analysis has yet been published.

The projected sensitivity of the coming IceCube analysis of 12 years of full-detector slow-trigger data is expected \cite{IceCubeSubRel2025} to improve, with respect to the 2014 result, by factors of approximately $10^3$ and $10^2$ at $\beta=10^{-2}$ and $10^{-3}$, respectively, for $\sigma_{\rm cat}=10^{-24}\,\mathrm{cm^2}$.  

IceCube-Gen2 is designed for an instrumented optical volume of about $8\,\mathrm{km^3}$ \cite{IceCubeGen2WhitePaper}.  For sufficiently bright signatures, the monopole sensitivity is expected to scale approximately with volume \cite{IceCubeGen2ExoticsSnowmass}.  The proposed High-energy Underwater Neutrino Telescope (HUNT) would instrument about $30\,\mathrm{km^3}$ \cite{Huang2023HUNT,Chen2026HUNT}.  Neither project has published a full monopole sensitivity.  In the slow channel, volume scaling alone is inadequate: the reach also depends on module spacing, optical noise, $\lambda_{\rm cat}$ and the recorded time window.

\subsection{Superheavy dark matter with high-energy neutrinos}
\label{sec:neutrino:shdm}

By superheavy dark matter (SHDM) one usually understands  non-thermal relics with masses well above
the electroweak scale.  The partial-wave unitarity bound
$m_X\lesssim\mathcal O(10^2)\,\mathrm{TeV}$ applies to conventional thermal freeze-out \cite{GriestKamionkowski1990}, but not to gravitational production at the end of inflation, for which $m_X\sim10^{13}$~GeV is possible
\cite{ChungKolbRiotto1998,KolbChungRiotto1999}.  Metastable relics were also studied as ``top-down'' sources of ultrahigh-energy cosmic rays
\cite{BerezinskyKachelriessVilenkin1997,KuzminRubakov1998,KuzminTkachev1999}.

For a decay into quarks, gauge or Higgs bosons, quantum-chromodynamic (QCD) and electroweak showers generate broad spectra of photons, nucleons and neutrinos \cite{BirkelSarkar1998}.  A two-body channel $X\to\nu\bar\nu$ has an endpoint near $E_\nu=m_X/2$, but radiative corrections still produce a continuum.  At the largest masses, muon- and pion-pair production in electromagnetic cascades also transfers energy to neutrinos \cite{BerghausHooperSimon2025}.
It is convenient to express the injection spectrum in terms of
$x=2E/m_X\leq1$. 
Approximate scaling in $x$ is broken by electroweak logarithms, particle thresholds and propagation.  

\subsubsection{From dark-matter decay to a neutrino flux}

For an SHDM particle $X$ of mass $m_X$, lifetime $\tau_X$, and neutrino yield
$\dd N_\nu/\dd E_\nu$ per decay, the Galactic flux in direction $\hat{\bm n}$ is
\begin{equation}
\frac{\dd\Phi_\nu^{\rm Gal}}{\dd E_\nu\,\dd\Omega}
=
\frac{1}{4\pi m_X\tau_X}
\frac{\dd N_\nu}{\dd E_\nu}
J_{\rm dec}(\hat{\bm n}).
\label{eq:galflux}
\end{equation}
Here, the superscript ``$\mathrm{Gal}$'' denotes the Galactic component, $\Omega$ is solid angle, $\ell$ is line-of-sight distance, $r$ is the Galactocentric radius, and $\rho_X$ is the SHDM mass density.  The decay $J$-factor is defined as
\begin{equation}
J_{\rm dec}(\hat{\bm n})=
\int \dd\ell\,\rho_X[r(\ell,\hat{\bm n})].
\end{equation}
Thus a flux upper limit gives a lower bound on $\tau_X$ for specified $m_X$,
spectrum, abundance and halo profile.

The density enhancement toward the Galactic center produces a broad anisotropy.
Its linear dependence on $\rho_X$, rather than the $\rho_X^2$ dependence of
annihilation, reduces sensitivity to the central cusp and substructure
\cite{EsmailiIbarraPeres2012}.
A common benchmark is the Navarro--Frenk--White profile \cite{NavarroProfile},
\begin{equation}
 \rho_X(r)=\frac{\rho_s}{(r/r_s)(1+r/r_s)^2},
 \label{eq:nfw_profile}
\end{equation}
where $r_s$ and $\rho_s$ are characteristic radius and density parameters. Real galaxies may have smoothed central cores, which affect the inner-Galaxy angular distribution. The full-sky normalization is comparatively insensitive to the details of the halo profile.

The extragalactic component is, schematically,
\begin{equation}
\frac{\dd\Phi_\nu^{\rm EG}}{\black\dd E_\nu\,\dd\Omega\black}
=
\frac{c\,\Omega_{\rm DM}\black\rho_{\rm crit}\black}{4\pi m_X\tau_X}
\int_0^\infty \frac{\dd z}{H(z)}
\frac{\dd N_\nu}{\dd E'_\nu(z)},
\label{eq:egflux}
\end{equation}
where $\Omega_{\rm DM}$ is the cosmological dark-matter density fraction, $\black\rho_{\rm crit}\black$ is the critical energy density, $E'_\nu=(1+z)E_\nu$ is the emission energy.  \black The expression gives the flux per unit solid angle, with the factor $1/(4\pi)$ appropriate to an isotropic extragalactic contribution. This flux is nearly isotropic and redshift-broadened; precision calculations include propagation effects \cite{EsmailiIbarraPeres2012,BerghausHooperSimon2025}.
At extreme energies, scattering on the cosmic neutrino background can no longer be neglected.  The $s$-channel $Z$ resonance occurs \cite{Weiler:Zburst} near 
\begin{equation}
 E_{\rm res}=\frac{M_Z^2}{2m_\nu}
 \simeq4.2\times10^{22}\,\mathrm{eV}
 \left(\frac{0.1\,\mathrm{eV}}{m_\nu}\right),
 \label{eq:z_resonance}
\end{equation}
where $M_Z$ is the $Z$-boson mass and $m_\nu$ the target-neutrino mass.  Away from resonance, electroweak scattering and regeneration still reshape spectra extending to sufficiently high $m_X$.

\subsubsection{Why neutrinos?}

At large $m_X$, the dark-matter number density $n_X=\rho_X/m_X$ is small, making direct production inaccessible and direct detection unlikely.  Rare decays of these particles nevertheless release energies of order $m_X$.

Spectral and directional fits can separate an SHDM contribution from the diffuse astrophysical flux.  Phenomenological fits considered a PeV-scale decay component \cite{RottKohriPark2015}; a dedicated IceCube track-and-cascade analysis found no significant contribution and obtained channel-dependent lifetime limits of order $10^{28}$~s over a broad mass interval \cite{IceCubeDM2018}. Combining the Antarctic Muon And Neutrino Detector Array (AMANDA), IceCube, Pierre Auger Observatory (Auger), and Antarctic Impulsive Transient Antenna (ANITA) results gives characteristic lifetime limits \cite{EsmailiIbarraPeres2012}
\begin{equation}
\tau_X\gtrsim10^{26}\text{--}10^{28}\ {\rm s},
\end{equation}
depending on mass and channel. The combined exposure of all experiments constrains any hard ultra-high-energy neutrino component after specifying its spectrum and halo distribution \cite{CombinedVHE}.

\subsubsection{Baikal-GVD and the multi-PeV window}

A Baikal-GVD cascade analysis using 2018--2024 data searched the interval $10^{15.5}$--$10^{20}$~eV \cite{BaikalGVDMultiPeV2025}.  Its highest-energy event had reconstructed energy $\simeq1200$~TeV, so no events were detected in this energy range. With negligible atmospheric background, the 90\% CL upper limit is $N_{90}\simeq2.3$ signal events.  Baikal-GVD reported differential diffuse limits in overlapping energy decades using an $E^{-1}$ spectrum within each bin.  The lowest ultra-high-energy bin is comparable in order of magnitude to the IceCube constraint \cite{BaikalGVDMultiPeV2025}.

A dedicated SHDM lifetime bound can be derived easily from this result by fixing both the halo profile and the decay channel, the sum of Eqs.~\eqref{eq:galflux} and \eqref{eq:egflux}, and the response fold in Eq.~\eqref{eq:telescope_counts}. The published effective areas permit such a model test, although Ref.~\cite{BaikalGVDMultiPeV2025} reports generic diffuse limits rather than an SHDM lifetime bound.

\subsubsection{Multimessenger constraints}

Generic SHDM fragmentation produces photons, electrons, nucleons and neutrinos. Auger photon, spectrum and composition data strongly constrain hadronic and gauge-boson channels \cite{DasMuraseFujii2023}.  Even for the primary $X\to\nu\bar\nu$ channel, electroweak radiation produces photons, so a nominally neutrino-only decay is not gamma-ray-free \cite{KalashevKuznetsov}.

At large $m_X$, Galactic synchrotron radiation from electron secondaries can dominate the constraint \cite{Deligny2025}.  Conversely, muon- and pion-pair production in electromagnetic cascades regenerates an ultra-high-energy neutrino component \cite{BerghausHooperSimon2025}.  Photon and neutrino limits
must therefore be calculated jointly.
At EeV energies and above, optical arrays are supplemented by radio detectors in ice, mountains and the atmosphere, see e.g.\ Ref.~\cite{CombinedVHE}.  In a background-free search, exposure increases translate approximately linearly into stronger lifetime bounds.

\subsection{Flavor composition as an observable}
\label{sec:flavor}

Many TeV--PeV neutrinos apparently arrive from megaparsec-to-gigaparsec distances, making them useful probes of production and small non-standard propagation effects \cite{Beacom2003Flavor,Ackermann2019Fundamental, Arguelles2023BSMFlavor}. The central observable is the flavor composition, measured as
\begin{equation}
    f_{\alpha} \equiv
    \frac{\Phi_{\nu_\alpha+\bar\nu_\alpha}}
    {\sum_{\beta=e,\mu,\tau}\Phi_{\nu_\beta+\bar\nu_\beta}},
    \qquad
    \sum_{\alpha}f_\alpha=1 .
\end{equation}
Here, $\Phi_{\nu_\alpha+\bar\nu_\alpha}$ is the flux of flavor
$\alpha=e,\mu,\tau$; each flavor flux includes neutrinos and antineutrinos.
An important exception to this experimental degeneracy is the Glashow
resonance,
\begin{equation}
 \bar\nu_e+e^-\to W^-,
 \qquad
 E_{\rm Gl}=\frac{M_W^2}{2m_e}\simeq6.32\ {\rm PeV},
 \label{eq:glashow_resonance}
\end{equation}
where $M_W$ is the $W$-boson mass.  Around $E_{\rm Gl}$, the event rate is sensitive specifically to the electron-antineutrino fraction. Away from such channel-specific features, the summed fractions $f_\alpha$ remain the appropriate observables.

\subsubsection{Standard scenarios}

In canonical astrophysical source models, high-energy neutrinos are produced through charged-pion decay,
\begin{align}
    \pi^+ &\rightarrow \mu^+ + \nu_\mu ,\\
    \mu^+ &\rightarrow e^+ + \nu_e+\bar\nu_\mu ,
\end{align}
and the charge-conjugated chain. Neglecting small corrections, this gives the composition at injection,
\begin{equation}
    (f_e:f_\mu:f_\tau)_{\rm S}
    =
    (1:2:0) .
    \label{eq:pion_source}
\end{equation}
The absence of a significant primary $\nu_\tau$ component is generic for conventional hadronic sources, because production of tau leptons or heavy-flavor states capable of producing energetic $\nu_\tau$ is comparatively suppressed.

Equation~(\ref{eq:pion_source}) should not, however, be regarded as universal. If high-energy muons lose energy before they decay, for example in a strongly magnetized source, the neutrinos from secondary muon decay are suppressed and the source approaches the ``muon-damped'' limit
\begin{equation}
    (f_e:f_\mu:f_\tau)_{\rm S}=(0:1:0).
\end{equation}
Conversely, neutrinos produced predominantly by beta decay of neutrons would initially have approximately
\begin{equation}
    (f_e:f_\mu:f_\tau)_{\rm S}=(1:0:0),
\end{equation}
although this mechanism is generally less efficient at producing neutrinos at the highest energies. The source composition may also evolve with neutrino energy. Thus, flavor is simultaneously a probe of neutrino physics and of the physical conditions in the sources \cite{BustamanteAhlers2019,Song2021Flavor}.

Muon damping is controlled by the competition between decay and radiative cooling.  For a source field $B_s$, the characteristic times are
\begin{equation}
 t_{\rm dec}=\tau_\mu\frac{E_\mu}{m_\mu c^2},
 \qquad
 t_{\rm syn}=\frac{6\pi m_\mu^4c^3}
 {\sigma_{\rm T}m_e^2B_s^2E_\mu},
 \label{eq:muon_cooling_times}
\end{equation}
where $m_\mu$, $\tau_\mu$ and $E_\mu$ are the muon mass, rest-frame lifetime and energy.  Above the energy at which $t_{\rm syn}<t_{\rm dec}$, neutrinos from muon decay are suppressed while the prompt $\nu_\mu$ from pion decay remains.  Source flavor can therefore move continuously from the pion to the muon-damped limit with energy, see e.g.\ \cite{Riabtsev:2022ynm}. 

The flavor and mass eigenstates are related by the Pontecorvo--Maki--Nakagawa--Sakata (PMNS) matrix $U$,
\begin{equation}
    |\nu_\alpha\rangle=
    \sum_i U_{\alpha i}^{*}|\nu_i\rangle .
\end{equation}
For eigenstates $i$ and $j$, the vacuum phase difference is
\begin{equation}
 \Delta\phi_{ij}=\frac{\Delta m_{ij}^2 D}{2E_\nu},
 \qquad \Delta m_{ij}^2=m_i^2-m_j^2,
 \label{eq:cosmic_oscillation_phase}
\end{equation}
where $D$ is propagation distance.  Over astrophysical baselines, the oscillation phase varies by many periods across the finite energy resolution and/or the distribution of source distances, so that the oscillations average to the transition probability from source flavor $\beta$ to detected flavor $\alpha$,
\begin{equation}
    P_{\beta\alpha}
    =
    \sum_{i=1}^{3}
    |U_{\beta i}|^2 |U_{\alpha i}|^2 .
    \label{eq:average_probability}
\end{equation}
The flavor composition observed at Earth is consequently
\begin{equation}
    f_{\alpha,\Earth}
    =
    \sum_{\beta=e,\mu,\tau}
    P_{\beta\alpha}\,
    f_{\beta,{\rm S}} .
    \label{eq:earth_flavor}
\end{equation}
where the subscripts ${\rm S}$ and $\Earth$ denote source and Earth.

For the pion-decay source composition, the result is close to
flavor equipartition, which is often assumed in interpretation of astrophysical data. The precise result depends on PMNS parameters and the source model \cite{Song2021Flavor,BustamanteLiuBarenboim2026}, see below. Accounting for these uncertainties, standard astrophysical sources populate a restricted region of observable flavor space.

\subsubsection{Cosmic baselines and neutrino decay}
\label{sec:decay}

New light states or interactions may permit heavier neutrino mass eigenstates to decay on astrophysical baselines \cite{Beacom2003Decay,Arguelles2023BSMFlavor,Valera2024Decay}.

For mass $m_i$, rest-frame lifetime $\tau_i$, energy $E_\nu$, and distance $D$, the survival factor is approximately
\begin{equation}
    \black P_i^{\rm dec}(E_\nu,D)\black
    =
    \exp\left[
       -\black\frac{D}{c}\black\frac{m_i}{E_\nu\tau_i}
    \right].
    \label{eq:decay_factor}
\end{equation}
Thus propagation constrains $\tau_i/m_i$; \black the factor $D/c$ is the light-travel time, \black and the large $D$ compensates the Lorentz boost.

For complete invisible neutrino decay, only the lightest mass eigenstate survives. In normal ordering (NO), $f_{\alpha,\Earth}=|U_{\alpha1}|^2$; in inverted ordering (IO), $f_{\alpha,\Earth}=|U_{\alpha3}|^2$.  The limits are electron-rich
and electron-poor, respectively \cite{Beacom2003Decay}.
For partial decay, Eq.~\eqref{eq:decay_factor} gives an energy-dependent flavor and spectral distortion; the standard mixture is recovered at high energy.

For cosmological propagation, redshift changes both elapsed proper time and energy.  For a source at $z_s$ and observed energy $E_\nu$,
\begin{equation}
 \black P_i^{\rm dec}(E_\nu,z_s)\black=\exp\left[-\frac{m_i}{\tau_iE_\nu}
 \int_0^{z_s}\frac{\dd z}{(1+z)^2H(z)}\right].
 \label{eq:cosmological_decay}
\end{equation}
A diffuse flux integrates this factor over the source-redshift distribution, smearing the transition.  Spectrum and flavor must therefore be fitted jointly
\cite{Valera2024Decay}.

Pseudo-Dirac oscillations, sterile mixing, LIV and non-standard interactions can also alter flavor \cite{Arguelles2023BSMFlavor}.  Their discrimination requires energy, distance and directional information in addition to an integrated flavor ratio.

\subsubsection{How flavor is measured}
\label{sec:measurement}

Flavor is inferred statistically from event morphologies.  If $\black R_{h\alpha,s}(E_\nu)\black$ is the probability that a selected event of flavor $\alpha$ and particle type $\black s\black$ is classified in topology $\black h\black$, one may write
\begin{equation}
 \black A_{{\rm eff},h}^{\alpha,s}(E_\nu,\Omega)=
 A_{\rm eff}^{\alpha,s}(E_\nu,\Omega)R_{h\alpha,s}(E_\nu).\black
 \label{eq:flavor_response}
\end{equation}
The topology counts in Eq.~\eqref{eq:telescope_counts} constrain the fractions
$f_\alpha$ only after simultaneous fitting of the spectrum, interaction type
and misclassification response.

Charged-current interactions of $\nu_\mu$ frequently produce long muon tracks,
\begin{equation}
    \nu_\mu + N\rightarrow \mu + X_{\rm had} ,
\end{equation}
where $N$ is a target nucleon and $X_{\rm had}$ is the hadronic final state,
and the track sample therefore provides the most direct handle on the muon-flavor component. Electron-neutrino charged-current interactions produce compact cascades, while neutral-current interactions of all flavors are also cascade-like. Most $\nu_\tau$ interactions at moderate energies likewise resemble cascades, leading to an important experimental degeneracy between electron and tau flavor.

At sufficiently high energy, however, the finite lifetime of the tau lepton becomes experimentally useful. Its mean decay length is approximately $d_\tau\simeq49\,\mathrm{m}\,(E_\tau/\mathrm{PeV})$.
A charged-current $\nu_\tau$ interaction may therefore produce two spatially separated showers: one at the neutrino interaction vertex and one when the tau decays. This ``double-cascade'' or ``double-bang'' topology provides a direct way to break the $\nu_e$--$\nu_\tau$ degeneracy \cite{LearnedPakvasa1995}. IceCube reported seven astrophysical $\nu_\tau$ candidates in 9.7 years, with about 0.5 expected background events, rejecting a zero-$\nu_\tau$ component at $5\sigma$ \cite{IceCube2024Tau}.

\begin{figure}[!ht]
    \centering
    \includegraphics[width=\columnwidth,trim=8 8 8 8,clip]{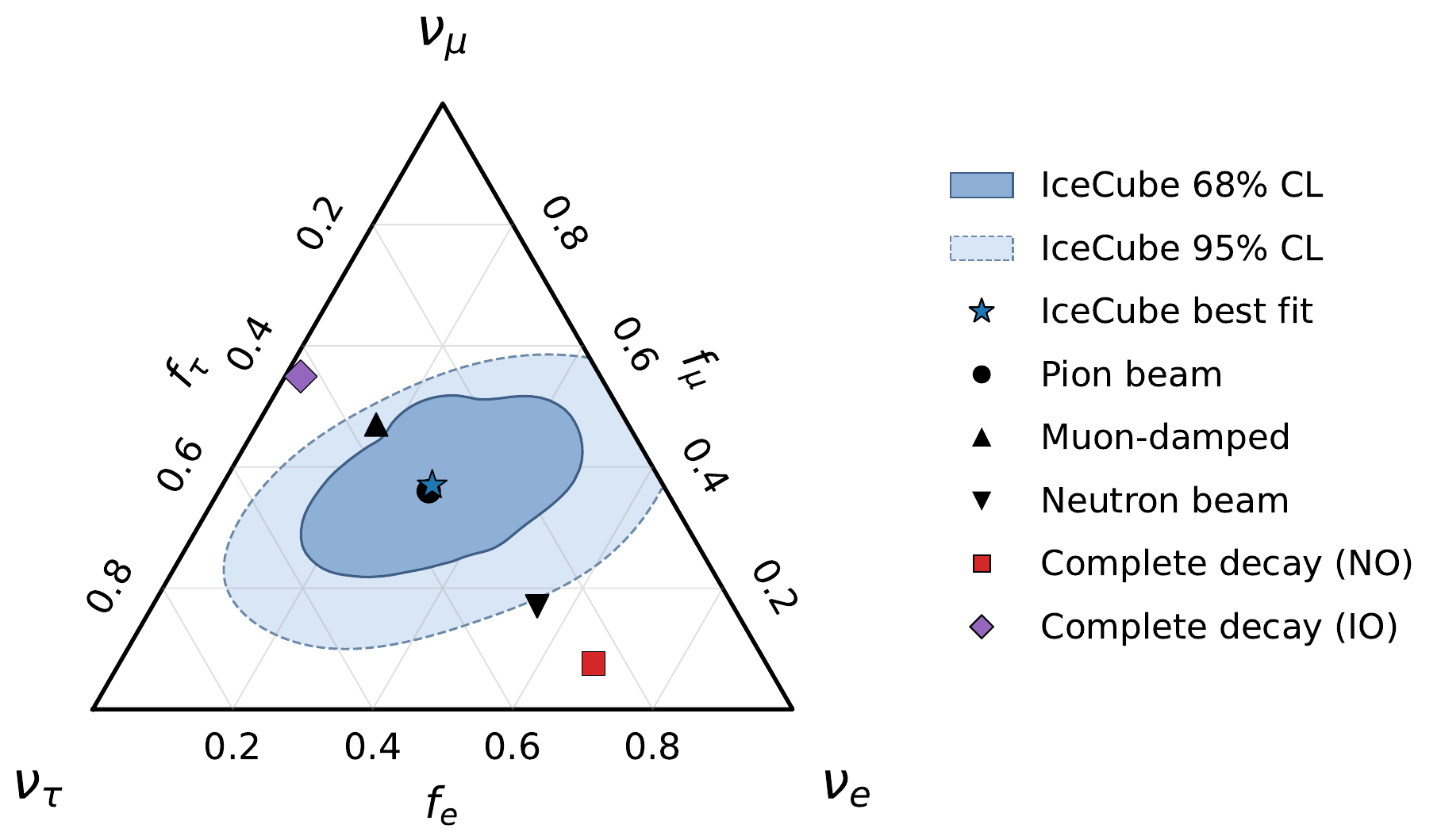}
    \caption{Astrophysical-neutrino flavor composition at Earth.  The star indicates the IceCube best-fit flavor composition; dark and light regions represent the 68\% and 95\% regions from Ref.~\cite{IceCube2026Flavor}.  Black markers show the standard images of pion, muon-damped and neutron-decay sources. The square and diamond indicate the NO and IO complete-decay limits.}
    \label{fig:flavor_triangle}
\end{figure}

\subsubsection{Reading the flavor triangle}
\label{subsec:triangle}

Figure~\ref{fig:flavor_triangle} gives a graphical representation of expected and observed neutrino flavor composition at high energies. Only two flavor fractions are independent because of $\sum_\alpha f_\alpha=1$; the figure therefore uses ternary coordinates.  Its lower-left, lower-right and upper vertices are pure $\nu_\tau$, $\nu_e$ and $\nu_\mu$, respectively; the center is $(1:1:1)$.
For the pion, muon-damped and neutron-decay benchmarks, standard mixing gives
\begin{align}
 (1:2:0)_{\rm S} &\rightarrow (0.30:0.36:0.34)_\Earth,\\
 (0:1:0)_{\rm S} &\rightarrow (0.17:0.47:0.36)_\Earth,\\
 (1:0:0)_{\rm S} &\rightarrow (0.55:0.17:0.28)_\Earth.
\end{align}
Thus mixing compresses the source-flavor domain and demands precise flavor-dependent detector response.

The 11.4-year IceCube analysis \cite{IceCube2026Flavor} gives the best fit of
\begin{equation}
 (f_e:f_\mu:f_\tau)_\Earth=(0.30:0.37:0.33),
\end{equation}
consistent with three-flavor mixing of neutrinos from pion decays at 68\% CL. The muon-damped benchmark remains within 95\%. In the collaboration likelihood, pure neutron-decay production lies outside the 95\% CL contour. The fitted range is approximately 5~TeV--10~PeV, and best-fit IceCube spectrum is assumed.

Both NO and IO complete-decay points, calculated with NuFIT~6.0 \cite{Esteban2024NuFIT6}, lie outside the 95\% CL IceCube contour. Partial decay remains allowed and may be energy-dependent. A recent dedicated analysis of the diffuse IceCube flux \cite{Valera2024Decay} disfavors  lifetime-to-mass ratios below roughly $\tau_i/m_i\sim(20\text{--}450)$~s/eV, depending on the unstable mass eigenstate $i$ and source-population assumptions. The source flavor composition, spectra and redshift distribution are nuisance parameters.

The uncertainties are nevertheless still large on the scale relevant for precision tests. Present data establish consistency with three-flavor mixing and exclude sufficiently extreme alternatives, but they do not yet turn TeV--PeV astrophysical neutrinos into a competitive precision determination of the individual PMNS parameters. The principal limitations are statistics, imperfect flavor identification, and degeneracy with the unknown source composition \cite{BustamanteLiuBarenboim2026}.
\subsubsection{Baikal-GVD and the next generation}
\label{sec:baikal}

Baikal-GVD is not yet a precision three-flavor experiment, and its present cascade results are not be presented as an independent measurement of the flavor triangle. Nevertheless, the experiment is already accumulating exactly the astrophysical event samples needed for such measurements.

Using cascade events recorded from April 2018 to March 2024, Baikal-GVD has detected the diffuse astrophysical neutrino component with a statistical significance of $5.1\sigma$ \cite{BaikalGVD2025Diffuse}. Together with the earlier $3\sigma$ Baikal-GVD result \cite{BaikalAstro2023}, this constitutes an independent experimental confirmation of the discovery of astrophysical neutrinos by IceCube \cite{IceCube-2013,IceCube-HESE-2020}, a discovery recognized by the 2026 Nobel Prize in Physics awarded to Francis Halzen.

The published Baikal-GVD diffuse-flux analysis assumes identical contributions from the three neutrino flavors rather than fitting three independent flavor fractions, so it should not itself be interpreted as a flavor measurement. It does demonstrate, however, that the cascade statistics of Baikal-GVD have entered the regime in which quantitative spectral measurements of the astrophysical component are possible.
Flavor inference combines cascades, muon tracks and observables separating $\nu_e$ from $\nu_\tau$.  Baikal-GVD and the Cubic Kilometre Neutrino Telescope (KM3NeT) operate in water, where lower optical scattering may preserve more topological information than in glacial ice; the quantitative gain is analysis-dependent \cite{BustamanteLiuBarenboim2026}.

Combined IceCube, Baikal-GVD and KM3NeT exposures provide independent media and sky coverage; future multi-km$^3$ instruments will increase the statistics \cite{Song2021Flavor,BustamanteLiuBarenboim2026}. Detailed  projections should however be interpreted with some caution. In the absence of public detector-response calculations for every planned installation, phenomenological forecasts may approximate future telescopes by scaling IceCube event rates with detector size. They therefore establish the scale of the physical opportunity rather than an official collaboration sensitivity for each instrument \cite{BustamanteLiuBarenboim2026}.

The qualitative conclusion is nevertheless robust. Present neutrino telescopes have reached the stage at which all three components of the cosmic-neutrino flux can be constrained experimentally. Increasing the exposure substantially, together with improved $\nu_\tau$ identification, can turn flavor from a consistency test into a precision observable. At that point the cosmic baseline itself becomes an essential experimental resource: standard source neutrinos can test whether the propagation and mixing of TeV--PeV neutrinos over megaparsecs and gigaparsecs follow the same three-flavor physics established in terrestrial experiments.

In this sense, measurements of high-energy astrophysical neutrino flavor provide a conceptually complementary search for new particle physics. Magnetic-monopole searches ask whether an entirely new particle crosses the detector; searches for superheavy dark matter ask whether a new particle elsewhere in the Universe produces neutrinos; flavor measurements instead ask whether apparently ordinary neutrinos behave unusually when propagating across the Universe.

\section{Conclusions}
\label{sec:concl}

Astrophysical tests extend the particle-physics apparatus beyond the detector.  Stellar bounds use the thermodynamic response and evolutionary time of a star; gamma-ray tests use radiation backgrounds and magnetic fields; neutrino-telescope searches use Galactic dark matter, cosmological baselines and kilometre-scale target media.
Large particle numbers, distances and exposures compensate weak microscopic rates.

The inference is indirect.  Its common structure is
\begin{equation}
\begin{gathered}
 \text{particle model}\longrightarrow
 \text{astrophysical response}\\
 \longrightarrow\text{detector observable}\longrightarrow\text{likelihood} .
\end{gathered}
\end{equation}
Each arrow has distinct nuisance parameters.  Controlled limits require a single forward model and, where possible, several observables with different dependencies on those parameters.

No example reviewed here points unambiguously to new physics.  Stellar data constrain additional cooling; GRB~221009A interpretations remain conditional on event association, source spectra and allowed ALP or LIV parameters; monopole and SHDM searches give upper limits; IceCube flavor data agree with averaged three-flavor oscillations.  The resulting constraints cover couplings, masses, lifetimes and symmetry-violation scales inaccessible to direct laboratory tests.

A discovery claim would require stability under alternative source and detector models, the predicted energy, distance or environmental dependence, and consistency with independent laboratory and astrophysical constraints. Increased exposure must therefore be accompanied by improved stellar microphysics, magnetic-field and radiation-background models, and precise 
detector response information.

This work was supported by the Russian Science Foundation, grant 22-12-00253-P. Publicly available large language models assisted in improving the language and style of this paper and in preparation of some of the graphs.
\bibliographystyle{ufn-en}
\bibliography{npa}

@article{Dirac1931,
  author  = {Dirac, P. A. M.},
  title   = {Quantised Singularities in the Electromagnetic Field},
  journal = {Proceedings of the Royal Society of London. Series A},
  volume  = {133},
  number  = {821},
  pages   = {60--72},
  year    = {1931},
  doi     = {10.1098/rspa.1931.0130}
}

@article{tHooft1974,
  author  = {{'t Hooft}, Gerard},
  title   = {Magnetic Monopoles in Unified Gauge Theories},
  journal = {Nuclear Physics B},
  volume  = {79},
  number  = {2},
  pages   = {276--284},
  year    = {1974},
  doi     = {10.1016/0550-3213(74)90486-6}
}

@article{Polyakov1974,
  author  = {Polyakov, A. M.},
  title   = {Particle Spectrum in Quantum Field Theory},
  journal = {JETP Letters},
  volume  = {20},
  pages   = {194--195},
  year    = {1974},
  note    = {Pis'ma Zh. Eksp. Teor. Fiz. 20, 430 (1974)}
}

@article{Tompkins1965,
  author  = {Tompkins, Donald R.},
  title   = {Total Energy Loss and \v{C}erenkov Emission from Monopoles},
  journal = {Physical Review},
  volume  = {138},
  pages   = {B248--B250},
  year    = {1965},
  doi     = {10.1103/PhysRev.138.B248}
}

@article{Tompkins1965Second,
  author  = {Tompkins, Donald R.},
  title   = {Second Method of Identifying Fast Monopoles},
  journal = {Physical Review},
  volume  = {140},
  pages   = {B443--B445},
  year    = {1965},
  doi     = {10.1103/PhysRev.140.B443}
}

@article{Parker1970,
  author  = {Parker, E. N.},
  title   = {The Origin of Magnetic Fields},
  journal = {The Astrophysical Journal},
  volume  = {160},
  pages   = {383--404},
  year    = {1970},
  doi     = {10.1086/150442}
}

@article{TurnerParkerBogdan1982,
  author  = {Turner, Michael S. and Parker, E. N. and Bogdan, T. J.},
  title   = {Magnetic Monopoles and the Survival of Galactic Magnetic Fields},
  journal = {Physical Review D},
  volume  = {26},
  pages   = {1296--1305},
  year    = {1982},
  doi     = {10.1103/PhysRevD.26.1296}
}

@article{Wick2003,
  author  = {Wick, Stuart D. and Kephart, Thomas W. and Weiler, Thomas J. and Biermann, Peter L.},
  title   = {Signatures for a Cosmic Flux of Magnetic Monopoles},
  journal = {Astroparticle Physics},
  volume  = {18},
  pages   = {663--687},
  year    = {2003},
  doi     = {10.1016/S0927-6505(02)00200-1},
  eprint  = {astro-ph/0001233},
  archivePrefix = {arXiv}
}

@article{KazamaYangGoldhaber1977,
  author  = {Kazama, Yoichi and Yang, Chen Ning and Goldhaber, Alfred S.},
  title   = {Scattering of a Dirac Particle with Charge by a Fixed Magnetic Monopole},
  journal = {Physical Review D},
  volume  = {15},
  pages   = {2287--2299},
  year    = {1977},
  doi     = {10.1103/PhysRevD.15.2287}
}

@article{Rubakov1981,
  author  = {Rubakov, V. A.},
  title   = {Superheavy Magnetic Monopoles and Proton Decay},
  journal = {JETP Letters},
  volume  = {33},
  pages   = {644--646},
  year    = {1981},
  note    = {Pis'ma Zh. Eksp. Teor. Fiz. 33, 658 (1981)}
}

@article{Rubakov1982,
  author  = {Rubakov, V. A.},
  title   = {Adler--Bell--Jackiw Anomaly and Fermion Number Breaking in the Presence of a Magnetic Monopole},
  journal = {Nuclear Physics B},
  volume  = {203},
  pages   = {311--348},
  year    = {1982},
  doi           = {10.1016/0550-3213(82)90034-7},
}

@article{Callan1982,
  author  = {Callan, Curtis G., Jr.},
  title   = {Dyon--Fermion Dynamics},
  journal = {Physical Review D},
  volume  = {26},
  pages   = {2058--2068},
  year    = {1982},
  doi     = {10.1103/PhysRevD.26.2058}
}

@article{Callan1983,
  author  = {Callan, Curtis G., Jr.},
  title   = {Monopole Catalysis of Baryon Decay},
  journal = {Nuclear Physics B},
  volume  = {212},
  pages   = {391--400},
  year    = {1983},
  doi           = {10.1016/0550-3213(83)90677-6},
}

@article{ArafuneFukugita1983,
  author  = {Arafune, J. and Fukugita, M.},
  title   = {Velocity-Dependent Factors for the Rubakov Process for Slowly Moving Magnetic Monopoles in Matter},
  journal = {Physical Review Letters},
  volume  = {50},
  pages   = {1901--1904},
  year    = {1983},
  doi     = {10.1103/PhysRevLett.50.1901}
}

@article{BezrukovBaikal1990,
  author  = {Bezrukov, L. B. and others},
  collaboration = {Baikal},
  title   = {Search for Superheavy Magnetic Monopoles in Deep Underwater Experiments at Lake Baikal},
  journal = {Soviet Journal of Nuclear Physics},
  volume  = {52},
  pages   = {54--59},
  year    = {1990},
  note    = {Russian original: Yad. Fiz. 52, 86--95 (1990)}
}

@misc{BaikalMonopoles1996,
  author  = {Bezrukov, L. B. and others},
  collaboration = {Baikal},
  title   = {Search for Magnetic Monopoles with Deep Underwater Cherenkov Detectors at Lake Baikal},
  year    = {1996},
  eprint  = {astro-ph/9601160},
  archivePrefix = {arXiv},
  note    = {Submitted to the Proceedings of the 2nd Workshop on ``The Dark Side of the Universe: Experimental Efforts and Theoretical Framework'', Rome, 1995}
}

@article{BaikalFast2008,
  author  = {Aynutdinov, V. and others},
  collaboration = {Baikal},
  title   = {Search for Relativistic Magnetic Monopoles with the Baikal Neutrino Telescope},
  journal = {Astroparticle Physics},
  volume  = {29},
  pages   = {366--372},
  year    = {2008},
  doi     = {10.1016/j.astropartphys.2008.03.006},
}

@inproceedings{BaikalGVDStatus2018,
  author       = {Avrorin, A. D. and others},
  collaboration = {Baikal-GVD},
  title        = {Baikal-GVD: Status and Prospects},
  booktitle    = {EPJ Web of Conferences},
  volume       = {191},
  pages        = {01006},
  year         = {2018},
  doi          = {10.1051/epjconf/201819101006},
  eprint       = {1808.10353},
  archivePrefix = {arXiv},
  primaryClass = {astro-ph.IM}
}

@article{Allakhverdyan2021,
  author  = {Allakhverdyan, V. A. and others},
  collaboration = {Baikal-GVD},
  title   = {Proposal for Fiber Optic Data Acquisition System for Baikal-GVD},
  journal = {Journal of Instrumentation},
  volume  = {16},
  number  = {11},
  pages   = {C11006},
  year    = {2021},
  doi     = {10.1088/1748-0221/16/11/C11006},
  eprint  = {2107.14183},
  archivePrefix = {arXiv},
  primaryClass = {astro-ph.IM}
}

@article{IceCubeRel2016,
  author  = {Aartsen, M. G. and others},
  collaboration = {IceCube},
  title   = {Searches for Relativistic Magnetic Monopoles in IceCube},
  journal = {European Physical Journal C},
  volume  = {76},
  pages   = {133},
  year    = {2016},
  doi     = {10.1140/epjc/s10052-016-3953-8},
  eprint  = {1511.01350},
  archivePrefix = {arXiv},
  primaryClass = {astro-ph.HE}
}

@article{IceCubeRel2022,
  author  = {Abbasi, R. and others},
  collaboration = {IceCube},
  title   = {Search for Relativistic Magnetic Monopoles with Eight Years of IceCube Data},
  journal = {Physical Review Letters},
  volume  = {128},
  pages   = {051101},
  year    = {2022},
  doi     = {10.1103/PhysRevLett.128.051101},
  eprint  = {2109.13719},
  archivePrefix = {arXiv},
  primaryClass = {astro-ph.HE}
}

@article{IceCubeSlow2014,
  author  = {Aartsen, M. G. and others},
  collaboration = {IceCube},
  title   = {Search for Non-Relativistic Magnetic Monopoles with IceCube},
  journal = {European Physical Journal C},
  volume  = {74},
  pages   = {2938},
  year    = {2014},
  doi     = {10.1140/epjc/s10052-014-2938-8},
  eprint  = {1402.3460},
  archivePrefix = {arXiv},
  note    = {Erratum: Eur. Phys. J. C 79, 124 (2019)}
}

@inproceedings{IceCubeSubRel2025,
  author       = {H{\"a}u{\ss}ler, Jonas and Moller, Nicolas and others},
  collaboration = {IceCube},
  title        = {Search for Sub-Relativistic Magnetic Monopoles with the IceCube Neutrino Observatory},
  booktitle    = {Proceedings of the 39th International Cosmic Ray Conference (ICRC2025)},
  series       = {Proceedings of Science},
  volume       = {501},
  pages        = {492},
  year         = {2025},
  doi          = {10.22323/1.501.0492},
  eprint       = {2507.05896},
  archivePrefix = {arXiv},
  primaryClass = {astro-ph.HE}
}

@article{ANTARESFullMonopoles2026,
  author  = {Albert, A. and others},
  collaboration = {ANTARES},
  title   = {Search for Magnetic Monopoles with the Complete ANTARES Dataset},
  journal = {Journal of High Energy Astrophysics},
  volume  = {50},
  pages   = {100500},
  year    = {2026},
  doi     = {10.1016/j.jheap.2025.100500},
  eprint  = {2505.23929},
  archivePrefix = {arXiv},
  primaryClass = {hep-ex}
}

@inproceedings{Pollmann2017,
  author       = {Obertacke Pollmann, Anna},
  collaboration = {IceCube},
  title        = {Luminescence of Water or Ice as a New Detection Method for Magnetic Monopoles},
  booktitle    = {EPJ Web of Conferences},
  volume       = {164},
  pages        = {07019},
  year         = {2017},
  doi          = {10.1051/epjconf/201716407019},
  eprint       = {1610.06397},
  archivePrefix = {arXiv},
  primaryClass = {astro-ph.IM}
}

@article{RadelWiebuschMuon2012,
  author  = {R{\"a}del, Leif and Wiebusch, Christopher},
  title   = {Calculation of the Cherenkov Light Yield from Low Energetic Secondary Particles Accompanying High-Energy Muons in Ice and Water with Geant4 Simulations},
  journal = {Astroparticle Physics},
  volume  = {38},
  pages   = {53--67},
  year    = {2012},
  doi     = {10.1016/j.astropartphys.2012.09.008},
  eprint  = {1206.5530},
  archivePrefix = {arXiv},
  primaryClass = {astro-ph.IM}
}

@article{RadelWiebuschCascade2013,
  author  = {R{\"a}del, Leif and Wiebusch, Christopher},
  title   = {Calculation of the Cherenkov Light Yield from Electromagnetic Cascades in Ice with Geant4},
  journal = {Astroparticle Physics},
  volume  = {44},
  pages   = {102--113},
  year    = {2013},
  doi     = {10.1016/j.astropartphys.2013.01.015},
  eprint  = {1210.5140},
  archivePrefix = {arXiv},
  primaryClass = {astro-ph.IM}
}

@article{Alameddine2024Proposal,
  author  = {Alameddine, Jean-Marco and Albrecht, Johannes and Dembinski, Hans and Gutjahr, Pascal and Kampert, Karl-Heinz and Rhode, Wolfgang and Sackel, Maximilian and Sandrock, Alexander and Soedingrekso, Jan},
  title   = {Improvements in Charged Lepton and Photon Propagation for the Software PROPOSAL},
  journal = {Computer Physics Communications},
  volume  = {302},
  pages   = {109243},
  year    = {2024},
  doi     = {10.1016/j.cpc.2024.109243},
  eprint  = {2311.13357},
  archivePrefix = {arXiv},
  primaryClass = {astro-ph.IM}
}

@article{Kondratyev2024Supernova,
  author  = {Kondratyev, V. N.},
  title   = {Neutrinos of Magnetorotational Supernovae},
  journal = {International Journal of Modern Physics E},
  volume  = {33},
  pages   = {2441024},
  year    = {2024},
  doi     = {10.1142/S0218301324410246}
}

@article{IceCubeGen2WhitePaper,
  author       = {Aartsen, M. G. and others},
  collaboration = {IceCube-Gen2},
  title        = {IceCube-Gen2: The Window to the Extreme Universe},
  journal      = {Journal of Physics G: Nuclear and Particle Physics},
  volume       = {48},
  pages        = {060501},
  year         = {2021},
  doi          = {10.1088/1361-6471/abbd48},
  eprint       = {2008.04323},
  archivePrefix = {arXiv},
  primaryClass = {astro-ph.HE}
}

@misc{IceCubeGen2ExoticsSnowmass,
  author       = {Pollmann, Anna and Taboada, Ignacio and others},
  collaboration = {IceCube and IceCube-Gen2},
  title        = {Searches for Exotic Particles with the IceCube Neutrino Observatory},
  year         = {2020},
  note         = {Snowmass 2021 Letter of Interest, NF03/NF10/CF01/CF07},
  howpublished = {\url{https://indico.fnal.gov/event/44987/contributions/197050/attachments/134483/166470/IceCube_Exotic_Particles_Snowmass.pdf}}
}

@inproceedings{Huang2023HUNT,
  author    = {Huang, Tian-Qi and Cao, Zhen and Chen, Mingjun and Liu, Jiali and Wang, Zike and You, Xiaohao and Qi, Ying},
  title     = {Proposal for the High Energy Neutrino Telescope},
  booktitle = {Proceedings of the 38th International Cosmic Ray Conference (ICRC2023)},
  series    = {Proceedings of Science},
  volume    = {444},
  pages     = {1080},
  year      = {2023},
  doi       = {10.22323/1.444.1080}
}

@article{Chen2026HUNT,
  author       = {Chen, Mingjun},
  collaboration = {HUNT},
  title        = {HUNT: An Ultra-Large-Scale Neutrino Astronomy Telescope},
  journal      = {Nuclear Instruments and Methods in Physics Research Section A},
  volume       = {1086},
  pages        = {171374},
  year         = {2026},
  doi          = {10.1016/j.nima.2026.171374}
}

@book{Raffelt1996,
  author    = {Georg G. Raffelt},
  title     = {Stars as Laboratories for Fundamental Physics},
  publisher = {University of Chicago Press},
  year      = {1996}
}

@book{Kippenhahn2012,
  author    = {Rudolf Kippenhahn and Alfred Weigert and Achim Weiss},
  title     = {Stellar Structure and Evolution},
  edition   = {2},
  publisher = {Springer},
  year      = {2012},
  doi       = {10.1007/978-3-642-30304-3}
}

@article{CaputoRaffelt2024,
  author        = {Andrea Caputo and Georg Raffelt},
  title         = {Astrophysical Axion Bounds: The 2024 Edition},
  journal       = {PoS},
  volume        = {COSMICWISPers},
  pages         = {041},
  year          = {2024},
  eprint        = {2401.13728},
  archivePrefix = {arXiv},
  primaryClass  = {hep-ph},
  doi           = {10.22323/1.454.0041},
}

@article{HardyLasenby2017,
  author        = {Edward Hardy and Robert Lasenby},
  title         = {Stellar cooling bounds on new light particles: plasma mixing effects},
  journal       = {JHEP},
  volume        = {02},
  pages         = {033},
  year          = {2017},
  eprint        = {1611.05852},
  archivePrefix = {arXiv},
  primaryClass  = {hep-ph},
  doi           = {10.1007/JHEP02(2017)033},
}

@article{DolanDarkPhoton2024,
  author        = {Matthew J. Dolan and Frederick J. Hiskens and Raymond R. Volkas},
  title         = {Constraining dark photons with self-consistent simulations of globular cluster stars},
  journal       = {JCAP},
  volume        = {05},
  pages         = {099},
  year          = {2024},
  eprint        = {2306.13335},
  archivePrefix = {arXiv},
  primaryClass  = {hep-ph},
  doi           = {10.1088/1475-7516/2024/05/099},
}

@article{Serenelli2016,
  author        = {Aldo Serenelli},
  title         = {Alive and well: a short review about standard solar models},
  journal       = {Eur. Phys. J. A},
  volume        = {52},
  pages         = {78},
  year          = {2016},
  eprint        = {1601.07179},
  archivePrefix = {arXiv},
  primaryClass  = {astro-ph.SR},
  doi           = {10.1140/epja/i2016-16078-1},
}

@article{Vinyoles2017,
  author        = {Nuria Vinyoles and Aldo M. Serenelli and Francesco L. Villante and Sarbani Basu and Johannes Bergstrom and M. C. Gonzalez-Garcia and Michele Maltoni and Carlos Pena-Garay and Ningqiang Song},
  title         = {A new generation of standard solar models},
  journal       = {Astrophys. J.},
  volume        = {835},
  pages         = {202},
  year          = {2017},
  eprint        = {1611.09867},
  archivePrefix = {arXiv},
  primaryClass  = {astro-ph.SR},
  doi           = {10.3847/1538-4357/835/2/202},
}

@article{Buldgen2019,
  author        = {Gael Buldgen and Arlette Noels and Sebastien Salmon},
  title         = {Progress in global helioseismology: a new light on the solar modelling problem and its implications for solar-like stars},
  journal       = {Front. Astron. Space Sci.},
  volume        = {6},
  pages         = {42},
  year          = {2019},
  eprint        = {1906.08213},
  archivePrefix = {arXiv},
  primaryClass  = {astro-ph.SR},
  doi           = {10.3389/fspas.2019.00042},
}

@inbook{VillanteSerenelli2020,
    author = "Villante, F. L. and Serenelli, A.",
    editor = "Meyer, Mikko and Zuber, Kai",
    title = "{An updated discussion of the solar abundance problem}",
    booktitle = "{Proceedings, 5th International Solar Neutrino Conference}: {Dresden, Germany, June 11-14, 2018}",
    eprint = "2004.06365",
    archivePrefix = "arXiv",
    primaryClass = "astro-ph.SR",
    doi = "10.1142/9789811204296_0005",
    pages = "103--120",
    year = "2019"
}

@article{Asplund2021,
  author        = {Martin Asplund and Anish M. Amarsi and Nicolas Grevesse},
  title         = {The chemical make-up of the Sun: A 2020 vision},
  journal       = {Astron. Astrophys.},
  volume        = {653},
  pages         = {A141},
  year          = {2021},
  eprint        = {2105.01661},
  archivePrefix = {arXiv},
  primaryClass  = {astro-ph.SR},
  doi           = {10.1051/0004-6361/202140445},
}

@article{Magg2022,
  author        = {Ekaterina Magg and Maria Bergemann and Aldo Serenelli and Manuel Bautista and Bertrand Plez and Ulrike Heiter and Jeffrey M. Gerber and Hans-Gunter Ludwig and Sarbani Basu and others},
  title         = {Observational constraints on the origin of the elements. IV: The standard composition of the Sun},
  journal       = {Astron. Astrophys.},
  volume        = {661},
  pages         = {A140},
  year          = {2022},
  eprint        = {2203.02255},
  archivePrefix = {arXiv},
  primaryClass  = {astro-ph.SR},
  doi           = {10.1051/0004-6361/202142971},
}

@article{Borexino2022,
  author        = {Appel, S. and others},
  collaboration = {Borexino},
  title         = {Improved measurement of solar neutrinos from the Carbon-Nitrogen-Oxygen cycle by Borexino and its implications for the Standard Solar Model},
  journal       = {Phys. Rev. Lett.},
  volume        = {129},
  pages         = {252701},
  year          = {2022},
  eprint        = {2205.15975},
  archivePrefix = {arXiv},
  primaryClass  = {hep-ex},
  doi           = {10.1103/PhysRevLett.129.252701},
}

@article{Borexino2023,
  author        = {Basilico, D. and others},
  collaboration = {Borexino},
  title         = {Final results of Borexino on CNO solar neutrinos},
  journal       = {Phys. Rev. D},
  volume        = {108},
  pages         = {102005},
  year          = {2023},
  eprint        = {2307.14636},
  archivePrefix = {arXiv},
  primaryClass  = {hep-ex},
  doi           = {10.1103/PhysRevD.108.102005},
}

@article{Bezrukov2023,
  author        = {L. B. Bezrukov and I. S. Karpikov and V. V. Sinev},
  title         = {The indication for $^{40}$K geo-antineutrino flux with Borexino phase-III data},
  journal       = {arXiv e-prints},
  pages         = {arXiv:2304.02747},
  year          = {2023},
  eprint        = {2304.02747},
  archivePrefix = {arXiv},
  primaryClass  = {hep-ex}
}

@article{Viaux2013,
  author        = {Nicolas Viaux and Marcio Catelan and Peter B. Stetson and Georg Raffelt and Javier Redondo and Aldo A. R. Valcarce and Achim Weiss},
  title         = {Particle-physics constraints from the globular cluster M5: Neutrino Dipole Moments},
  journal       = {Astron. Astrophys.},
  volume        = {558},
  pages         = {A12},
  year          = {2013},
  eprint        = {1308.4627},
  archivePrefix = {arXiv},
  primaryClass  = {astro-ph.SR},
  doi           = {10.1051/0004-6361/201322004},
}

@article{Straniero2020,
  author        = {O. Straniero and C. Pallanca and E. Dalessandro and I. Dominguez and F. R. Ferraro and M. Giannotti and A. Mirizzi and L. Piersanti},
  title         = {The RGB tip of galactic globular clusters and the revision of the bound of the axion-electron coupling},
  journal       = {Astron. Astrophys.},
  volume        = {644},
  pages         = {A166},
  year          = {2020},
  eprint        = {2010.03833},
  archivePrefix = {arXiv},
  primaryClass  = {astro-ph.SR},
  doi           = {10.1051/0004-6361/202038775},
}

@article{Fung2024,
  author        = {Anthony Fung and Sukannya Heeba and Qianshu Liu and V. Muralidharan and Katelin Schutz and Aaron C. Vincent},
  title         = {New bounds on light millicharged particles from the tip of the red-giant branch},
  journal       = {Phys. Rev. D},
  volume        = {109},
  pages         = {083011},
  year          = {2024},
  eprint        = {2309.06465},
  archivePrefix = {arXiv},
  primaryClass  = {hep-ph},
  doi           = {10.1103/PhysRevD.109.083011},
}

@article{Ayala2014,
  author        = {Adrian Ayala and Inma Dominguez and Maurizio Giannotti and Alessandro Mirizzi and Oscar Straniero},
  title         = {Revisiting the bound on axion-photon coupling from Globular Clusters},
  journal       = {Phys. Rev. Lett.},
  volume        = {113},
  pages         = {191302},
  year          = {2014},
  eprint        = {1406.6053},
  archivePrefix = {arXiv},
  primaryClass  = {astro-ph.SR},
  doi           = {10.1103/PhysRevLett.113.191302},
}

@article{Dolan2022R2,
  author        = {Matthew J. Dolan and Frederick J. Hiskens and Raymond R. Volkas},
  title         = {Advancing Globular Cluster Constraints on the Axion-Photon Coupling},
  journal       = {JCAP},
  volume        = {10},
  pages         = {096},
  year          = {2022},
  eprint        = {2207.03102},
  archivePrefix = {arXiv},
  primaryClass  = {hep-ph},
  doi           = {10.1088/1475-7516/2022/10/096},
}

@article{MillerBertolami2014,
  author        = {Marcelo M. Miller Bertolami and Bruno E. Melendez and Leandro G. Althaus and Jordi Isern},
  title         = {Revisiting the axion bounds from the Galactic white dwarf luminosity function},
  journal       = {JCAP},
  volume        = {10},
  pages         = {069},
  year          = {2014},
  eprint        = {1406.7712},
  archivePrefix = {arXiv},
  primaryClass  = {hep-ph},
  doi           = {10.1088/1475-7516/2014/10/069},
}

@article{Giannotti2016,
  author        = {Maurizio Giannotti and Igor Irastorza and Javier Redondo and Andreas Ringwald},
  title         = {Cool WISPs for stellar cooling excesses},
  journal       = {JCAP},
  volume        = {05},
  pages         = {057},
  year          = {2016},
  eprint        = {1512.08108},
  archivePrefix = {arXiv},
  primaryClass  = {astro-ph.HE},
  doi           = {10.1088/1475-7516/2016/05/057},
}

@article{HegerWoosley2002,
  author        = {Alexander Heger and Stan E. Woosley},
  title         = {The nucleosynthetic signature of population III},
  journal       = {Astrophys. J.},
  volume        = {567},
  pages         = {532--543},
  year          = {2002},
  eprint        = {astro-ph/0107037},
  archivePrefix = {arXiv},
  doi           = {10.1086/338487},
}

@article{Woosley2017,
  author        = {Stan E. Woosley},
  title         = {Pulsational Pair-instability Supernovae},
  journal       = {Astrophys. J.},
  volume        = {836},
  pages         = {244},
  year          = {2017},
  eprint        = {1608.08939},
  archivePrefix = {arXiv},
  primaryClass  = {astro-ph.HE},
  doi           = {10.3847/1538-4357/836/2/244},
}

@article{Gontcharov2026TRGB,
  author        = {G. A. Gontcharov and A. M. Kudashov and O. S. Ryutina and S. V. Troitsky},
  title         = {Isochrone fitting of Galactic globular clusters -- IX. Tip of the red-giant branch for 27 clusters and constraints on new particle physics},
  journal       = {JCAP},
  year          = {2026}, 
  note          = {in press},
  eprint        = {2608.04801},
  archivePrefix = {arXiv},
  primaryClass  = {hep-ph}
}

@article{Gontcharov2026Clusters,
  author        = {G. A. Gontcharov and O. S. Ryutina and C. J. Bonatto and S. S. Savchenko and S. V. Troitsky},
  title         = {Isochrone Fitting of Galactic Globular Clusters -- VIII. Homogeneous estimates of parameters for 27 clusters},
  journal       = {Research in Astronomy and Astrophysics},
  year          = {2026},
  note          = {in press},
  eprint        = {2605.28505},
  archivePrefix = {arXiv},
  primaryClass  = {astro-ph.GA}
}

@article{Paxton2011,
  author        = {Bill Paxton and Lars Bildsten and Aaron Dotter and Falk Herwig and Pierre Lesaffre and Frank X. Timmes},
  title         = {Modules for Experiments in Stellar Astrophysics (MESA)},
  journal       = {Astrophys. J. Suppl.},
  volume        = {192},
  pages         = {3},
  year          = {2011},
  eprint        = {1009.1622},
  archivePrefix = {arXiv},
  primaryClass  = {astro-ph.SR},
  doi           = {10.1088/0067-0049/192/1/3},
}

@article{Jermyn2023,
  author        = {Adam S. Jermyn and Evan B. Bauer and Josiah Schwab and others},
  title         = {Modules for Experiments in Stellar Astrophysics (MESA): Time-Dependent Convection, Energy Conservation, Automatic Differentiation, and Infrastructure},
  journal       = {Astrophys. J. Suppl.},
  volume        = {265},
  pages         = {15},
  year          = {2023},
  eprint        = {2208.03651},
  archivePrefix = {arXiv},
  primaryClass  = {astro-ph.SR},
  doi           = {10.3847/1538-4365/acae8d},
}

@article{GonzalezGarcia2024,
  author        = {M. C. Gonzalez-Garcia and Michele Maltoni and Joao Paulo Pinheiro and Aldo M. Serenelli},
  title         = {Status of direct determination of solar neutrino fluxes after Borexino},
  journal       = {JHEP},
  volume        = {02},
  pages         = {064},
  year          = {2024},
  eprint        = {2311.16226},
  archivePrefix = {arXiv},
  primaryClass  = {hep-ph},
  doi           = {10.1007/JHEP02(2024)064}
}

@article{Rubakov1983MonopoleDetection,
  author  = {Rubakov, V. A.},
  title   = {Structure of the Vacuum in Gauge Theories and
             Monopole Catalysis of Proton Decay},
  journal = {Sov. Phys. Usp.},
  volume  = {26},
  pages   = {1111--1112},
  year    = {1983},
  doi     = {10.1070/PU1983v026n12ABEH004592},
  note    = {Russian original: Usp. Fiz. Nauk 141, 714--715 (1983)}
}

@article{Nikishov1962,
  author = {Nikishov, A. I.},
  title = {Absorption of high-energy photons in the Universe},
  journal = {Sov. Phys. JETP},
  volume = {14},
  pages = {393--394},
  year = {1962},
  note = {[Zh. Eksp. Teor. Fiz. 41, 549 (1961)]}
}

@article{GouldSchreder1967,
  author = {Gould, Robert J. and Schreder, Gerald P.},
  title = {Opacity of the Universe to High-Energy Photons},
  journal = {Phys. Rev.},
  volume = {155},
  pages = {1408--1411},
  year = {1967},
  doi = {10.1103/PhysRev.155.1408}
}

@article{SaldanaLopez2021,
  author = {Saldana-Lopez, A. and Dominguez, A. and Perez-Gonzalez, P. G. and Finke, J. D. and Ajello, M. and Primack, J. R. and Paliya, V. S. and Desai, A.},
  title = {An observational determination of the evolving extragalactic background light from the multiwavelength HST/CANDELS survey in the Fermi and CTA era},
  journal = {Mon. Not. R. Astron. Soc.},
  volume = {507},
  pages = {5144--5160},
  year = {2021},
  doi = {10.1093/mnras/stab2393},
  eprint = {2012.03035},
  archivePrefix = {arXiv}
}

@article{MAGIC2019,
  author = {Acciari, V. A. and others},
  collaboration = {MAGIC, Fermi-LAT},
  title = {Measurement of the extragalactic background light using MAGIC and Fermi-LAT gamma-ray observations of blazars up to z = 1},
  journal = {Mon. Not. R. Astron. Soc.},
  volume = {486},
  pages = {4233--4251},
  year = {2019},
  doi = {10.1093/mnras/stz943},
  eprint = {1904.00134},
  archivePrefix = {arXiv}
}

@article{Banerjee2026,
  author = {Banerjee, Anuvab and Finke, Justin D. and Ajello, Marco and Dominguez, Alberto and Desai, Abhishek and Baxter, Joshua and Hartmann, Dieter and Paliya, Vaidehi S.},
  title = {A New Measurement of the Extragalactic Background Light Using 15 yr of Fermi-Large Area Telescope Data},
  journal = {Astrophys. J.},
  volume = {1003},
  pages = {86},
  year = {2026},
  doi = {10.3847/1538-4357/ae581a},
  eprint = {2604.09139},
  archivePrefix = {arXiv}
}

@article{HornsMeyer2012,
  author = {Horns, Dieter and Meyer, Manuel},
  title = {Indications for a pair-production anomaly from the propagation of VHE gamma-rays},
  journal = {JCAP},
  volume = {02},
  pages = {033},
  year = {2012},
  doi = {10.1088/1475-7516/2012/02/033},
  eprint = {1201.4711},
  archivePrefix = {arXiv}
}

@article{Troitsky2017,
  author = {Troitsky, S. V.},
  title = {Axion-like particles and the propagation of gamma rays over astronomical distances},
  journal = {JETP Lett.},
  volume = {105},
  pages = {55--59},
  year = {2017},
  doi = {10.1134/S0021364017010052},
  eprint = {1612.01864},
  archivePrefix = {arXiv}
}

@article{RaffeltStodolsky1988,
  author = {Raffelt, Georg and Stodolsky, Leo},
  title = {Mixing of the photon with low-mass particles},
  journal = {Phys. Rev. D},
  volume = {37},
  pages = {1237--1249},
  year = {1988},
  doi = {10.1103/PhysRevD.37.1237}
}

@article{GalantiRoncadelli2022,
  author = {Galanti, Giorgio and Roncadelli, Marco},
  title = {Axion-like Particles Implications for High-Energy Astrophysics},
  journal = {Universe},
  volume = {8},
  pages = {253},
  year = {2022},
  doi = {10.3390/universe8050253},
  eprint = {2205.00940},
  archivePrefix = {arXiv}
}

@article{Troitsky2024,
  author = {Troitsky, S. V.},
  title = {Towards a model of photon-axion conversion in the host galaxy of GRB 221009A},
  journal = {JCAP},
  volume = {01},
  pages = {016},
  year = {2024},
  doi = {10.1088/1475-7516/2024/01/016},
  eprint = {2307.08313},
  archivePrefix = {arXiv}
}

@article{LibanovTroitsky2020,
  author = {Libanov, M. and Troitsky, S.},
  title = {On the impact of magnetic-field models in galaxy clusters on constraints on axion-like particles from the lack of irregularities in high-energy spectra of astrophysical sources},
  journal = {Phys. Lett. B},
  volume = {802},
  pages = {135252},
  year = {2020},
  doi = {10.1016/j.physletb.2020.135252},
  eprint = {1908.03084},
  archivePrefix = {arXiv}
}

@article{CAST2024,
  author = {Altenm{\"u}ller, K. and others},
  collaboration = {CAST},
  title = {New Upper Limit on the Axion-Photon Coupling with an Extended CAST Run with a Xe-Based Micromegas Detector},
  journal = {Phys. Rev. Lett.},
  volume = {133},
  pages = {221005},
  year = {2024},
  doi = {10.1103/PhysRevLett.133.221005},
  eprint = {2406.16840},
  archivePrefix = {arXiv}
}

@article{Troitsky2025Stars,
  author = {Troitsky, S. V.},
  title = {Stellar Evolution and ALPs: New Constraints and Hints from Globular Clusters in Gaia DR3 Data},
  journal = {JETP Lett.},
  volume = {121},
  pages = {159--165},
  year = {2025},
  doi = {10.1134/S0021364024603798},
  eprint        = {2410.02266},
  archivePrefix = {arXiv},
}

@article{Dessert2022,
  author = {Dessert, Christopher and Dunsky, David and Safdi, Benjamin R.},
  title = {Upper limit on the axion-photon coupling from magnetic white dwarf polarization},
  journal = {Phys. Rev. D},
  volume = {105},
  pages = {103034},
  year = {2022},
  doi = {10.1103/PhysRevD.105.103034},
  eprint = {2203.04319},
  archivePrefix = {arXiv}
}

@article{Addazi2022,
  author = {Addazi, Andrea and others},
  title = {Quantum gravity phenomenology at the dawn of the multi-messenger era---A review},
  journal = {Prog. Part. Nucl. Phys.},
  volume = {125},
  pages = {103948},
  year = {2022},
  doi = {10.1016/j.ppnp.2022.103948},
  eprint = {2111.05659},
  archivePrefix = {arXiv}
}

@article{Carmona2024,
  author = {Carmona, J. M. and Cortes, J. L. and Rescic, F. and Reyes, M. A. and Terzic, T. and Vrban, F. I.},
  title = {Approaches to photon absorption in a Lorentz invariance violation scenario},
  journal = {Phys. Rev. D},
  volume = {110},
  pages = {063035},
  year = {2024},
  doi = {10.1103/PhysRevD.110.063035},
  eprint = {2404.07842},
  archivePrefix = {arXiv}
}

@article{LHAASOLIV2024,
  author = {Cao, Z. and others},
  collaboration = {LHAASO},
  title = {Stringent Tests of Lorentz Invariance Violation from LHAASO Observations of GRB 221009A},
  journal = {Phys. Rev. Lett.},
  volume = {133},
  pages = {071501},
  year = {2024},
  doi = {10.1103/PhysRevLett.133.071501},
  eprint        = {2402.06009},
  archivePrefix = {arXiv},
}

@article{Rubtsov2012,
  author = {Rubtsov, G. and Satunin, P. and Sibiryakov, S.},
  title = {On calculation of cross sections in Lorentz violating theories},
  journal = {Phys. Rev. D},
  volume = {86},
  pages = {085012},
  year = {2012},
  doi = {10.1103/PhysRevD.86.085012},
  eprint = {1204.5782},
  archivePrefix = {arXiv}
}

@mastersthesis{Vrban2022,
  author = {Vrban, Frane Ivan},
  title = {Pair Production in a Lorentz Invariance Violation Scenario},
  school = {University of Rijeka},
  year = {2022},
  url = {https://repository.phy.uniri.hr/islandora/object/phyri:224}
}

@article{Derbin2014,
  author = {Derbin, A. V.},
  title = {Solar neutrino experiments},
  journal = {Phys. Usp.},
  volume = {57},
  pages = {512--523},
  year = {2014},
  doi = {10.3367/UFNe.0184.201405j.0555},
  note = {Usp. Fiz. Nauk 184, 555--567 (2014)}
}

@article{Sandron2026,
  author        = {Sandron, T. and Buldgen, G. and Noels, A. and Dupret, M. A. and Scuflaire, R.},
  title         = {Investigating the impact of Solar Fusion III reaction rates on helioseismic constraints and solar neutrino fluxes},
  year          = {2026},
  eprint        = {2605.20366},
  archivePrefix = {arXiv},
  primaryClass  = {astro-ph.SR}
}

@article{Satunin2021,
  author = {Satunin, Petr},
  title = {New constraints on Lorentz Invariance violation from Crab Nebula spectrum beyond 100 TeV},
  journal = {Eur. Phys. J. C},
  volume = {81},
  pages = {750},
  year = {2021},
  doi = {10.1140/epjc/s10052-021-09506-w},
  eprint = {2106.06393},
  archivePrefix = {arXiv}
}

@article{Martynenko2025,
  author = {Martynenko, N. S. and Rubtsov, G. I. and Satunin, P. S. and Sharofeev, A. K. and Troitsky, S. V.},
  title = {Hypothetical Lorentz invariance violation and muon content of extensive air showers},
  journal = {Phys. Rev. D},
  volume = {111},
  pages = {063010},
  year = {2025},
  doi = {10.1103/PhysRevD.111.063010},
  eprint = {2412.08349},
  archivePrefix = {arXiv}
}

@article{MyersPospelov2003,
  author = {Myers, Robert C. and Pospelov, Maxim},
  title = {Ultraviolet modifications of dispersion relations in effective field theory},
  journal = {Phys. Rev. Lett.},
  volume = {90},
  pages = {211601},
  year = {2003},
  doi = {10.1103/PhysRevLett.90.211601},
  eprint = {hep-ph/0301124},
  archivePrefix = {arXiv}
}

@article{Goetz2013,
  author = {Goetz, D. and Covino, S. and Fernandez-Soto, A. and Laurent, P. and Bosnjak, Z.},
  title = {The polarized gamma-ray burst GRB 061122},
  journal = {Mon. Not. R. Astron. Soc.},
  volume = {431},
  pages = {3550--3556},
  year = {2013},
  doi = {10.1093/mnras/stt439},
  eprint = {1303.4186},
  archivePrefix = {arXiv}
}

@article{LHAASO2023,
  author = {Cao, Z. and others},
  collaboration = {LHAASO},
  title = {Very high-energy gamma-ray emission beyond 10 TeV from GRB 221009A},
  journal = {Sci. Adv.},
  volume = {9},
  pages = {eadj2778},
  year = {2023},
  doi = {10.1126/sciadv.adj2778},
  eprint = {2310.08845},
  archivePrefix = {arXiv}
}

@article{Carpet2025,
  author = {Dzhappuev, D. D. and Dzaparova, I. M. and Dzhatdoev, T. A. and Gorbacheva, E. A. and Karpikov, I. S. and Khadzhiev, M. M. and Klimenko, N. F. and Kudzhaev, A. U. and Kurenya, A. N. and others},
  collaboration = {Carpet-3},
  title = {Carpet-3 detection of a photonlike air shower with estimated primary energy above 100 TeV in a spatial and temporal coincidence with GRB 221009A},
  journal = {Phys. Rev. D},
  volume = {111},
  pages = {102005},
  year = {2025},
  doi = {10.1103/PhysRevD.111.102005},
  eprint = {2502.02425},
  archivePrefix = {arXiv}
}

@article{SatuninTroitsky2026,
  author = {Satunin, P. S. and Troitsky, S. V.},
  title = {Testing New-Physics Scenarios with the Combined LHAASO and Carpet-3 Fluence Spectrum of GRB 221009A: Axion-Like Particles and Lorentz-Invariance Violation},
  journal = {JETP Lett.},
  volume = {123},
  pages = {73--79},
  year = {2026},
  doi = {10.1134/S0021364025609297},
  eprint = {2510.07234},
  archivePrefix = {arXiv}
}

@article{GalantiRoncadelli2026,
  author = {Galanti, Giorgio and Roncadelli, Marco},
  title = {Lorentz-violating scenarios for the highest-energy photons from GRB 221009A},
  journal = {Phys. Rev. Lett.},
  year = {2026},
  note = {accepted 24 June 2026},
  doi = {10.1103/zjh2-mc47},
  eprint = {2504.01830},
  archivePrefix = {arXiv}
}

@article{OfengeimPiran2025,
  author = {Ofengeim, D. D. and Piran, T.},
  title = {GRB 221009A: Lorentz invariance violation from its 300 TeV photon},
  journal = {Phys. Rev. D},
  volume = {112},
  pages = {083055},
  year = {2025},
  doi = {10.1103/cydc-6j4z},
  eprint = {2508.07153},
  archivePrefix = {arXiv}
}

@article{Gorbunov2004,
  author = {Gorbunov, D. S. and Tinyakov, P. G. and Tkachev, I. I. and Troitsky, S. V.},
  title = {Testing the correlations between ultrahigh-energy cosmic rays and BL Lac type objects with HiRes stereoscopic data},
  journal = {JETP Lett.},
  volume = {80},
  pages = {145--148},
  year = {2004},
  doi = {10.1134/1.1808838},
  eprint = {astro-ph/0406654},
  archivePrefix = {arXiv}
}

@inproceedings{KudenkoTA2025,
  author = {Kudenko, Maria and Kuznetsov, M. and Rubtsov, G. and Troitsky, S. and others},
  collaboration = {Telescope Array},
  title = {Tests of anomalous correlations between ultra-high-energy cosmic rays and BL Lac type objects with the Telescope Array data},
  booktitle = {Proceedings of the 39th International Cosmic Ray Conference (ICRC2025)},
  series = {PoS},
  volume = {501},
  pages = {311},
  year = {2025},
  doi = {10.22323/1.501.0311},
}

@article{Pshirkov2011,
  author = {Pshirkov, M. S. and Tinyakov, P. G. and Kronberg, P. P. and Newton-McGee, K. J.},
  title = {Deriving the Global Structure of the Galactic Magnetic Field from Faraday Rotation Measures of Extragalactic Sources},
  journal = {Astrophys. J.},
  volume = {738},
  pages = {192},
  year = {2011},
  doi = {10.1088/0004-637X/738/2/192},
  eprint = {1103.0814},
  archivePrefix = {arXiv}
}

@article{Galanti2023,
  author = {Galanti, Giorgio and Nava, Lara and Roncadelli, Marco and Tavecchio, Fabrizio and Bonnoli, Giacomo},
  title = {Observability of the Very-High-Energy Emission from GRB 221009A},
  journal = {Phys. Rev. Lett.},
  volume = {131},
  pages = {251001},
  year = {2023},
  doi = {10.1103/PhysRevLett.131.251001},
  eprint = {2210.05659},
  archivePrefix = {arXiv}
}

@article{GriestKamionkowski1990,
  author  = {Griest, Kim and Kamionkowski, Marc},
  title   = {Unitarity Limits on the Mass and Radius of Dark-Matter Particles},
  journal = {Phys. Rev. Lett.},
  volume  = {64},
  pages   = {615--618},
  year    = {1990},
  doi     = {10.1103/PhysRevLett.64.615},
  url     = {https://doi.org/10.1103/PhysRevLett.64.615}
}

@article{BerezinskyKachelriessVilenkin1997,
  author        = {Berezinsky, V. and Kachelriess, M. and Vilenkin, A.},
  title         = {Ultrahigh-Energy Cosmic Rays without {GZK} Cutoff},
  journal       = {Phys. Rev. Lett.},
  volume        = {79},
  pages         = {4302--4305},
  year          = {1997},
  doi           = {10.1103/PhysRevLett.79.4302},
  eprint        = {astro-ph/9708217},
  archivePrefix = {arXiv},
  url           = {https://arxiv.org/abs/astro-ph/9708217}
}

@article{KuzminRubakov1998,
  author        = {Kuzmin, V. A. and Rubakov, V. A.},
  title         = {Ultrahigh-Energy Cosmic Rays: A Window to Postinflationary Reheating Epoch of the Universe?},
  journal       = {Phys. Atom. Nucl.},
  volume        = {61},
  pages         = {1028--1030},
  year          = {1998},
  eprint        = {astro-ph/9709187},
  archivePrefix = {arXiv},
  url           = {https://arxiv.org/abs/astro-ph/9709187}
}

@article{ChungKolbRiotto1998,
  author        = {Chung, Daniel J. H. and Kolb, Edward W. and Riotto, Antonio},
  title         = {Superheavy Dark Matter},
  journal       = {Phys. Rev. D},
  volume        = {59},
  pages         = {023501},
  year          = {1998},
  doi           = {10.1103/PhysRevD.59.023501},
  eprint        = {hep-ph/9802238},
  archivePrefix = {arXiv},
  url           = {https://arxiv.org/abs/hep-ph/9802238}
}

@inproceedings{KolbChungRiotto1999,
  author        = {Kolb, Edward W. and Chung, Daniel J. H. and Riotto, Antonio},
  title         = {{WIMPZILLAS!}},
  booktitle     = {Trends in Theoretical Physics II},
  series        = {AIP Conference Proceedings},
  volume        = {484},
  pages         = {91--105},
  year          = {1999},
  eprint        = {hep-ph/9810361},
  archivePrefix = {arXiv},
  url           = {https://arxiv.org/abs/hep-ph/9810361},
  doi           = {10.1063/1.59655},
}

@article{BirkelSarkar1998,
  author        = {Birkel, Michael and Sarkar, Subir},
  title         = {Extremely High Energy Cosmic Rays from Relic Particle Decays},
  journal       = {Astropart. Phys.},
  volume        = {9},
  pages         = {297--309},
  year          = {1998},
  doi           = {10.1016/S0927-6505(98)00028-0},
  eprint        = {hep-ph/9804285},
  archivePrefix = {arXiv},
  url           = {https://arxiv.org/abs/hep-ph/9804285}
}

@article{KuzminTkachev1999,
  author        = {Kuzmin, V. A. and Tkachev, I. I.},
  title         = {Matter Creation via Vacuum Fluctuations in the Early Universe and Observed Ultrahigh-Energy Cosmic Ray Events},
  journal       = {Phys. Rev. D},
  volume        = {59},
  pages         = {123006},
  year          = {1999},
  doi           = {10.1103/PhysRevD.59.123006},
  eprint        = {hep-ph/9809547},
  archivePrefix = {arXiv},
  url           = {https://arxiv.org/abs/hep-ph/9809547}
}

@article{EsmailiIbarraPeres2012,
  author        = {Esmaili, Arman and Ibarra, Alejandro and Peres, Orlando L. G.},
  title         = {Probing the Stability of Superheavy Dark Matter Particles with High-Energy Neutrinos},
  journal       = {JCAP},
  volume        = {2012},
  number        = {11},
  pages         = {034},
  year          = {2012},
  doi           = {10.1088/1475-7516/2012/11/034},
  eprint        = {1205.5281},
  archivePrefix = {arXiv},
  primaryClass  = {hep-ph},
  url           = {https://arxiv.org/abs/1205.5281}
}

@article{RottKohriPark2015,
  author        = {Rott, Carsten and Kohri, Kazunori and Park, Seong Chan},
  title         = {Superheavy Dark Matter and IceCube Neutrino Signals: Bounds on Decaying Dark Matter},
  journal       = {Phys. Rev. D},
  volume        = {92},
  pages         = {023529},
  year          = {2015},
  doi           = {10.1103/PhysRevD.92.023529},
  eprint        = {1408.4575},
  archivePrefix = {arXiv},
  primaryClass  = {hep-ph},
  url           = {https://arxiv.org/abs/1408.4575}
}

@article{IceCubeDM2018,
  author        = {Aartsen, M. G. and others},
  collaboration = {IceCube},
  title         = {Search for Neutrinos from Decaying Dark Matter with IceCube},
  journal       = {Eur. Phys. J. C},
  volume        = {78},
  pages         = {831},
  year          = {2018},
  doi           = {10.1140/epjc/s10052-018-6273-3},
  eprint        = {1804.03848},
  archivePrefix = {arXiv},
  primaryClass  = {astro-ph.HE},
  url           = {https://arxiv.org/abs/1804.03848}
}

@article{DasMuraseFujii2023,
  author        = {Das, Saikat and Murase, Kohta and Fujii, Toshihiro},
  title         = {Revisiting Ultrahigh-Energy Constraints on Decaying Superheavy Dark Matter},
  journal       = {Phys. Rev. D},
  volume        = {107},
  pages         = {103013},
  year          = {2023},
  doi           = {10.1103/PhysRevD.107.103013},
  eprint        = {2302.02993},
  archivePrefix = {arXiv},
  primaryClass  = {astro-ph.HE},
  url           = {https://arxiv.org/abs/2302.02993}
}

@article{BaikalGVDMultiPeV2025,
  author        = {Allakhverdyan, V. A. and others},
  collaboration = {Baikal-GVD},
  title         = {Constraints on the Diffuse Flux of Multi-PeV Astrophysical Neutrinos Obtained with the Baikal Gigaton Volume Detector},
  journal       = {Phys. Rev. D},
  volume        = {112},
  pages         = {083025},
  year          = {2025},
  eprint        = {2507.05769},
  archivePrefix = {arXiv},
  primaryClass  = {astro-ph.HE},
  url           = {https://arxiv.org/abs/2507.05769},
  doi           = {10.1103/jlz3-26lw},
}

@article{BerghausHooperSimon2025,
  author        = {Berghaus, Kim V. and Hooper, Dan and Simon, Emily R.},
  title         = {Searching for Superheavy Decaying Particles with Ultra-High-Energy Neutrino Observatories},
  journal       = {JCAP},
  volume        = {2025},
  number        = {07},
  pages         = {067},
  year          = {2025},
  eprint        = {2502.12238},
  archivePrefix = {arXiv},
  primaryClass  = {hep-ph},
  url           = {https://arxiv.org/abs/2502.12238},
  doi           = {10.1088/1475-7516/2025/07/067},
}

@article{Deligny2025,
  author        = {Deligny, Olivier},
  title         = {Constraints on Superheavy Dark Matter Decaying into $h\nu$, $Z\nu$ and $W\ell$: Benchmark Example within an Extended Seesaw Framework},
  journal       = {Eur. Phys. J. C},
  doi           = {10.1140/epjc/s10052-025-14736-3},
  volume        = {85},
  pages         = {985},
  year          = {2025},
  eprint        = {2408.17111},
  archivePrefix = {arXiv},
  primaryClass  = {hep-ph},
  url           = {https://arxiv.org/abs/2408.17111}
}

@article{Beacom2003Flavor,
  author        = {Beacom, John F. and Bell, Nicole F. and Hooper, Dan and Pakvasa, Sandip and Weiler, Thomas J.},
  title         = {Measuring Flavor Ratios of High-Energy Astrophysical Neutrinos},
  journal       = {Phys. Rev. D},
  volume        = {68},
  pages         = {093005},
  year          = {2003},
  doi           = {10.1103/PhysRevD.68.093005},
  eprint        = {hep-ph/0307025},
  archivePrefix = {arXiv}
}

@article{Ackermann2019Fundamental,
  author        = {Ackermann, Markus and others},
  title         = {Fundamental Physics with High-Energy Cosmic Neutrinos},
  journal       = {Bull. Am. Astron. Soc.},
  volume        = {51},
  pages         = {215},
  year          = {2019},
  eprint        = {1903.04333},
  archivePrefix = {arXiv},
  primaryClass  = {astro-ph.HE}
}

@article{BustamanteAhlers2019,
  author        = {Bustamante, Mauricio and Ahlers, Markus},
  title         = {Inferring the Flavor of High-Energy Astrophysical Neutrinos at Their Sources},
  journal       = {Phys. Rev. Lett.},
  volume        = {122},
  pages         = {241101},
  year          = {2019},
  doi           = {10.1103/PhysRevLett.122.241101},
  eprint        = {1901.10087},
  archivePrefix = {arXiv},
  primaryClass  = {astro-ph.HE}
}

@article{Song2021Flavor,
  author        = {Song, Ningqiang and Li, Shirley Weishi and Arg{\"u}elles, Carlos A. and Bustamante, Mauricio and Vincent, Aaron C.},
  title         = {The Future of High-Energy Astrophysical Neutrino Flavor Measurements},
  journal       = {JCAP},
  volume        = {2021},
  number        = {04},
  pages         = {054},
  year          = {2021},
  doi           = {10.1088/1475-7516/2021/04/054},
  eprint        = {2012.12893},
  archivePrefix = {arXiv},
  primaryClass  = {hep-ph}
}

@article{Beacom2003Decay,
  author        = {Beacom, John F. and Bell, Nicole F. and Hooper, Dan and Pakvasa, Sandip and Weiler, Thomas J.},
  title         = {Decay of High-Energy Astrophysical Neutrinos},
  journal       = {Phys. Rev. Lett.},
  volume        = {90},
  pages         = {181301},
  year          = {2003},
  doi           = {10.1103/PhysRevLett.90.181301},
  eprint        = {hep-ph/0211305},
  archivePrefix = {arXiv}
}

@article{Arguelles2023BSMFlavor,
  author        = {Arg{\"u}elles, Carlos A. and others},
  title         = {Snowmass White Paper: Beyond the Standard Model Effects on Neutrino Flavor},
  journal       = {Eur. Phys. J. C},
  volume        = {83},
  pages         = {15},
  year          = {2023},
  doi           = {10.1140/epjc/s10052-022-11049-7},
  eprint        = {2203.10811},
  archivePrefix = {arXiv},
  primaryClass  = {hep-ph}
}

@article{Valera2024Decay,
  author        = {Valera, Victor B. and Fiorillo, Damiano F. G. and Esteban, Ivan and Bustamante, Mauricio},
  title         = {New Limits on Neutrino Decay from High-Energy Astrophysical Neutrinos},
  journal       = {Phys. Rev. D},
  volume        = {110},
  pages         = {043004},
  year          = {2024},
  doi           = {10.1103/PhysRevD.110.043004},
  eprint        = {2405.14826},
  archivePrefix = {arXiv},
  primaryClass  = {astro-ph.HE}
}

@article{LearnedPakvasa1995,
  author        = {Learned, John G. and Pakvasa, Sandip},
  title         = {Detecting Tau-Neutrino Oscillations at PeV Energies},
  journal       = {Astropart. Phys.},
  volume        = {3},
  pages         = {267--274},
  year          = {1995},
  eprint        = {hep-ph/9405296},
  archivePrefix = {arXiv},
  doi           = {10.1016/0927-6505(94)00043-3},
}

@article{IceCube2024Tau,
  author        = {Abbasi, R. and others},
  collaboration = {IceCube},
  title         = {Observation of Seven Astrophysical Tau Neutrino Candidates with IceCube},
  journal       = {Phys. Rev. Lett.},
  volume        = {132},
  pages         = {151001},
  year          = {2024},
  doi           = {10.1103/PhysRevLett.132.151001},
  eprint        = {2403.02516},
  archivePrefix = {arXiv},
  primaryClass  = {astro-ph.HE}
}

@article{IceCube2026Flavor,
  author        = {Abbasi, R. and others},
  collaboration = {IceCube},
  title         = {Characterization of the Three-Flavor Composition of Cosmic Neutrinos with IceCube},
  journal       = {Phys. Rev. Lett.},
  year          = {2026},
  note          = {accepted for publication},
  doi           = {10.1103/vntd-m133},
  eprint        = {2510.24957},
  archivePrefix = {arXiv},
  primaryClass  = {astro-ph.HE}
}

@article{BaikalGVD2025Diffuse,
  author        = {Allakhverdyan, V. A. and others},
  collaboration = {Baikal-GVD},
  title         = {Measurement of the Diffuse Astrophysical Neutrino Flux over Six Seasons Using Cascade Events from the Baikal-GVD Expanding Telescope},
  year          = {2025},
  eprint        = {2507.01893},
  archivePrefix = {arXiv},
  primaryClass  = {astro-ph.HE},
  url           = {https://arxiv.org/abs/2507.01893}
}

@misc{BustamanteLiuBarenboim2026,
  author        = {Bustamante, Mauricio and Liu, Qinrui and Barenboim, Gabriela},
  title         = {Measuring Neutrino Mixing above 1 TeV with Astrophysical Neutrinos},
  year          = {2026},
  eprint        = {2602.14308},
  archivePrefix = {arXiv},
  primaryClass  = {hep-ph},
  url           = {https://arxiv.org/abs/2602.14308}
}

@article{Esteban2024NuFIT6,
  author        = {Esteban, Ivan and Gonzalez-Garcia, M. C. and Maltoni, Michele and Martinez-Soler, Ivan and Pinheiro, Jo{\~a}o Paulo and Schwetz, Thomas},
  title         = {{NuFit-6.0}: Updated Global Analysis of Three-Flavor Neutrino Oscillations},
  journal       = {JHEP},
  volume        = {2024},
  number        = {12},
  pages         = {216},
  year          = {2024},
  doi           = {10.1007/JHEP12(2024)216},
  eprint        = {2410.05380},
  archivePrefix = {arXiv},
  primaryClass  = {hep-ph}
}

@article{Troitsky2012UFN,
  author        = {Troitsky, S. V.},
  title         = {Unsolved problems in particle physics},
  journal       = {Phys. Usp.},
  volume        = {55},
  pages         = {72--95},
  year          = {2012},
  doi           = {10.3367/UFNe.0182.201201d.0077},
  eprint        = {1112.4515},
  archivePrefix = {arXiv},
  primaryClass  = {hep-ph}
}

@article{Troitsky2021UFN,
  author        = {Troitsky, S. V.},
  title         = {Constraints on models of the origin of high-energy astrophysical neutrinos},
  journal       = {Phys. Usp.},
  volume        = {64},
  pages         = {1261--1285},
  year          = {2021},
  doi           = {10.3367/UFNe.2021.09.039062},
  eprint        = {2112.09611},
  archivePrefix = {arXiv},
  primaryClass  = {astro-ph.HE}
}

@article{Troitsky2023UFN,
  author        = {Troitsky, S. V.},
  title         = {The origin of high-energy astrophysical neutrinos: new results and prospects},
  journal       = {Phys. Usp.},
  volume        = {67},
  pages         = {349--360},
  year          = {2024},
  doi           = {10.3367/UFNe.2023.04.039581},
  eprint        = {2311.00281},
  archivePrefix = {arXiv},
  primaryClass  = {astro-ph.HE}
}

@article{Hirata1987SN,
  author  = {Hirata, K. and others},
  title   = {Observation of a neutrino burst from the supernova SN1987A},
  journal = {Phys. Rev. Lett.},
  volume  = {58},
  pages   = {1490--1493},
  year    = {1987},
  doi     = {10.1103/PhysRevLett.58.1490}
}

@article{Bionta1987SN,
  author  = {Bionta, R. M. and others},
  title   = {Observation of a neutrino burst in coincidence with supernova 1987A in the Large Magellanic Cloud},
  journal = {Phys. Rev. Lett.},
  volume  = {58},
  pages   = {1494--1496},
  year    = {1987},
  doi     = {10.1103/PhysRevLett.58.1494}
}

@article{Alekseev1987SN,
  author  = {Alekseev, E. N. and Alekseeva, L. N. and Volchenko, V. I. and Krivosheina, I. V.},
  title   = {Possible detection of a neutrino signal on 23 February 1987 at the Baksan Underground Scintillation Telescope of the Institute of Nuclear Research},
  journal = {JETP Lett.},
  volume  = {45},
  pages   = {589--592},
  year    = {1987},
  note    = {Russian original: Pis'ma Zh. Eksp. Teor. Fiz. 45, 461--464 (1987)}
}

@article{breaks,
    author = "Rubtsov, G. I. and Troitsky, S. V.",
    title = "{Breaks in gamma-ray spectra of distant blazars and transparency of the Universe}",
    eprint = "1406.0239",
    archivePrefix = "arXiv",
    primaryClass = "astro-ph.HE",
    reportNumber = "INR-TH-2014-014",
    doi = "10.7868/S0370274X14180015",
    journal = "JETP Lett.",
    volume = "100",
    number = "6",
    pages = "355--359",
    year = "2014"
}

@article{breaks2,
    author = "Korochkin, Alexander and Rubtsov, Grigory and Troitsky, Sergey",
    title = "{Search for anomalous features in gamma-ray blazar spectra corrected for the absorption on the extragalactic background light}",
    eprint = "1810.03443",
    archivePrefix = "arXiv",
    primaryClass = "astro-ph.HE",
    reportNumber = "INR-TH-2018-025",
    doi = "10.1088/1475-7516/2019/12/002",
    journal = "JCAP",
    volume = "12",
    pages = "002",
    year = "2019"
}

@ARTICLE{Carpet-ATel,
       author = {{Dzhappuev}, D.~D. and {Afashokov}, Yu. Z. and {Dzaparova}, I.~M. and {Dzhatdoev}, T.~A. and {Gorbacheva}, E.~A. and {Karpikov}, I.~S. and {Khadzhiev}, M.~M. and {Klimenko}, N.~F. and {Kudzhaev}, A.~U. and {Kurenya}, A.~N. and {Lidvansky}, A.~S. and {Mikhailova}, O.~I. and {Petkov}, V.~B. and {Podlesnyi}, E.~I. and {Pozdnukhov}, N.~A. and {Romanenko}, V.~S. and {Rubtsov}, G.~I. and {Troitsky}, S.~V. and {Unatlokov}, I.~B. and {Vaiman}, I.~A. and {Yanin}, A.~F. and {Zhuravleva}, K.~V.},
        title = "{Swift J1913.1+1946/GRB 221009A: detection of a 250-TeV photon-like air shower by Carpet-2}",
      journal = {The Astronomer's Telegram},
         year = 2022,
        month = oct,
       volume = {15669},
        pages = {1},
       adsurl = {https://ui.adsabs.harvard.edu/abs/2022ATel15669....1D}
}

@article{HiRes:BL,
    author = "Abbasi, R. U. and others",
    collaboration = "HiRes",
    title = "{Search for cross-correlations of ultrahigh-energy cosmic rays with BL Lacertae objects}",
    eprint = "astro-ph/0507120",
    archivePrefix = "arXiv",
    doi = "10.1086/498142",
    journal = "Astrophys. J.",
    volume = "636",
    pages = "680--684",
    year = "2006"
}

@misc{Cowan:2011PCL,
    author = {Cowan, Glen and Cranmer, Kyle and Gross, Eilam and Vitells, Ofer},
    title = {Power-Constrained Limits},
    eprint = {1105.3166},
    archivePrefix = {arXiv},
    primaryClass = {physics.data-an},
    year = {2011}
}

@article{Spiering2026,
    author = "Spiering, Christian",
    title = "{Neutrino Astronomy at High Energies - An Experimental Review}",
    eprint = "2609.26436",
    archivePrefix = "arXiv",
    primaryClass = "astro-ph.HE",
    doi = "10.3390/universe12090273",
    journal = "Universe",
    volume = "2026",
    pages = "12",
    month = "9",
    year = "2026"
}

@article{NavarroProfile,
   title={A Universal Density Profile from Hierarchical Clustering},
   volume={490},
   ISSN={1538-4357},
   url={http://dx.doi.org/10.1086/304888},
   DOI={10.1086/304888},
   number={2},
   journal={Astrophys. J.},
   publisher={American Astronomical Society},
   author={Navarro, Julio F. and Frenk, Carlos S. and White, Simon D. M.},
   year={1997},
   month=dec, pages={493–508},
  eprint        = {astro-ph/9611107},
  archivePrefix = {arXiv},
}

@article{Riabtsev:2022ynm,
    author = "Riabtsev, Kirill and Troitsky, Sergey",
    title = "{Energy-dependent flavor ratios, cascade/track spectrum tension and high-energy neutrinos from magnetospheres of supermassive black holes}",
    eprint = "2204.09339",
    archivePrefix = "arXiv",
    primaryClass = "astro-ph.HE",
    reportNumber = "INR-TH-2022-009",
    doi = "10.1016/j.physletb.2023.137758",
    journal = "Phys. Lett. B",
    volume = "839",
    pages = "137758",
    year = "2023"
}

@article{Ahmad2002SNO,
  author        = {Ahmad, Q. R. and others},
  collaboration = {SNO},
  title         = {Direct Evidence for Neutrino Flavor Transformation
                   from Neutral-Current Interactions in the Sudbury
                   Neutrino Observatory},
  journal       = {Phys. Rev. Lett.},
  volume        = {89},
  pages         = {011301},
  year          = {2002},
  doi           = {10.1103/PhysRevLett.89.011301},
  eprint        = {nucl-ex/0204008},
  archivePrefix = {arXiv}
}

@article{ArnettRosner1987,
  author  = {Arnett, W. David and Rosner, Jonathan L.},
  title   = {Neutrino Mass Limits from SN1987A},
  journal = {Phys. Rev. Lett.},
  volume  = {58},
  pages   = {1906},
  year    = {1987},
  doi     = {10.1103/PhysRevLett.58.1906}
}

@article{RaffeltSeckel1988,
  author  = {Raffelt, Georg and Seckel, David},
  title   = {Bounds on Exotic-Particle Interactions from SN1987A},
  journal = {Phys. Rev. Lett.},
  volume  = {60},
  pages   = {1793--1796},
  year    = {1988},
  doi     = {10.1103/PhysRevLett.60.1793}
}

@article{Aglietta1987LSD,
  author  = {Aglietta, M. and others},
  title   = {On the Event Observed in the Mont Blanc Underground
             Neutrino Observatory during the Occurrence of
             Supernova 1987A},
  journal = {Europhys. Lett.},
  volume  = {3},
  pages   = {1315--1320},
  year    = {1987},
  doi     = {10.1209/0295-5075/3/12/011}
}

@article{Davis1968Solar,
  author  = {Davis, Raymond, Jr. and Harmer, Don S. and Hoffman, Kenneth C.},
  title   = {Search for Neutrinos from the Sun},
  journal = {Phys. Rev. Lett.},
  volume  = {20},
  pages   = {1205--1209},
  year    = {1968},
  doi     = {10.1103/PhysRevLett.20.1205}
}

@article{Hirata1989Solar,
  author  = {Hirata, K. S. and others},
  title   = {Observation of $^{8}${B} Solar Neutrinos in the
             {Kamiokande-II} Detector},
  journal = {Phys. Rev. Lett.},
  volume  = {63},
  pages   = {16--19},
  year    = {1989},
  doi     = {10.1103/PhysRevLett.63.16}
}

@article{Anselmann1992GALLEX,
  author        = {Anselmann, P. and others},
  collaboration = {GALLEX},
  title         = {Solar Neutrinos Observed by {GALLEX} at Gran Sasso},
  journal       = {Phys. Lett. B},
  volume        = {285},
  pages         = {376--389},
  year          = {1992},
  doi           = {10.1016/0370-2693(92)91521-A}
}

@article{Abdurashitov1994SAGE,
  author        = {Abdurashitov, J. N. and others},
  collaboration = {SAGE},
  title         = {Results from {SAGE} (The Russian-American Gallium
                   Solar Neutrino Experiment)},
  journal       = {Phys. Lett. B},
  volume        = {328},
  pages         = {234--248},
  year          = {1994},
  doi           = {10.1016/0370-2693(94)90454-5}
}

@article{FairbairnRashbaTroitsky2011,
    author = "Fairbairn, Malcolm and Rashba, Timur and Troitsky, Sergey V.",
    title = "{Photon-axion mixing and ultra-high-energy cosmic rays from BL Lac type objects - Shining light through the Universe}",
    eprint = "0901.4085",
    archivePrefix = "arXiv",
    primaryClass = "astro-ph.HE",
    doi = "10.1103/PhysRevD.84.125019",
    journal = "Phys. Rev. D",
    volume = "84",
    pages = "125019",
    year = "2011"
}

@article{Weiler:Zburst,
    author = "Weiler, Thomas J.",
    title = "{Resonant Absorption of Cosmic Ray Neutrinos by the Relic Neutrino Background}",
    reportNumber = "SLAC-PUB-2869",
    doi = "10.1103/PhysRevLett.49.234",
    journal = "Phys. Rev. Lett.",
    volume = "49",
    pages = "234",
    year = "1982"
}

@article{CombinedVHE,
    author = "Troitsky, S. V.",
    title = "{Combined constraints on the diffuse flux of cosmic neutrinos between 10{\textasciicircum}16 eV and 10{\textasciicircum}26 eV}",
    eprint = "2609.36632",
    archivePrefix = "arXiv",
    primaryClass = "astro-ph.HE",
    reportNumber = "INR-TH-2026-013",
    month = "9",
    year = "2026",
    url           = {https://arxiv.org/abs/2609.36632}
}

@article{IceCube-2013,
    author = "Aartsen, M. G. and others",
    collaboration = "IceCube",
    title = "{Evidence for High-Energy Extraterrestrial Neutrinos at the IceCube Detector}",
    eprint = "1311.5238",
    archivePrefix = "arXiv",
    primaryClass = "astro-ph.HE",
    doi = "10.1126/science.1242856",
    journal = "Science",
    volume = "342",
    pages = "1242856",
    year = "2013"
}

@article{IceCube-HESE-2020,
    author = "Abbasi, R. U. and others",
    collaboration = "IceCube",
    title = "{The IceCube high-energy starting event sample: Description and flux characterization with 7.5 years of data}",
    eprint = "2011.03545",
    archivePrefix = "arXiv",
    primaryClass = "astro-ph.HE",
    doi = "10.1103/PhysRevD.104.022002",
    journal = "Phys. Rev. D",
    volume = "104",
    pages = "022002",
    year = "2021"
}

@article{BaikalAstro2023,
    author = "Allakhverdyan, V. A. and others",
    collaboration = "Baikal-GVD",
    title = "{Diffuse neutrino flux measurements with the Baikal-GVD neutrino telescope}",
    eprint = "2211.09447",
    archivePrefix = "arXiv",
    primaryClass = "astro-ph.HE",
    reportNumber = "Published: Feb 15, 2023",
    doi = "10.1103/PhysRevD.107.042005",
    journal = "Phys. Rev. D",
    volume = "107",
    number = "4",
    pages = "042005",
    year = "2023"
}

@article{KalashevKuznetsov,
    author = "Kalashev, O. K. and Kuznetsov, M. Yu.",
    title = "{Constraining heavy decaying dark matter with the high energy gamma-ray limits}",
    eprint = "1606.07354",
    archivePrefix = "arXiv",
    primaryClass = "astro-ph.HE",
    reportNumber = "INR-TH-2016-019",
    doi = "10.1103/PhysRevD.94.063535",
    journal = "Phys. Rev. D",
    volume = "94",
    number = "6",
    pages = "063535",
    year = "2016"
}

\end{document}